\documentclass[journal=jacsat,manuscript=article]{achemso}
\usepackage{amssymb}
\usepackage{amsmath}
\usepackage{graphicx}
\usepackage{dcolumn}
\usepackage{bm}

\usepackage{subfigure}
\usepackage{multirow}
\usepackage{tabularx}
\usepackage{natbib}
\usepackage{mathtools}
\usepackage[table,svgnames]{xcolor}

\author{J. A. Reyes-Retana}
\email{angelreyes@fisica.unam.mx}
\affiliation{Departamento de F\'{i}sica-Qu\'{i}mica, Instituto de F\'{i}sica, 
Universidad Nacional Aut\'{o}noma de M\'{e}xico (UNAM), Apartado Postal 20-364, 01000 M\'{e}xico, Distrito Federal, M\'{e}xico}
\affiliation{Departamento de F\'isica y Matem\'aticas, Universidad Iberoamericana, Prolongac\'on Paseo de la Reforma 880, Lomas de Santa Fe, 01219, DF, M\'exico}

\author{G. G. Naumis}
\affiliation{Departamento de F\'{i}sica-Qu\'{i}mica, Instituto de F\'{i}sica, 
Universidad Nacional Aut\'{o}noma de M\'{e}xico (UNAM), Apartado Postal 20-364, 01000 M\'{e}xico, Distrito Federal, M\'{e}xico}

\author{Felipe Cervantes-Sodi}
\affiliation{Departamento de F\'isica y Matem\'aticas, Universidad Iberoamericana, Prolongaci\'on Paseo de la Reforma 880, Lomas de Santa Fe, 01219, DF, M\'exico}

\title{Centered honeycomb NiSe$_2$ nanoribbons, structure and electronic properties}

\begin{document}

\begin{abstract}

Quasi one-dimensional nanoribbons are excellent candidates for nanoelectronics and as electrocatalysts in hydrogen evolution reactions, therefore here we investigate by means of density functional theory the structure and electronic properties of a new kind of 1D ribbons, namely: centered honeycomb NiSe$_2$ nanoribbons. Depending on the crystallography and atomic composition of the edges, these ribbons can belong to one of six (two) zigzag (armchair) families. In the zigzag families, after edge reconstruction, all the bare ribbons are metallic. The influence of edge hydrogen passivation produces band gaps in two of the six families. For the armchair nanoribbons, the geometrical reconstruction leads to semiconductors with small band gap and the hydrogen passivation of the edges increases the band gap up to $\sim$0.6 eV.

\end{abstract}

\section{Introduction}

Common bulk materials present emerging properties when isolated as atom thick 2D crystals \cite{PhysRevB.79.115409,NovoselovPNAS}, having in graphene an undoubtful landmark example\cite{NovoselovPNAS,novoselov2004electric}. However recent development on other 2D materials has shown that feasible electronics involving graphene will include the integration of different materials\cite{britnellScience} among which silicene, germanene\cite{VogtSilicenePRL},  BN honeycomb sheets \cite{NovoselovPNAS,DeanNNano}, III-V binary compounds, metallic oxides and 2D transition-metal dichacogenides (2D--MX$_2$) have attracted attention in the filed, with their existence facilitated by different experimental techniques as mechanical exfoliation\cite{OsadaJMC,NovoselovPNAS}, chemical vapor deposition (CVD)\cite{RuoffScience,LiAdvMat} or liquid exfoliation\cite{ColemanScience,NicolosiScience}. 

The last of the former materials, 2D--MX$_2$, is obtained from transition-metal dichalcogenides\cite{Wilson, Mattheis} which are 3D versatile layered compounds with a wide range of electrical and optical properties of the narrow \textit{d} band type. As graphite, composed by graphene layers, transition-metal dichalcogenides are composed by layers of sandwich-type basic building blocks consisting of a sheet of hexagonal close-packed transition-metal atoms between two sheets of hexagonal close-packed chalcogen atoms\cite{ataca2012stable}. MoS$_2$ and MoSe$_2$ single layers have been successfully isolated attracting much attention due to their direct band gaps of 1.90 eV\cite{mak2010atomically} and 1.55 eV\cite{tongay2012thermally} respectively. Recent studies show that these systems could be promising for novel optoelectronics devices, such as two-dimensional light detectors and emitters\cite{sundaram2012electroluminescence}. 

More importantly, Lukowski\textit{et al.}\cite{Lukowski} and Voiry \textit{et al.}\cite{Voiry} have recently shown that both MoS$_2$ and WS$_2$ exfoliated nanosheets in the strained metallic centered honeycomb T structure, are highly effective as electrocatalists in the hydrogen evolution reaction. This effect is due to the high density of active sites at the edges of the nanosheets in the T structure, representing the first application of the metallic T polymorph of layered metal chaccogenides in catalysts\cite{Lukowski,Voiry}. 

%
%
As for graphene and other 2D materials\cite{novoselov2004electric} that can be cut as quasi one-dimensional (1D) structures in the form of nanoribbons\cite{Han2007PRL, son2006PRL, Nakada1996PRB}, some 2D- MX$_2$ and metal oxides also exist as 1D structures or nanoribbons\cite{botello2009metallic}. Due to quantum confinement effects, particularly at the edges\cite{Nakada1996PRB, son2006PRL},  1D compounds present different properties compared with bulk MX$_2$ and 2D-MX$_2$, and thus, it is important to study the electronic properties of nanoribbons\cite{Lukowksi}.

Some properties of 1D-MX$_2$ have been recently studied. For example, MoS$_2$ nanoribbons and their defects have been explored to give different functionalizations depending on the edge, pressure and electric field\cite{yue2012bandgap, PhysRevB.87.144105, ataca2011mechanical}. Other metallic-chalcogenide nanoribbons have been experimental obtained, such as Bi$_2$Se$_3$\cite{koski2012chemical} and CuTe\cite{she2008template}. 

According to density functional theory (DFT), several of these 2D-MX$_2$ compounds are stable or metastable in one of two possible crystallographic structures; honeycomb (H) and centered honeycomb (T)\cite{podberezskaya2001crystal, ataca2012stable, Ding, YunPRB2012}.

Up to now, most of the studies on 2D-MX$_2$ and their ribbons are on H-structure compounds. However, according to Ataca {\it et al.}      \cite{ataca2012stable}, SeO$_2$, SeS$_2$, ScSe$_2$, ScTe$_2$, TiS$_2$, TiSe$_2$, TiTe$_2$, VS$_2$, VSe$_2$, VTe$_2$, MnO$_2$, MnS$_2$, MnSe$_2$, MnTe$_2$, NiO$_2$, NiS$_2$, NiSe$_2$, NiTe$_2$, NbS$_2$, NbSe$_2$, NbTe$_2$ are stable in the T-structure. In fact, the prediction indicates that some of them do not exist in the commonly reported H-structure but only in the T one. 

To the best of our knowledge, we report the first ever DFT structure and electronic study of 1D-MX$_2$ ribbons with stable T-structure. Here we focus on a representative MX$_2$ compound, the NiSe$_2$. Theoretically, 2D-NiSe$_2$ can occur in both the T and H structure, with the T-structure being $\sim$0.5 eV energetically more favorable\cite{ataca2012stable}. Whereas NiSe$_2$ is a nonmagnetic metal in the H structure, the same compound is a narrow  indirect-band-gap semiconductor in the T-structure\cite{ataca2012stable}, making it more attractive to possible electronic applications. 

Although 1D-T-NiSe$_2$ ribbons have not yet been experimentally produced, other nickel-selenide nanocompounds have. Moloto {\it et al.} had synthesized nickel selenide nanoparticles of different sizes and shapes using a modified solvothermal method\cite{moloto2011synthesis}, also Sobhanbi {\it et al.}  had prepared nanoparticles using simple hydrothermal reduction process\cite{sobhani2012shape}. Hankare {\it et al.} had  deposited Ni-Se thin films using chemical bath method on non-conducting glass substrates in a tartarate bath containing nickel sulphate octahydrate, hydrazine hydrate, sodium seleno-sulphate in an aqueous alkaline medium \cite{hankare2010synthesis}. The direct optical band gap of the thin film was 1.61 eV and electrical resistivity of thin film was in the order of 103 ($\Omega$cm) with a p-type conduction mechanism. Recently, the Qian group \cite{fan2009hydrothermal,zhao2005synthesis} reported the existence of  NiSe$_2$ hexagonal tubular nanocrystals, nanotubes and nanocables using 
hydrothermal growth and procedure that combines self-sacrificing template and hydrothermal methods. The experimental production of 1D-T-NiSe$_2$ or other 1D-T-MX$_2$ ribbons is just a matter of time, thus the relevance of our present work.

The layout of this manuscript is the following.  We start with the reproduction of the centered honeycomb NiSe$_2$ single layer. We demonstrate the stability of ribbons of NiSe$_2$ with widths from 9 to 35\AA. Our ribbons present 6 different edge terminations for zigzag like edges and two for armchair edges. The electronic properties of these 8 families are presented. All these families were hydrogen passivated. For the semiconductor ribbons we present the variation in band gap with the ribbon width.  Electron densities and specific orbitals are also analyzed. 

\section{Methodology}

In this work, we perform {\it ab initio} calculations with the Quantum ESPRESSO\cite{giannozzi2009quantum}  plane wave DFT and density functional perturbation theory (DFPT) code, available under the GNU Public License\cite{GNU}.

Scalar relativistic, spin polarized and non-spin polarized calculation were performed. A plane-wave basis set with kinetic energy of  612 eV was used. Also, a ultrasoft pseudo-potential\cite{PhysRevB.41.7892} from the standard distribution generated using a modified RRKJ\cite{ PhysRevB.41.1227} approach , and the generalized gradient approximation \cite{PhysRevB.46.6671} (GGA) for the exchange-correlation functional in its PBE parametrization \cite{PhysRevLett.77.3865} was used.

Calculations were done for 2D structures and both nanoribbons, bare and hydrogen passivated. All atomic positions and lattice parameters were optimized using the conjugate gradient method. The convergence for energy was chosen as 10$^{-7}$ eV between two consecutive steps and the maximum forces acting are smaller than 0.05 eV/\AA. The stress in the periodic direction is lower than 0.01GPa in all cases.

Rectangular supercells were build for the zigzag and armchair nanoribbons, periodic in the $x-$direction, with lattice parameters of $a$=3.51 and $a$=6.08 \AA\ respectively.
To simulate isolated ribbons, the inplane and perpendicular distances between ribbons in adjacent supercells have to be larger than 10\AA.
In the case of ribbons, the cell optimizations were just in the x-direction.
The Brillouin zones of the GNR unit cells are sampled by Monkhorst-Pack\cite{Monkhorst} grids of the form $16\times16\times1$,  in the 2D structures and $16\times1\times1$ for the 1D structures.

Depending on their width, ribbons present different edge terminations, defining ribbon's \textit{families}. Six possible edge terminations are found for the zigzag ribbons and two for the armchair, therefore here we studied 6 zigzag families and 2 armchair families. All ribbons were also hydrogen passivated. 

\begin{figure}
 \subfigure[T-NiSe$_2$ structure]{\label{Fig1a}\includegraphics[width=0.20\textwidth,height=0.15\textheight]{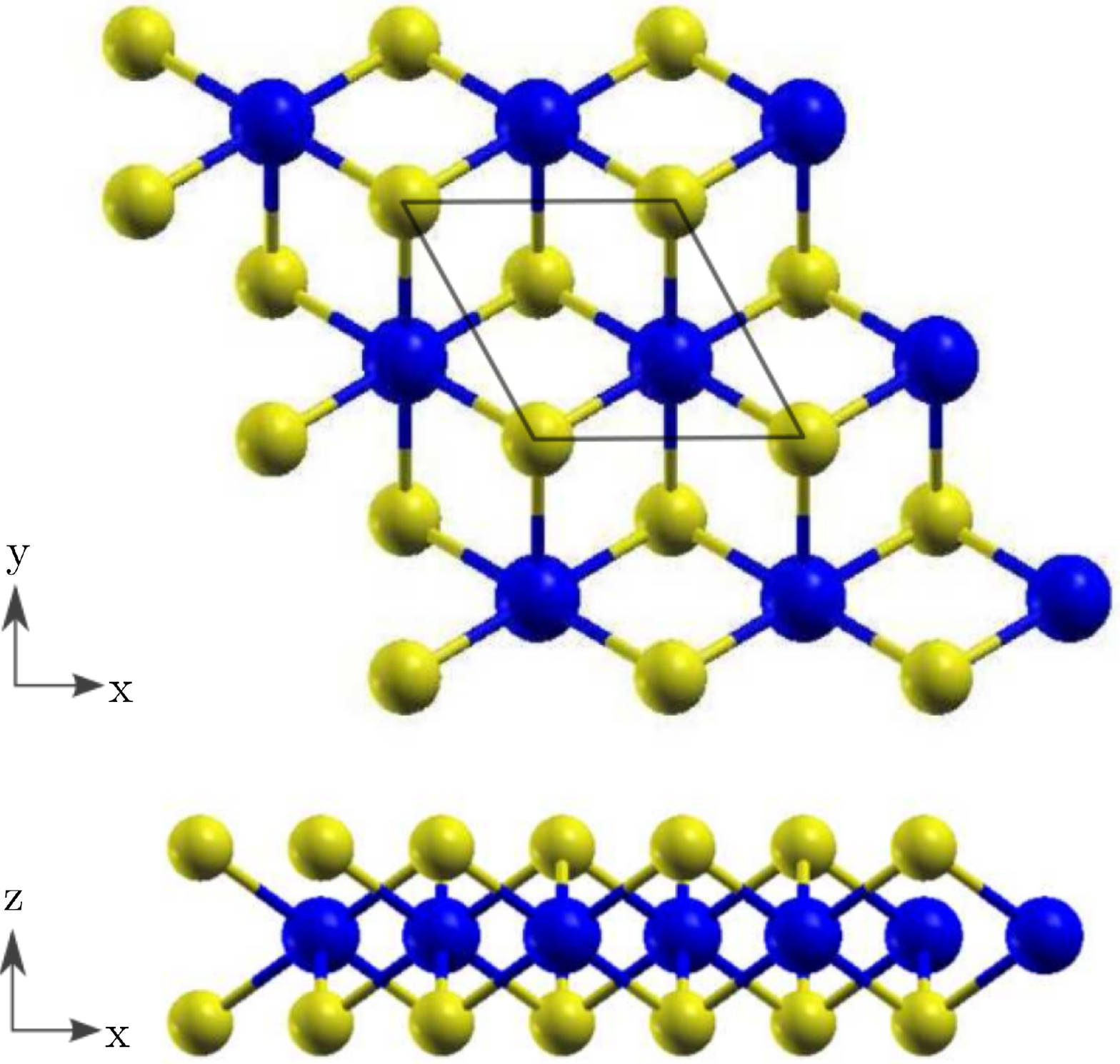}}
 \subfigure[Zigzag ribbon]{\label{Fig1b}\includegraphics[width=0.24\textwidth,height=0.15\textheight]{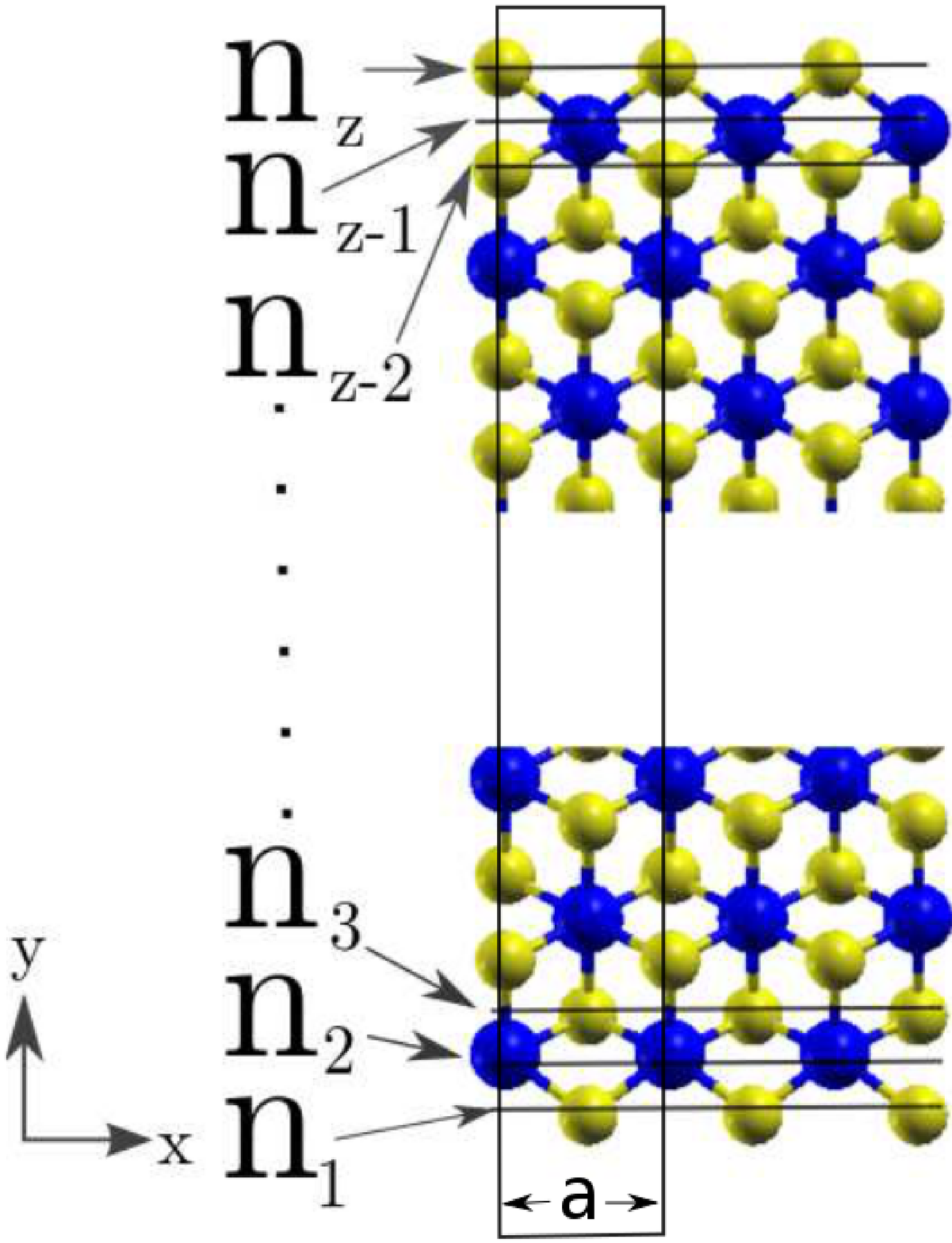}}
 \subfigure[Armchair Ni-centered-Se]{\label{Fig1c}\includegraphics[width=0.23\textwidth,height=0.15\textheight]{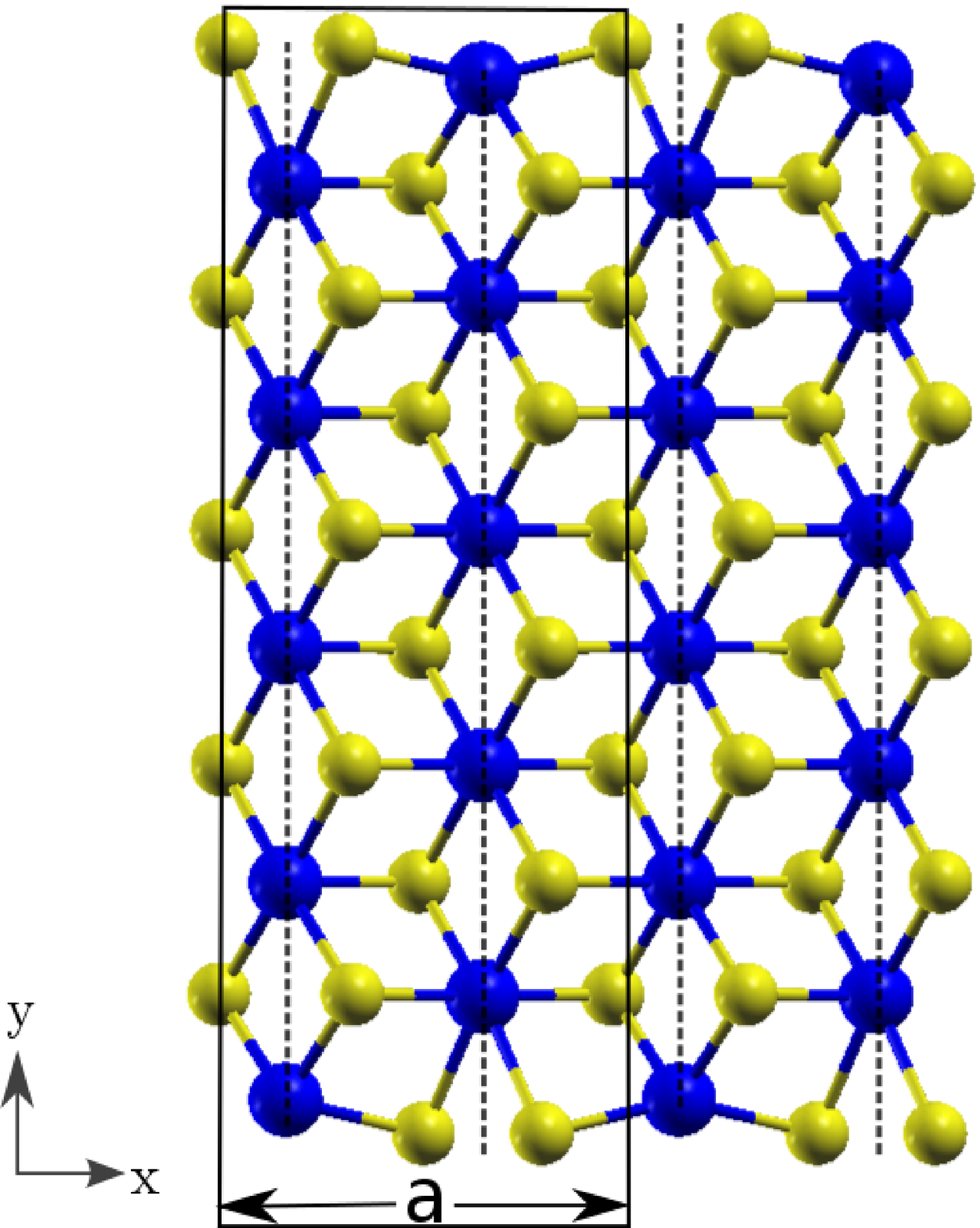}}
 \subfigure[Armchair Ni-aligned-Ni]{\label{Fig1d}\includegraphics[width=0.23\textwidth,height=0.15\textheight]{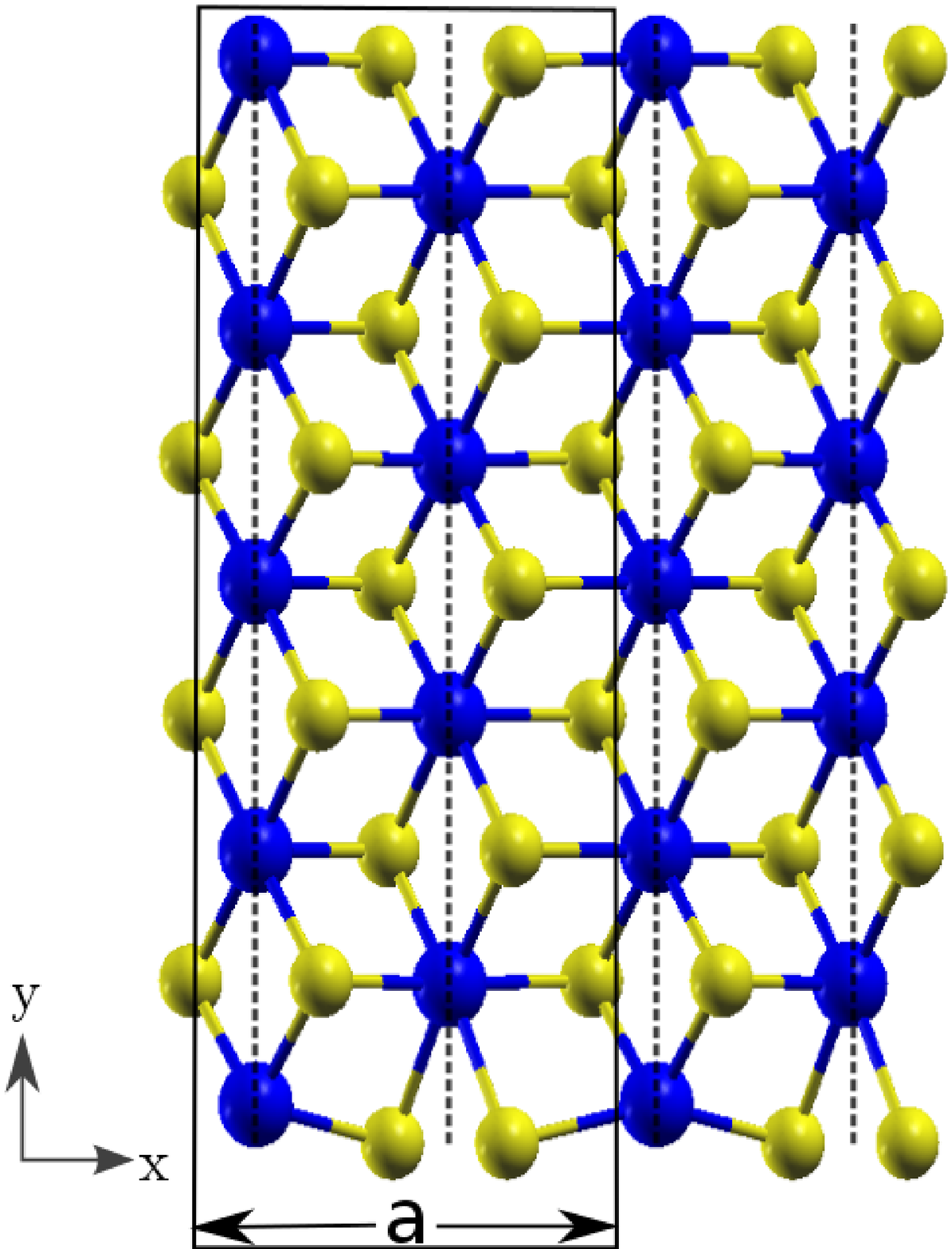}}
 \caption{(Color online)  Large (blue) and  small (yellow) circles represent Ni and Se atoms. \subref{Fig1a} $xy$ and $zx$ planes of the 2D systems in the stable T configuration. \subref{Fig1b}, \subref{Fig1c} and \subref{Fig1d} represent the structure of 1D-T-NiSe$_2$ zigzag and armchair nanorribons respectively. In \subref{Fig1b}, the $n$'s indicates the three outer atomic rows at each ribbon edge, with $z$ the total number of atomic rows. Black solid lines indicate the unit cells of the 2D, zigzag and armchair ribbons. $a$ is the periodic lattice parameter of the supercell in the corresponding ribbons. For the armchair ribbons, dotted lines guides the identification of the two possible edge configurations, the Ni-centered-Se and the Ni-aligned-Ni}\label{Fig1}
\end{figure}

\begin{figure*}
\subfigure[Bare NiSeSe--SeSeNi]{\label{NiSeSe-SeSeNi}\includegraphics[width=0.15\textwidth]{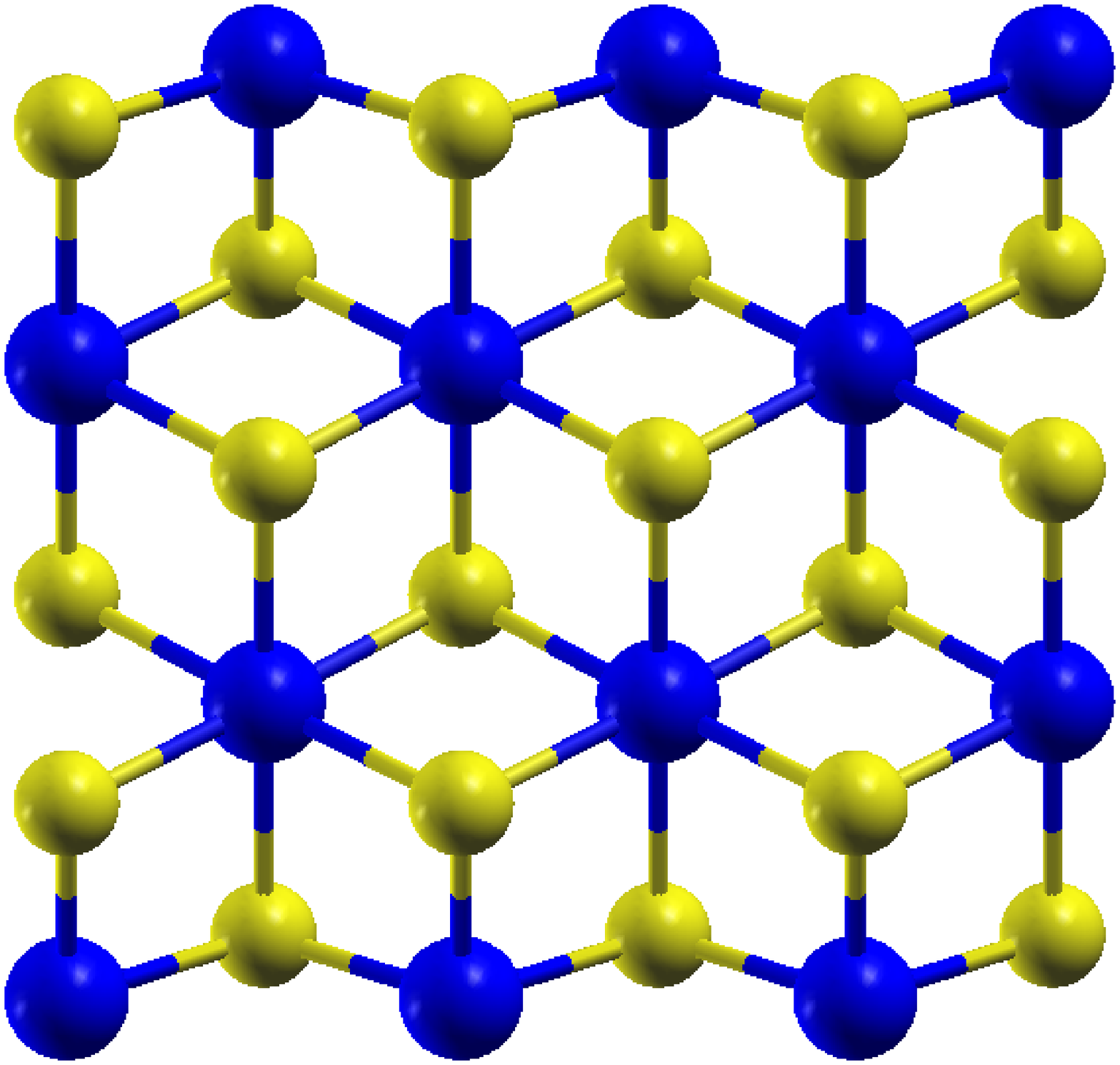}
\includegraphics[width=0.16\textwidth]{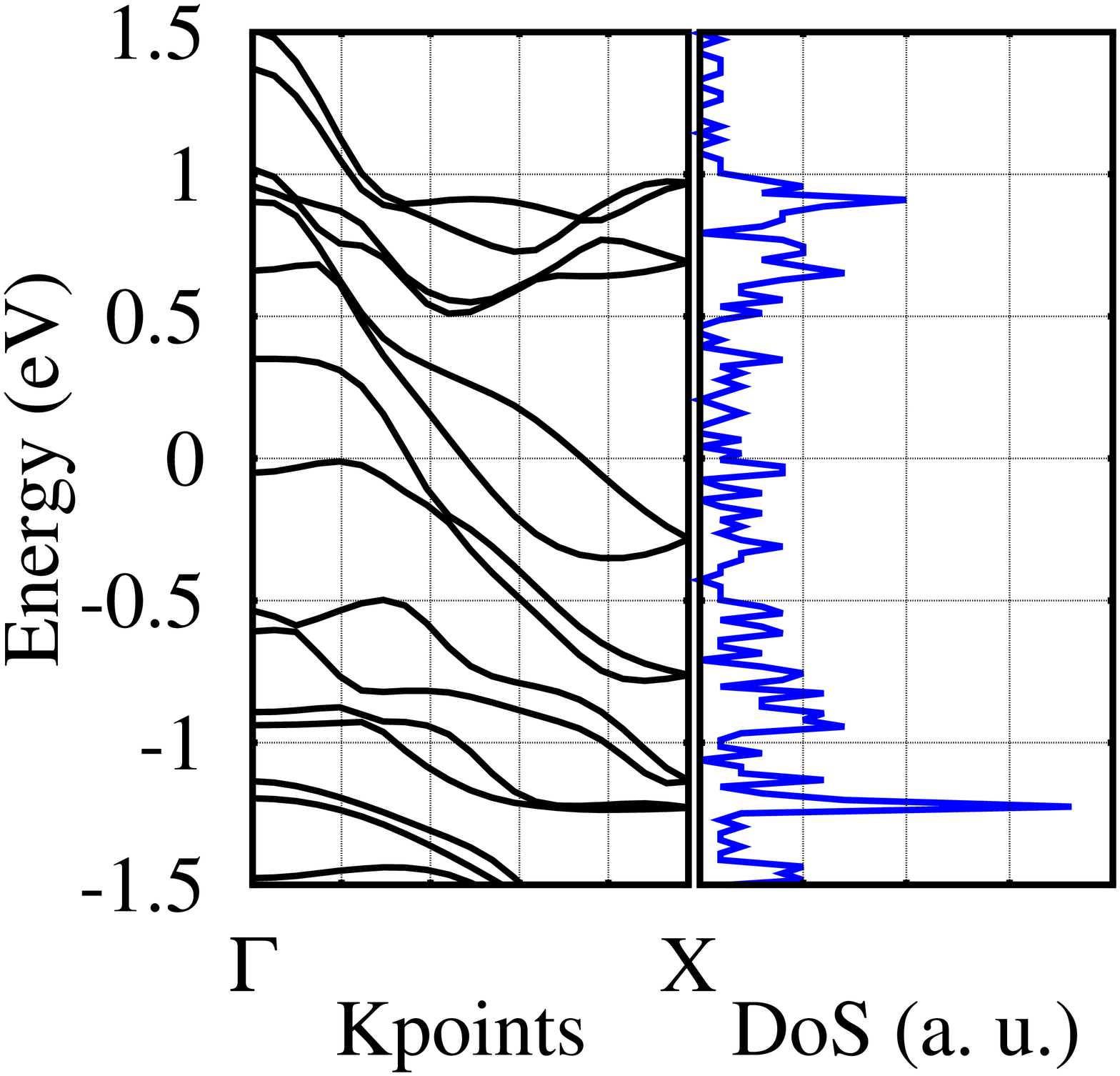}}
\subfigure[Passivated NiSeSe--SeSeNi]{\label{HNiSeSe-SeSeNi}
\includegraphics[width=0.15\textwidth]{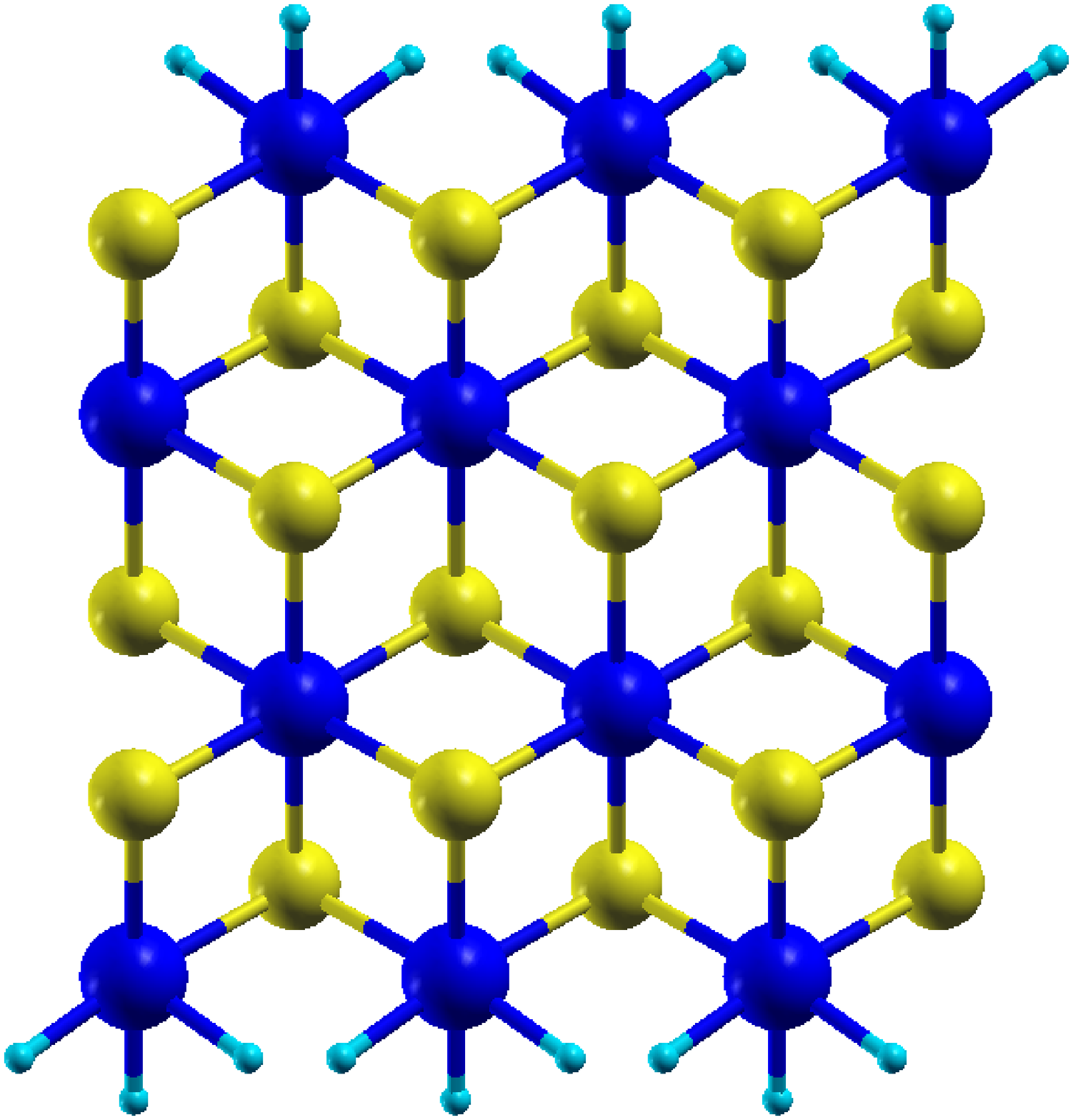}
\includegraphics[width=0.16\textwidth]{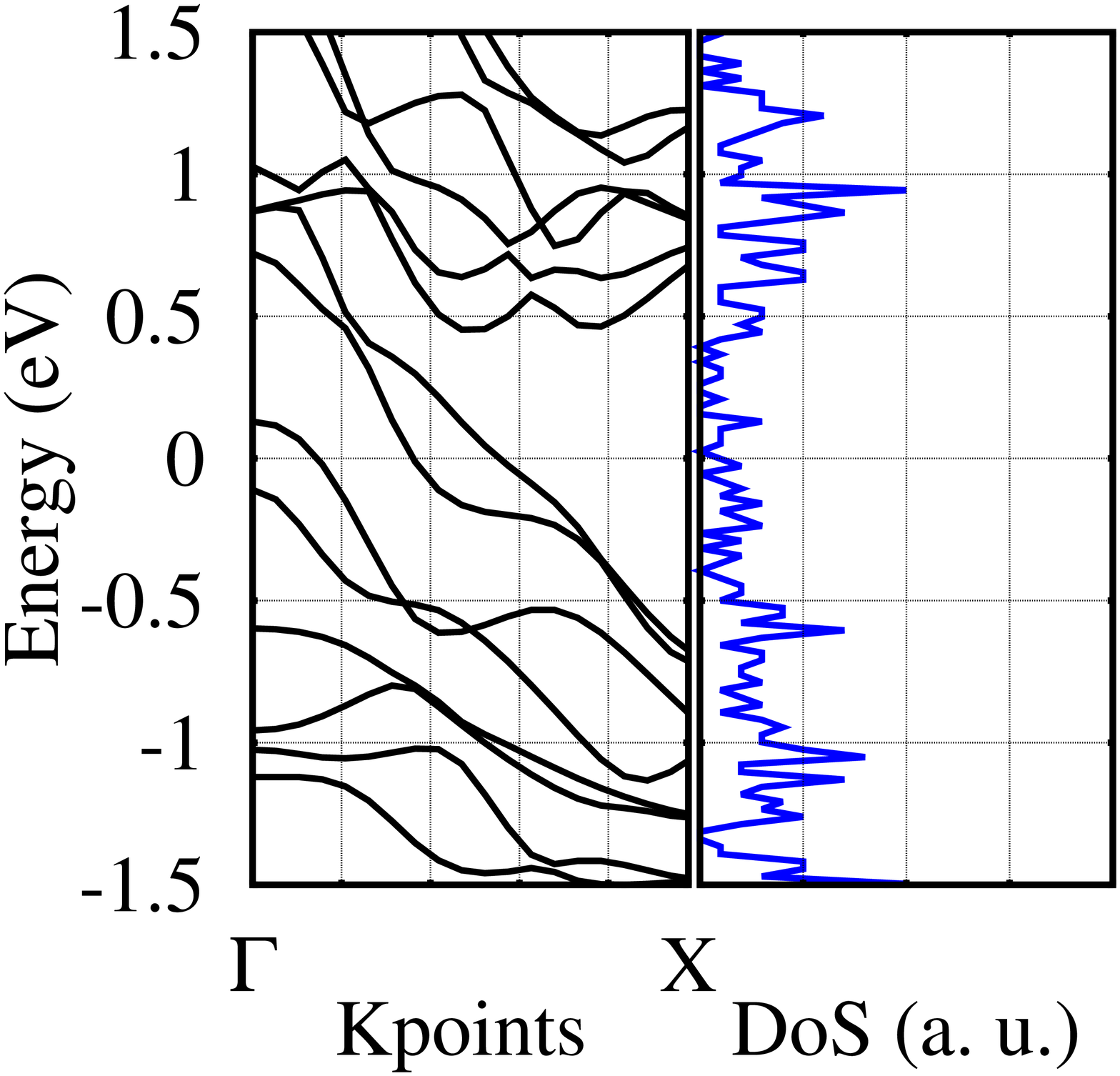}}\\
\subfigure[Bare NiSeSe--SeNiSe]{\label{NiSeSe-SeNiSe}
\includegraphics[width=0.15\textwidth]{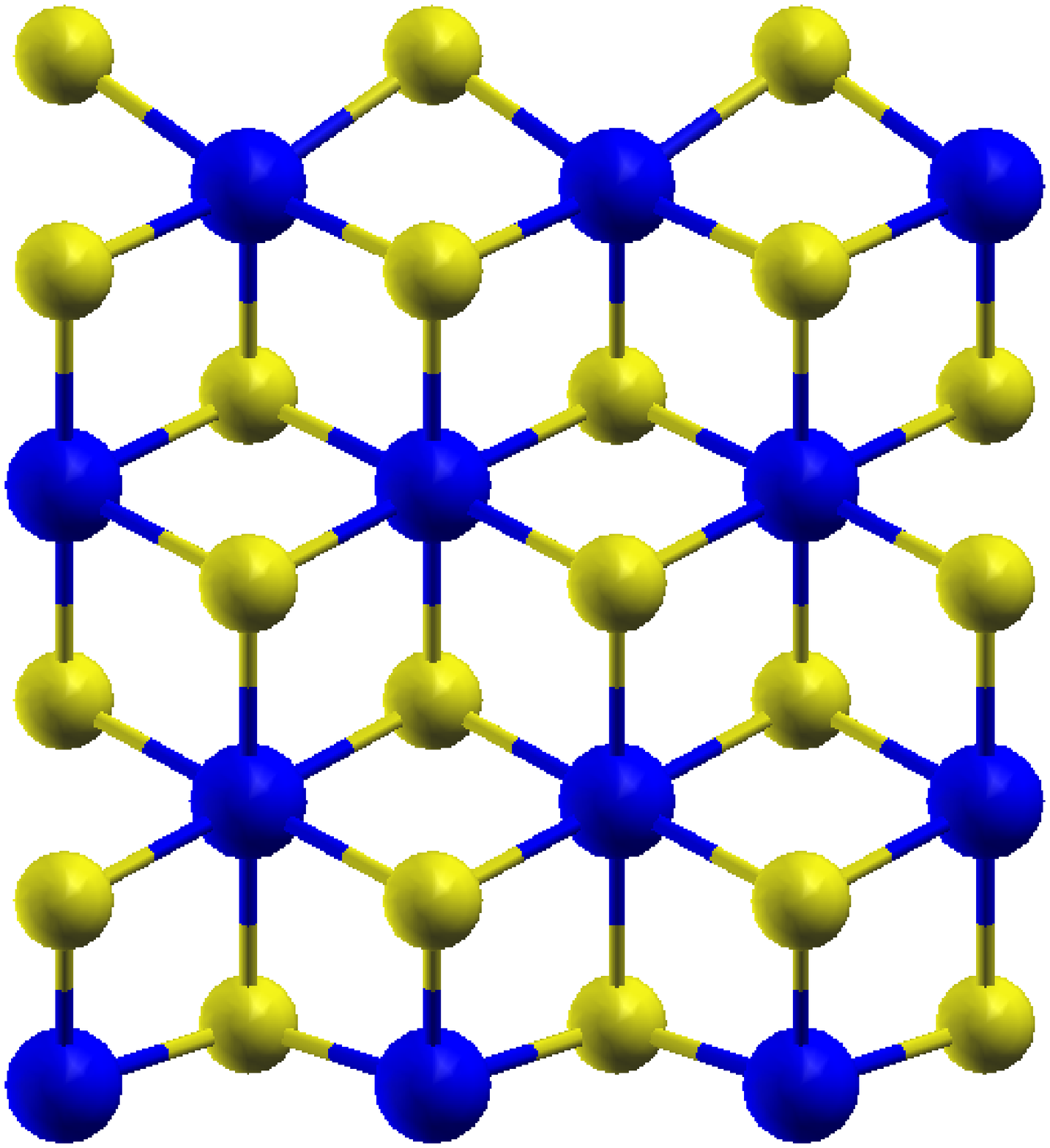}
\includegraphics[width=0.16\textwidth]{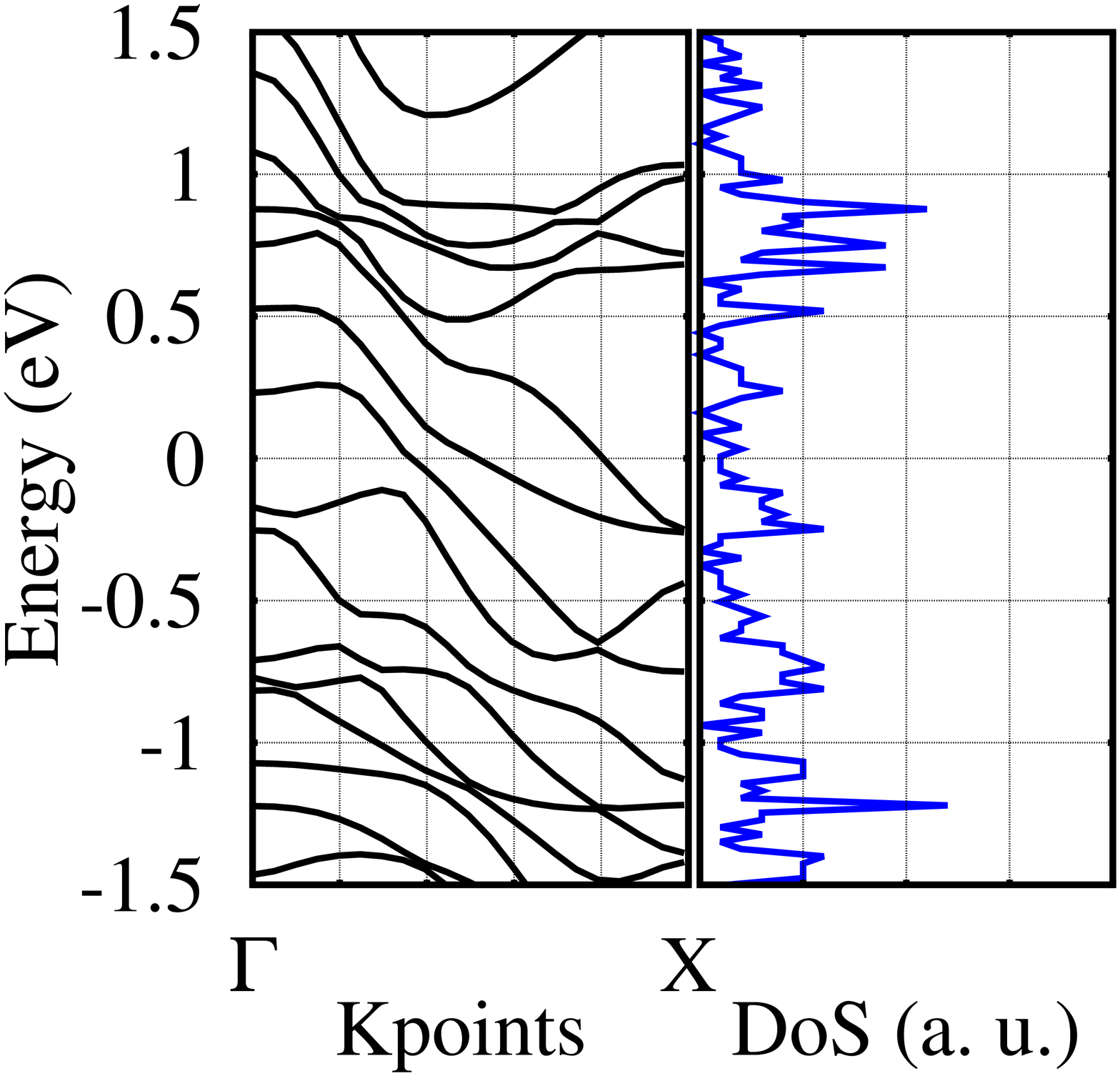}}
\subfigure[Passivated NiSeSe--SeNiSe]{\label{HNiSeSe-SeNiSe}\includegraphics[width=0.15\textwidth]{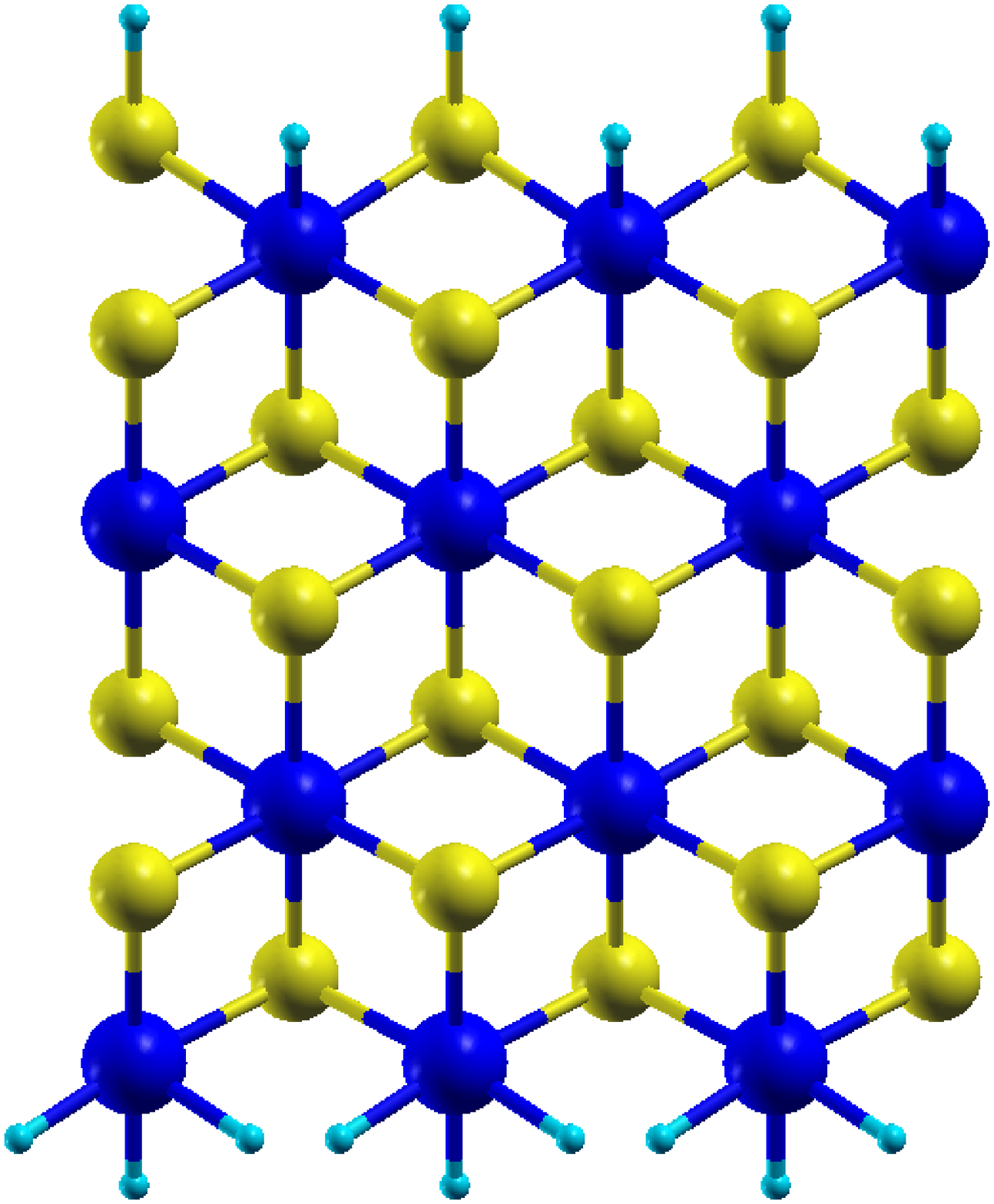}
\includegraphics[width=0.16\textwidth]{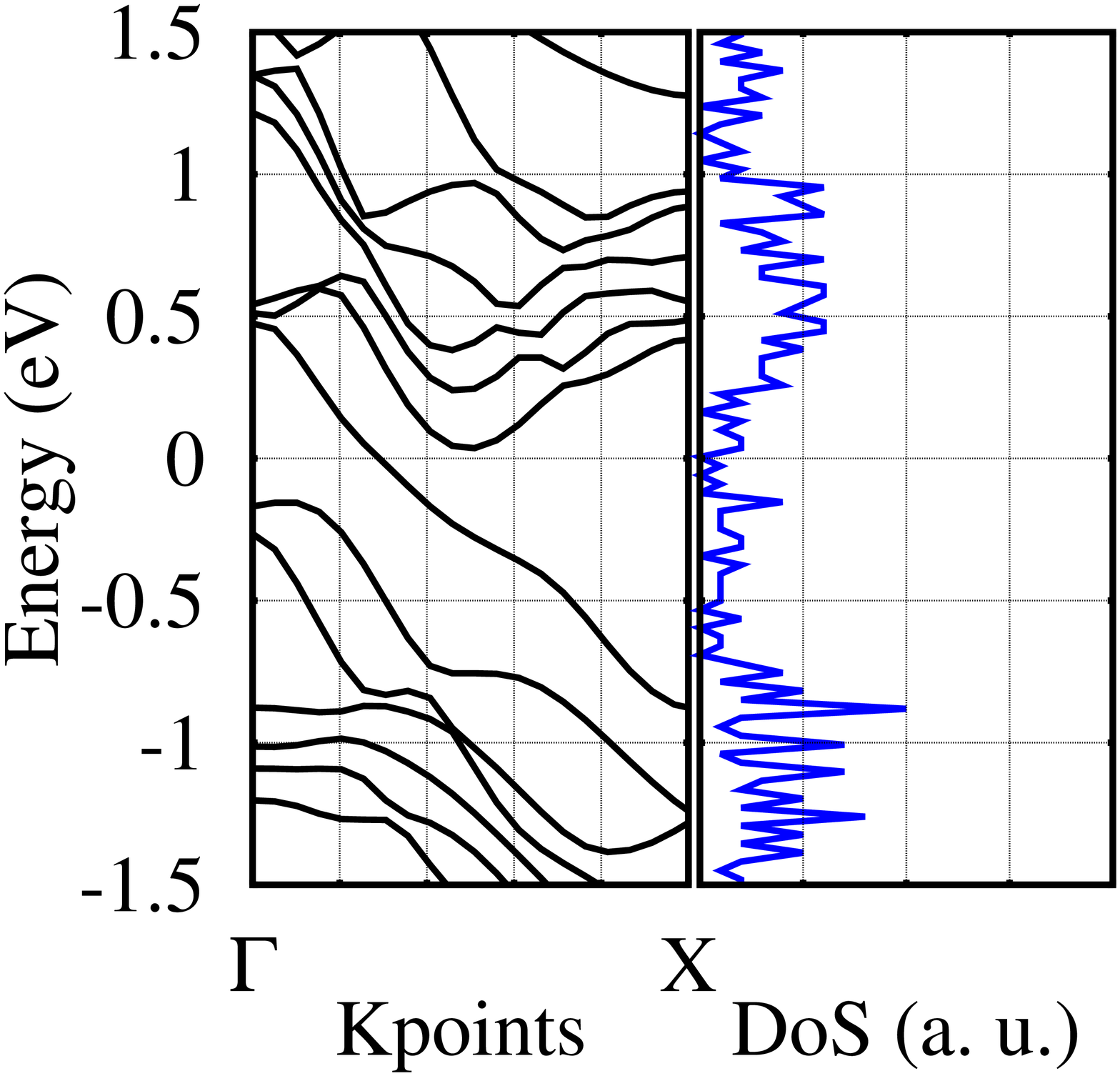}}\\
\subfigure[Bare NiSeSe--NiSeSe]{\label{NiSeSe-NiSeSe}
\includegraphics[width=0.15\textwidth]{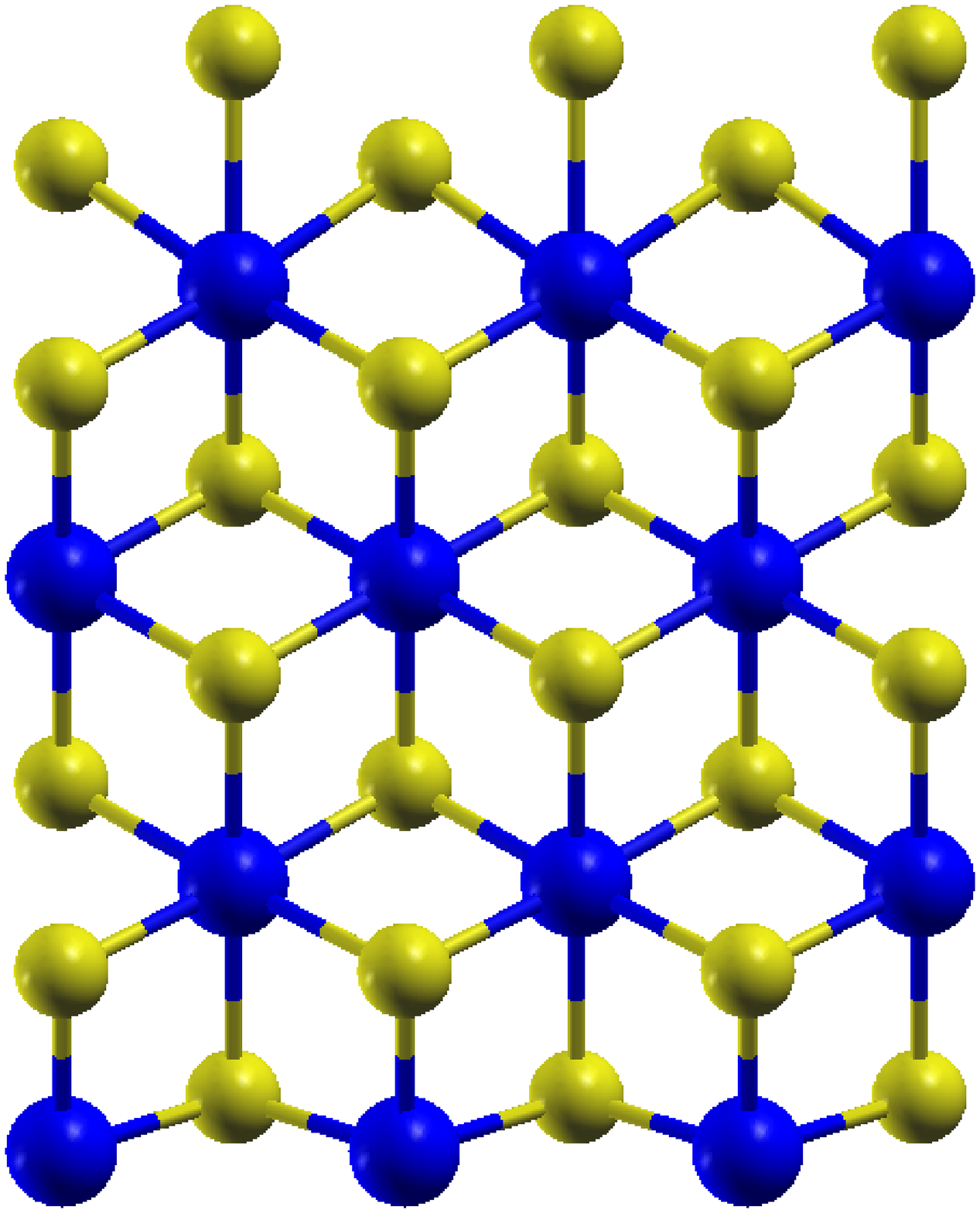}
\includegraphics[width=0.16\textwidth]{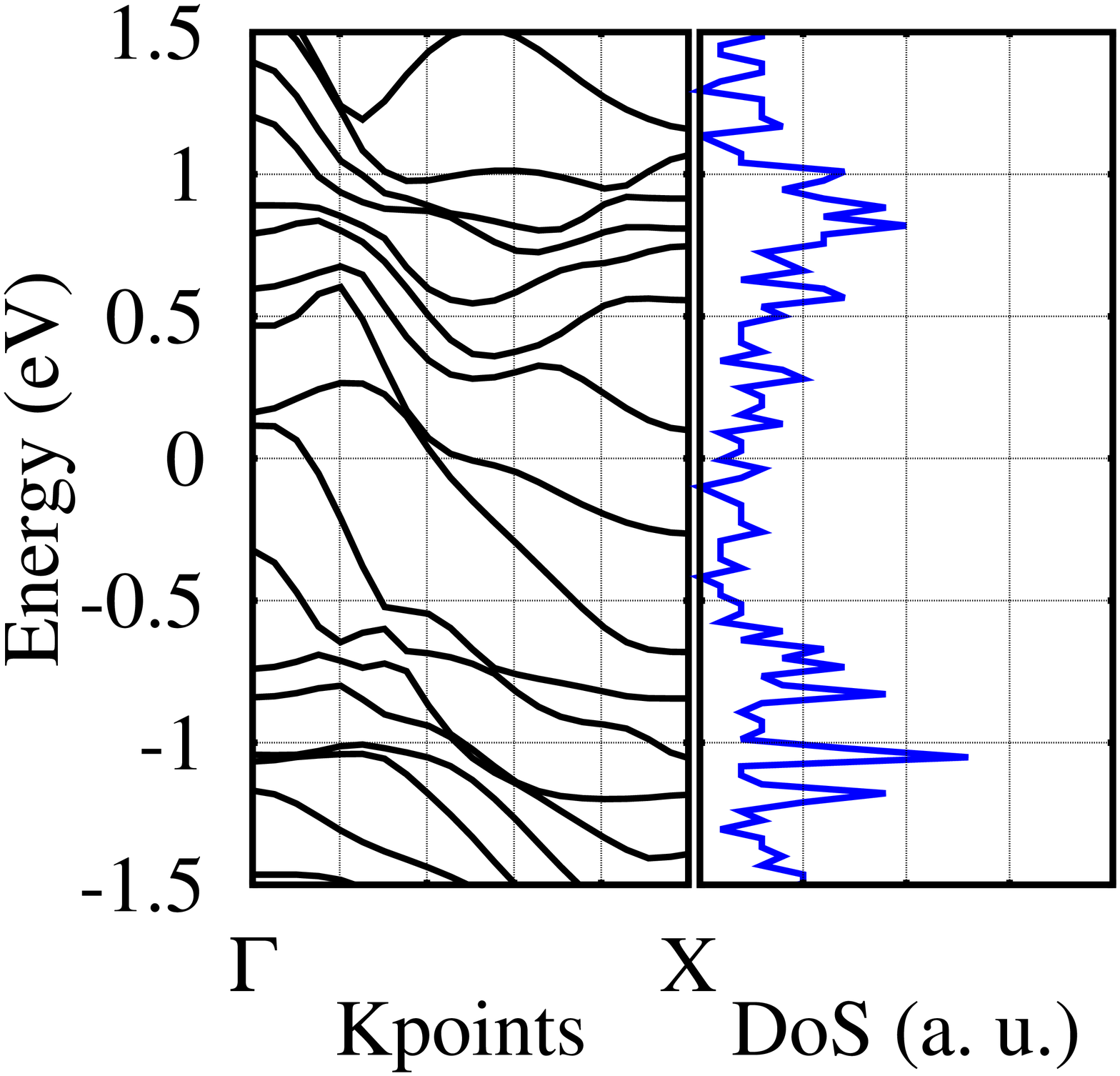}}
\subfigure[Passivated NiSeSe--NiSeSe]{\label{HNiSeSe-NiSeSe}
\includegraphics[width=0.15\textwidth]{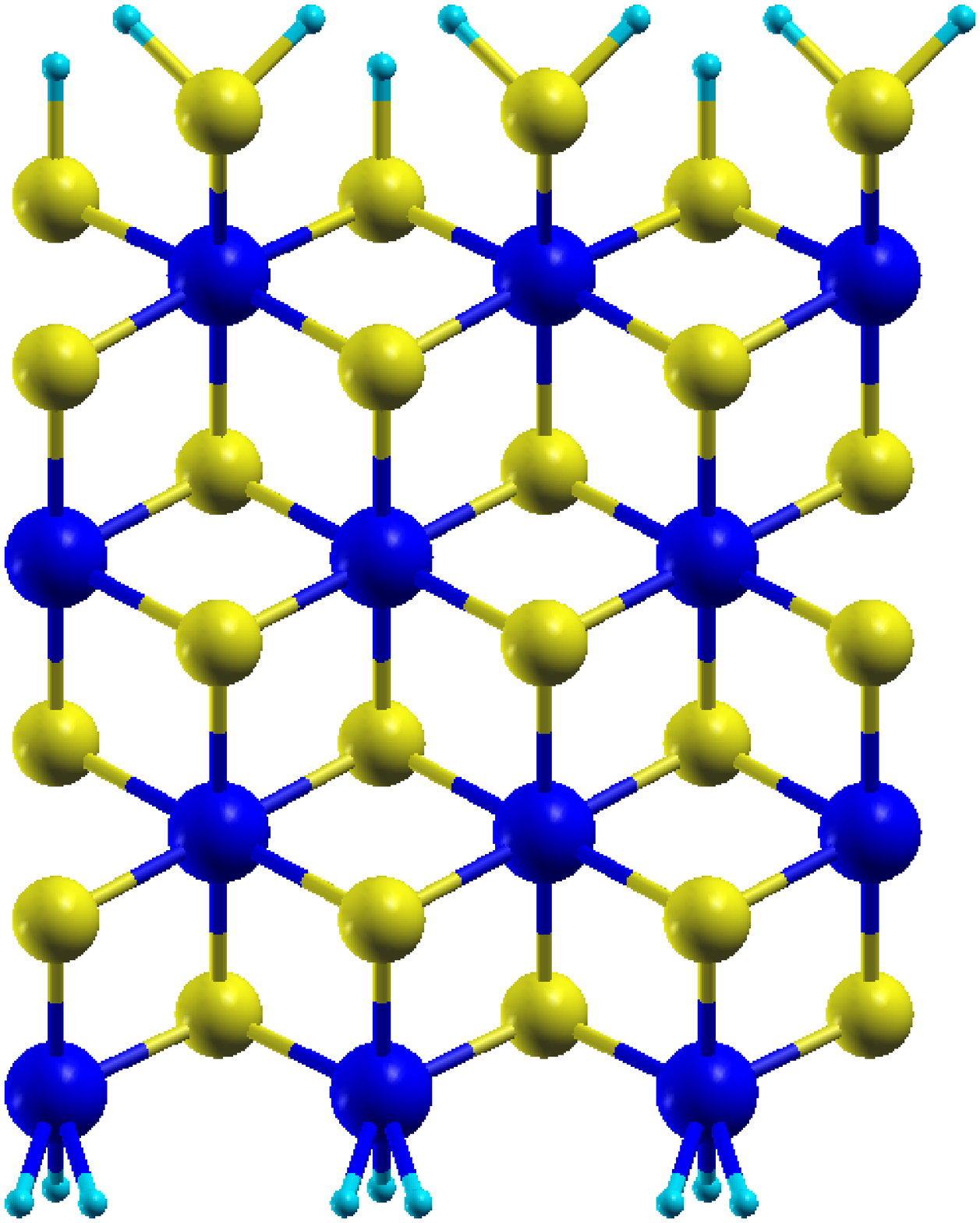}
\includegraphics[width=0.16\textwidth]{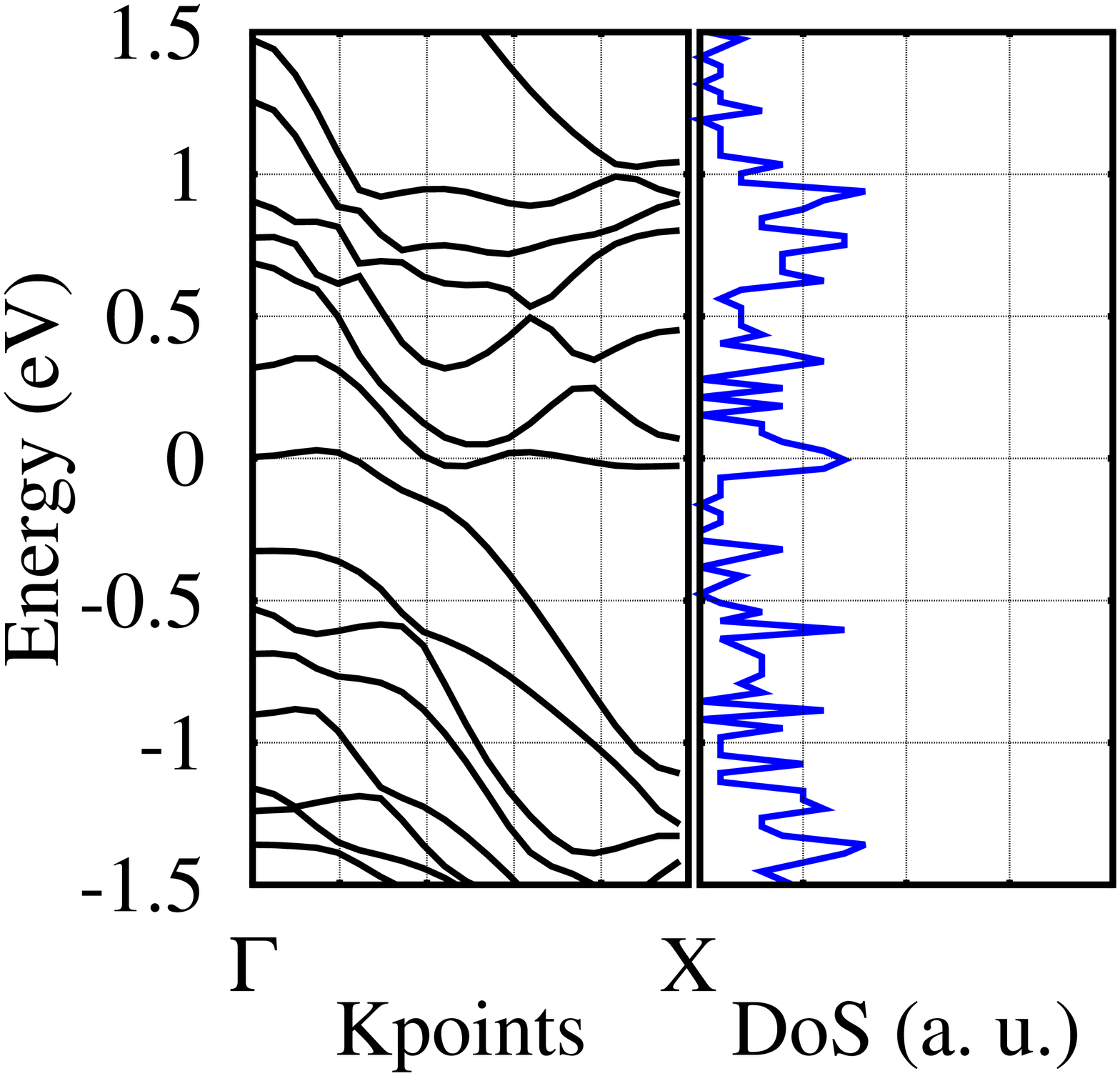}}\\
\subfigure[Bare SeNiSe--SeNiSe]{\label{SeNiSe-SeNiSe}\includegraphics[width=0.15\textwidth]{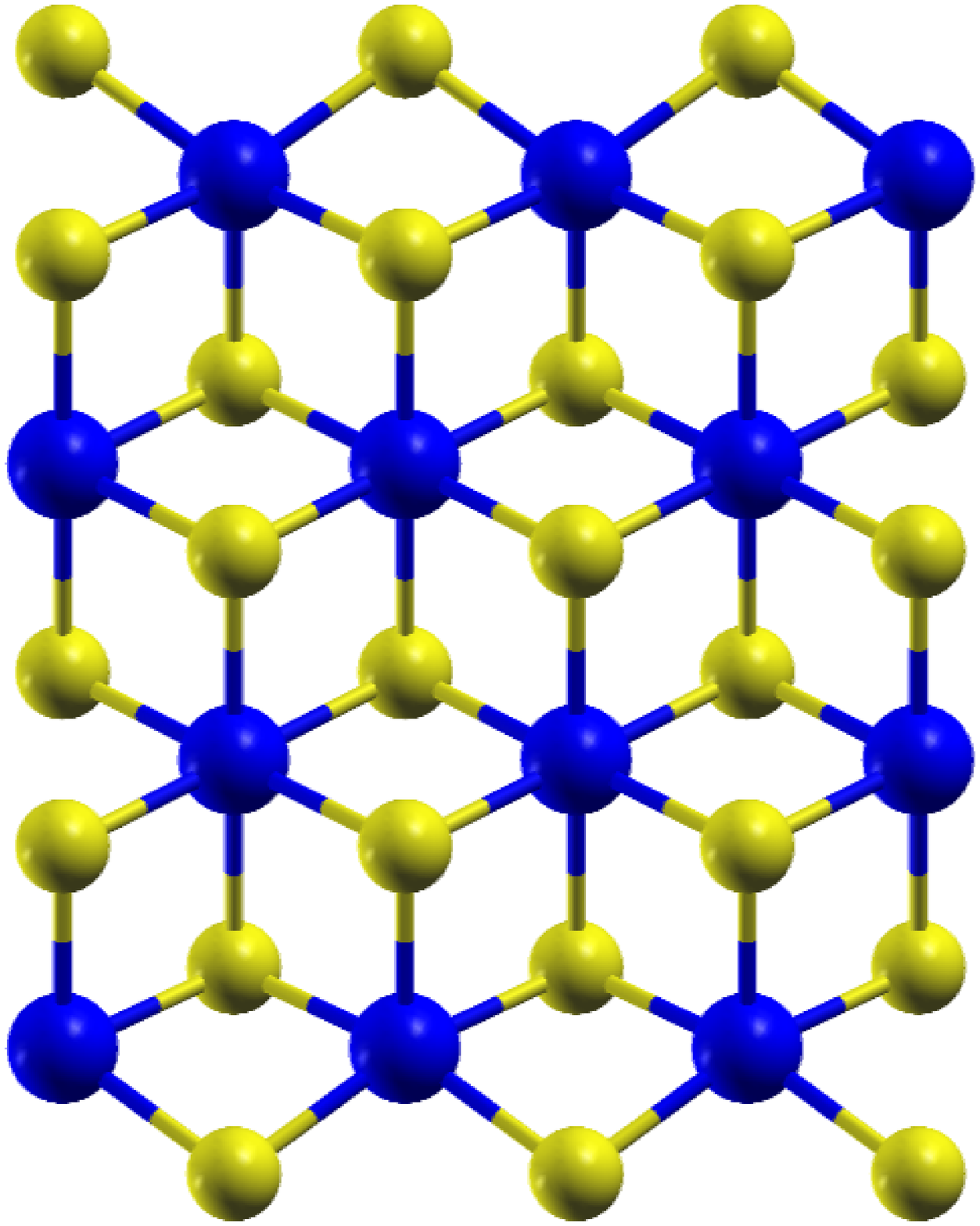}
\includegraphics[width=0.16\textwidth]{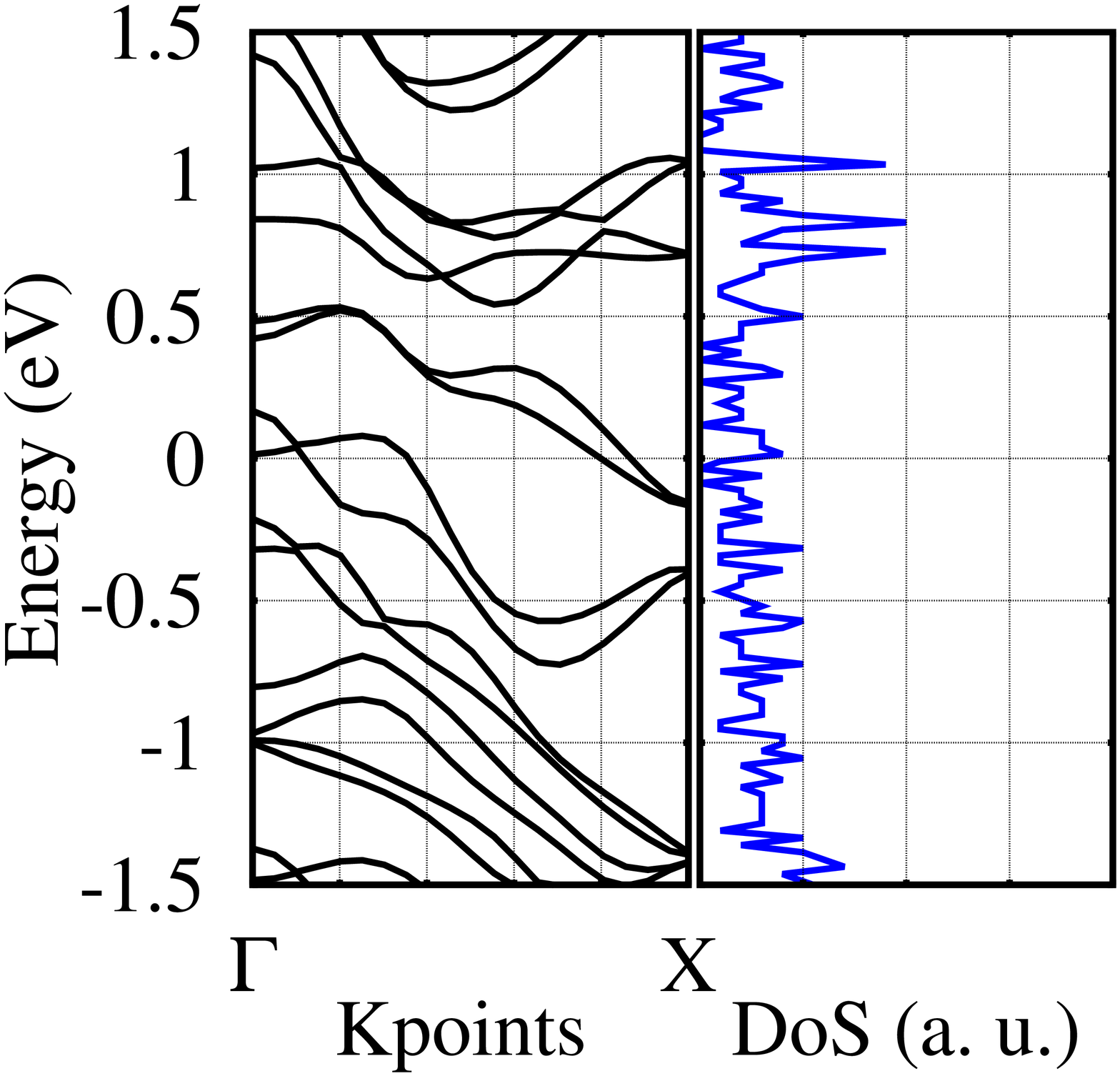}}
\subfigure[Passivated SeNiSe--SeNiSe]{\label{HSeNiSe-SeNiSe}
\includegraphics[width=0.15\textwidth]{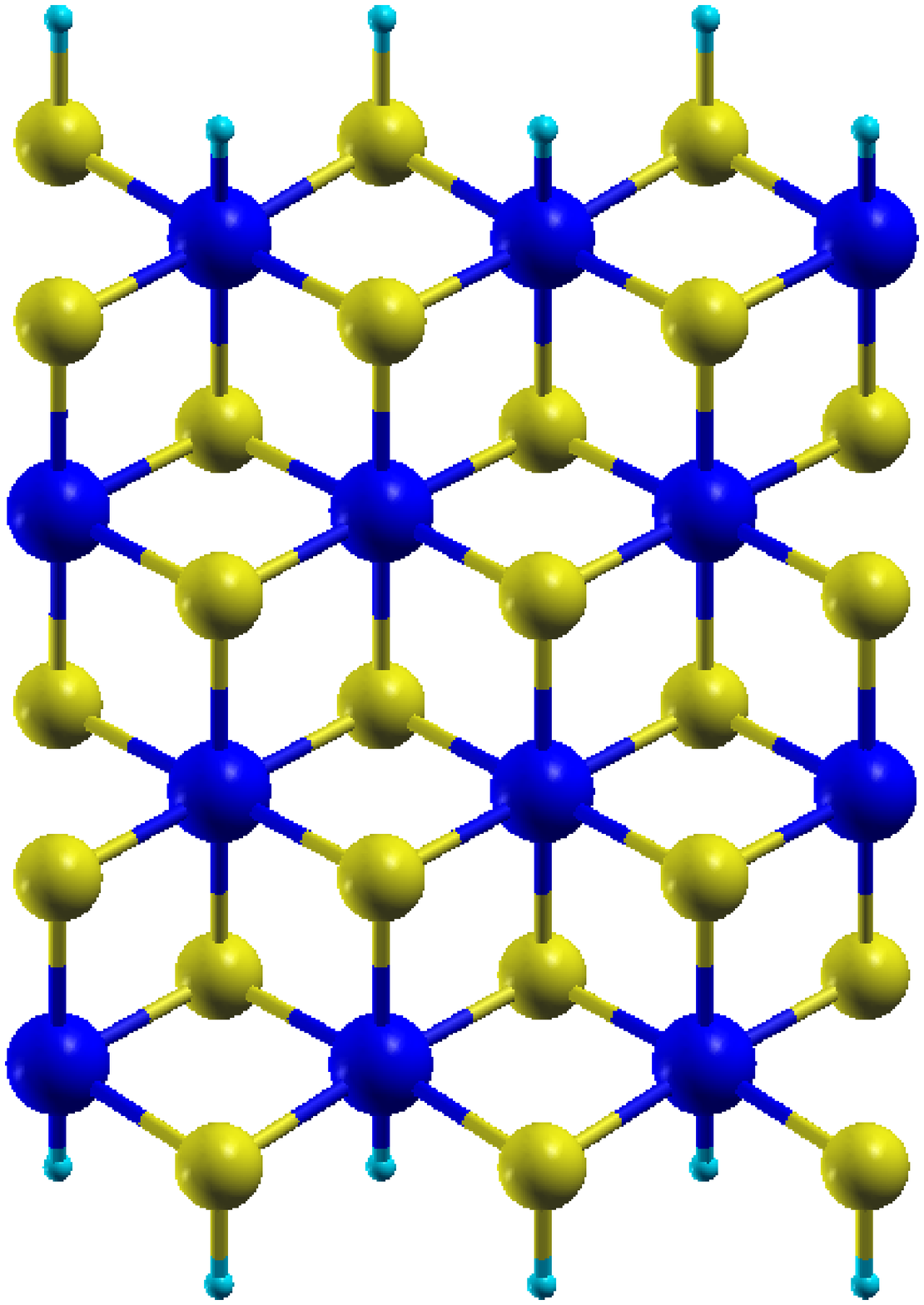}
\includegraphics[width=0.16\textwidth]{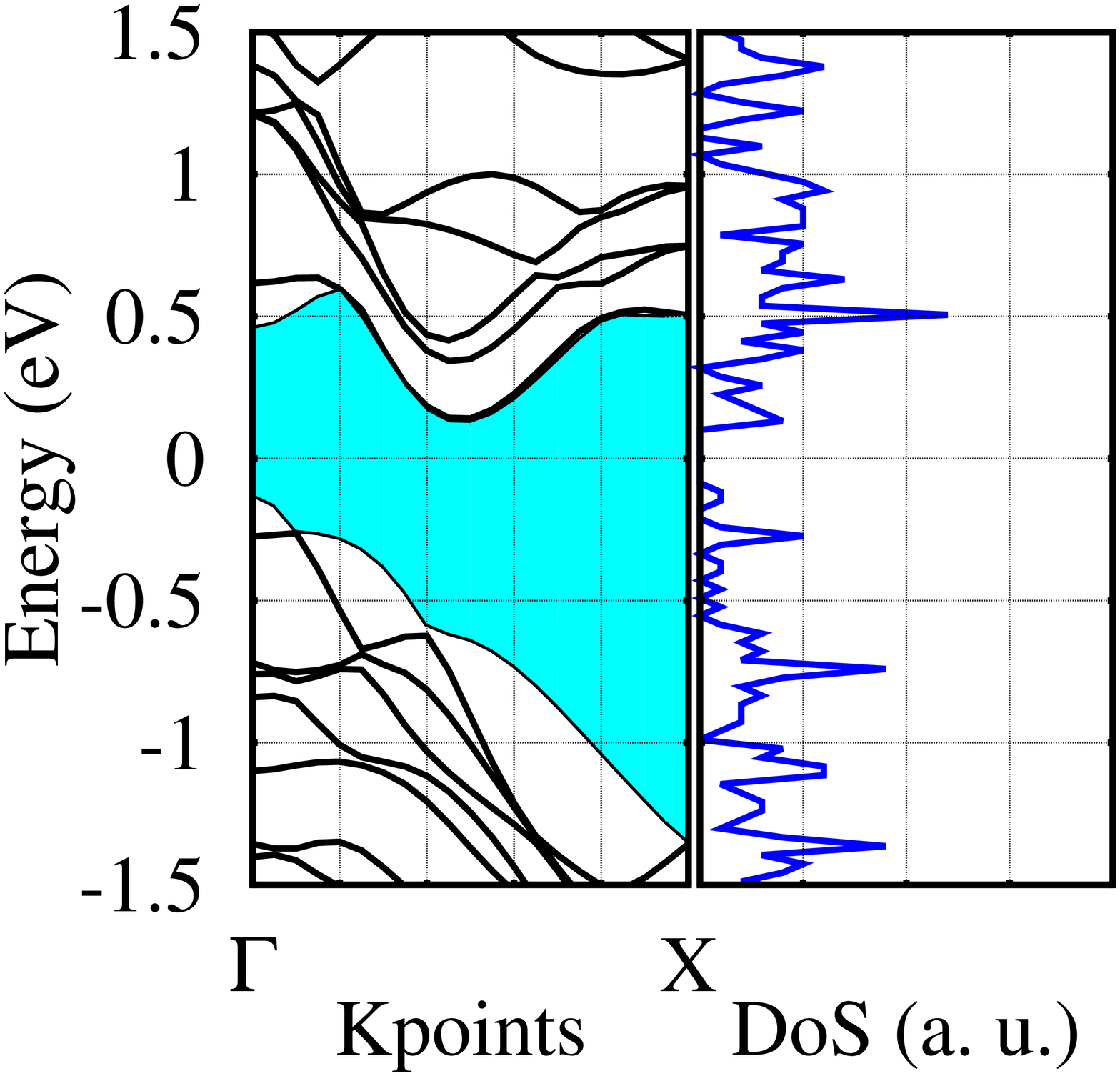}}\\
\subfigure[Bare SeNiSe--NiSeSe]{\label{SeNiSe-NiSeSe}
\includegraphics[width=0.15\textwidth]{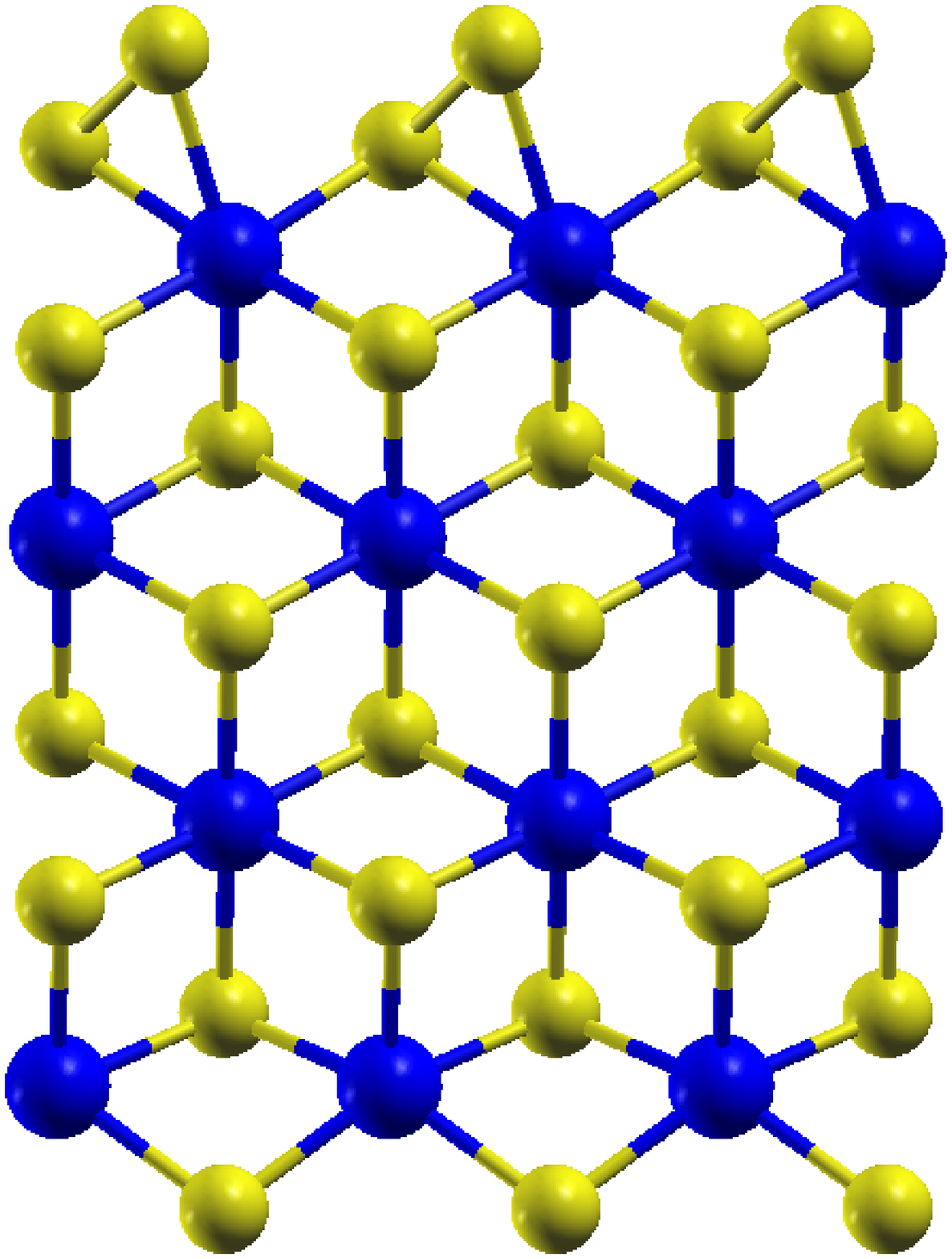}
\includegraphics[width=0.16\textwidth]{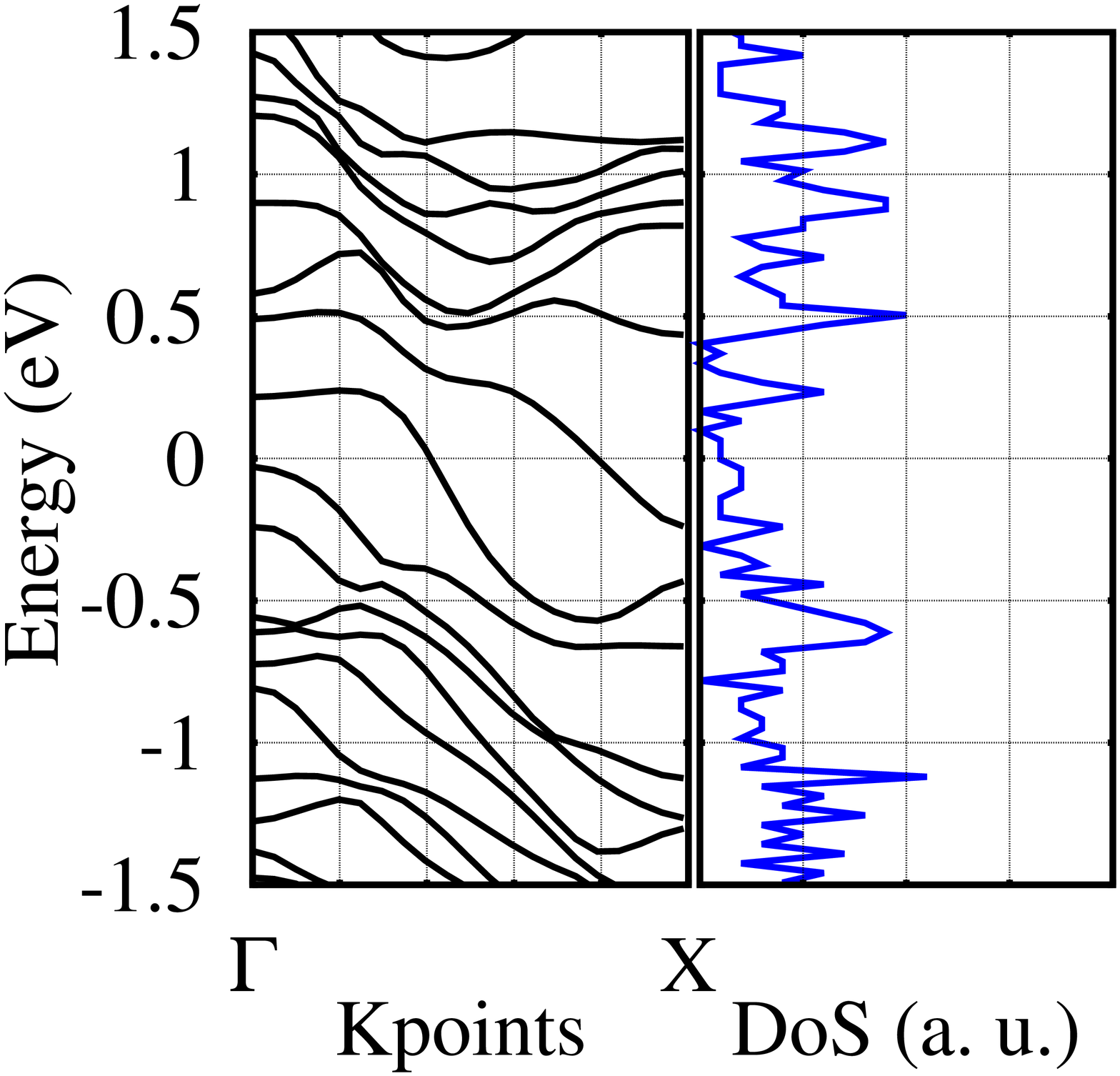}}
\subfigure[Passivated SeNiSe--NiSeSe]{\label{HSeNiSe-NiSeSe}\includegraphics[width=0.15\textwidth]{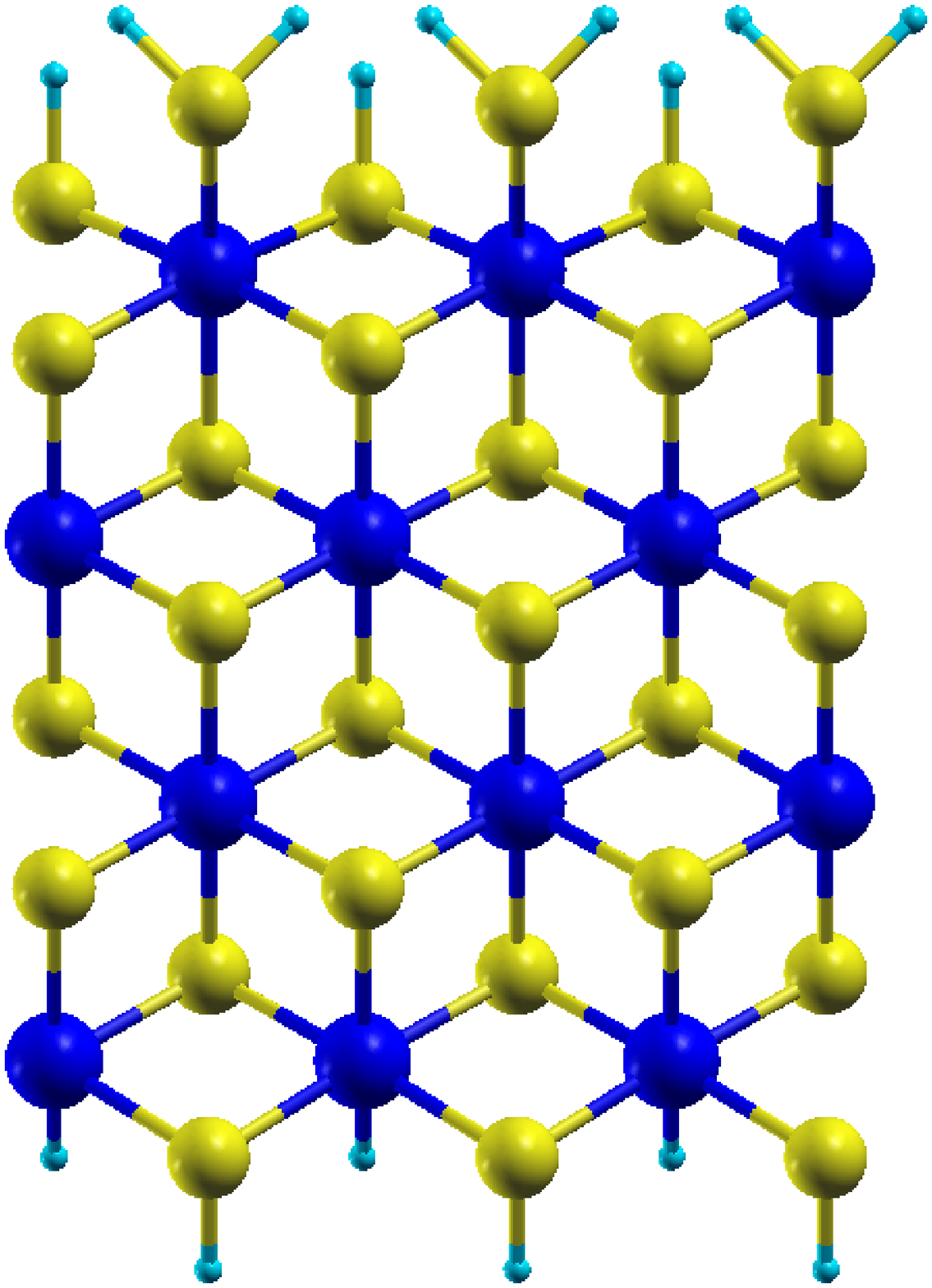}
\includegraphics[width=0.16\textwidth]{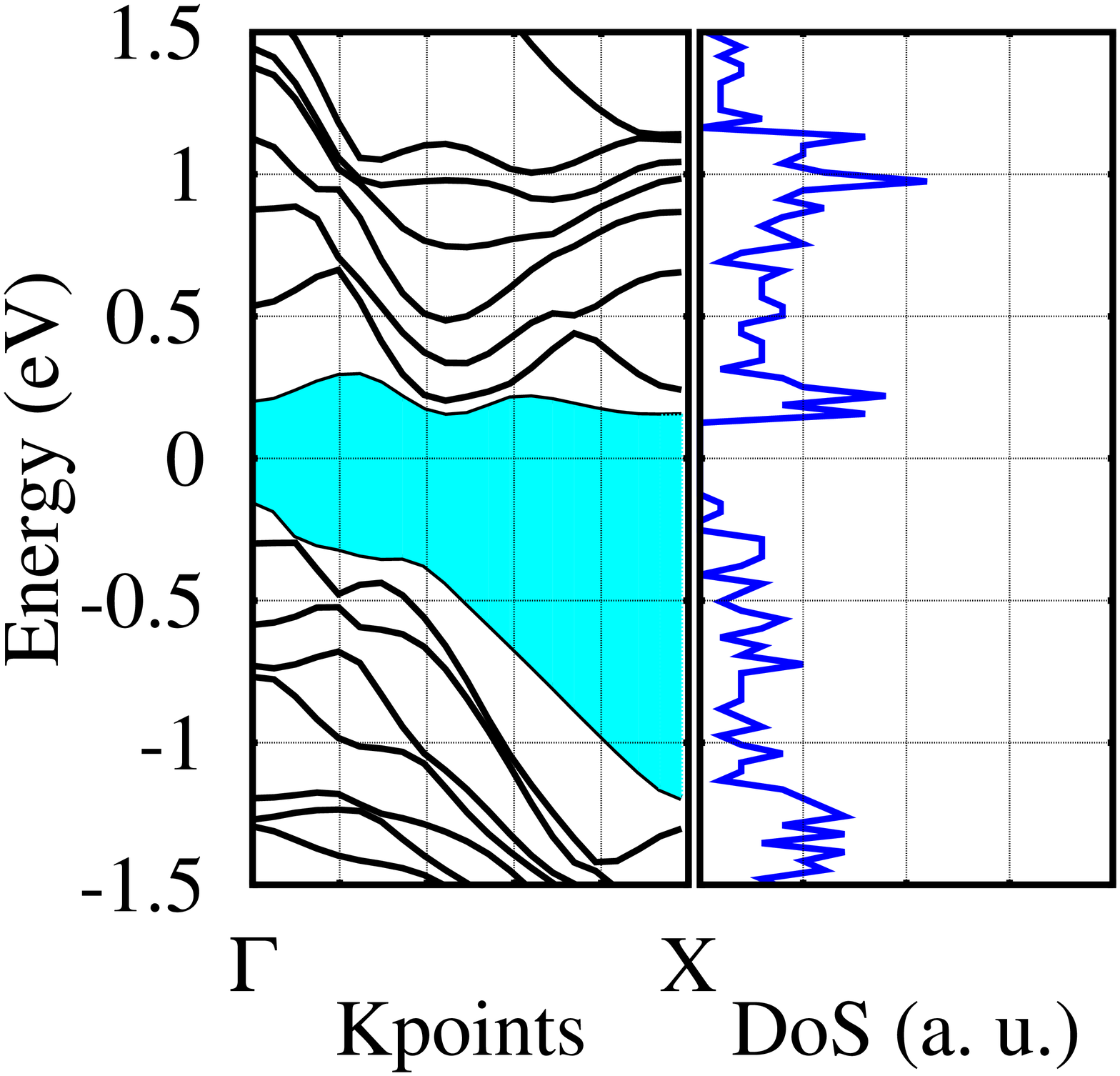}}\\
\subfigure[Bare SeSeNi--NiSeSe]{\label{SeSeNi-NiSeSe}
\includegraphics[width=0.15\textwidth]{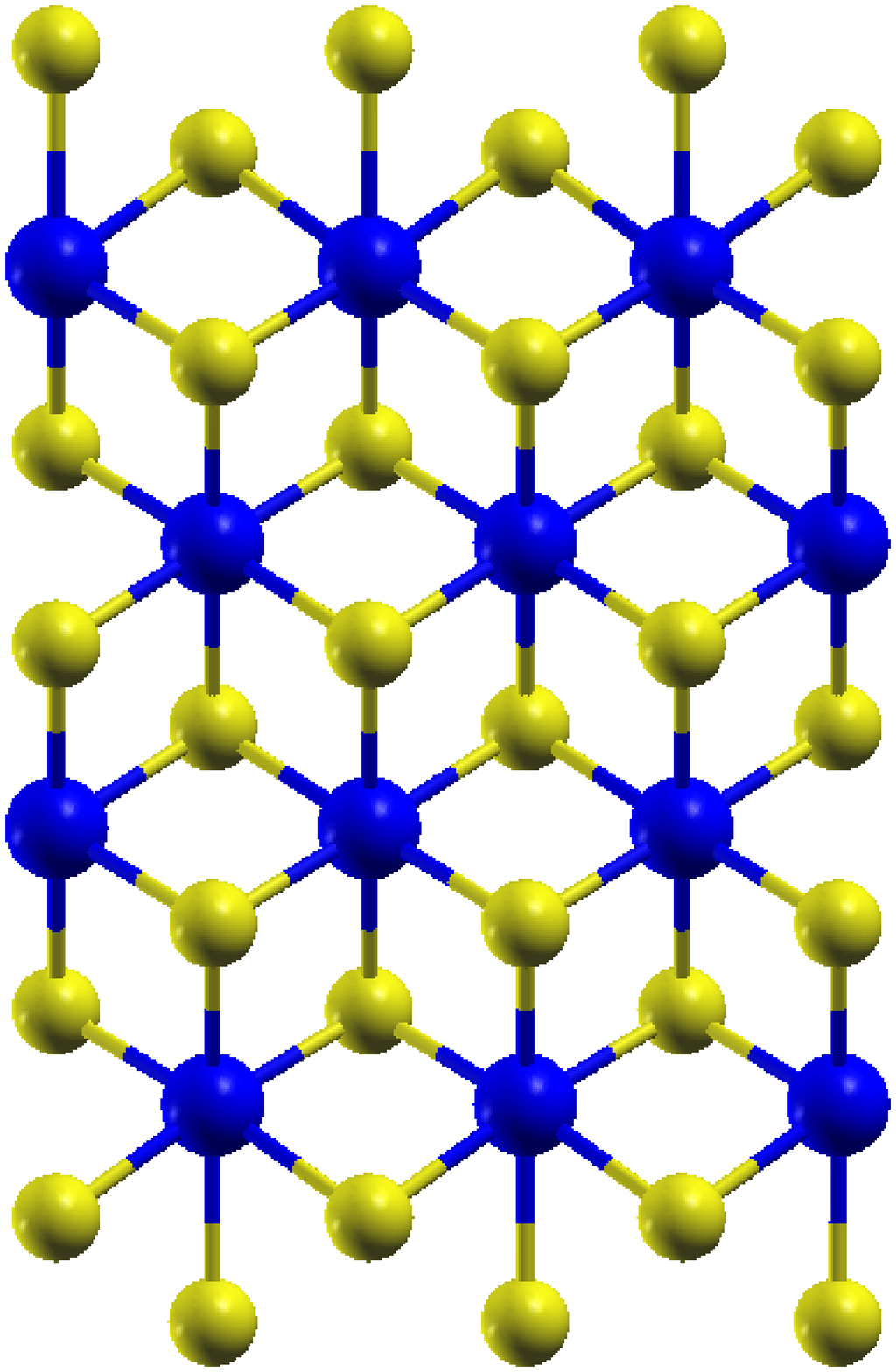}
\includegraphics[width=0.16\textwidth]{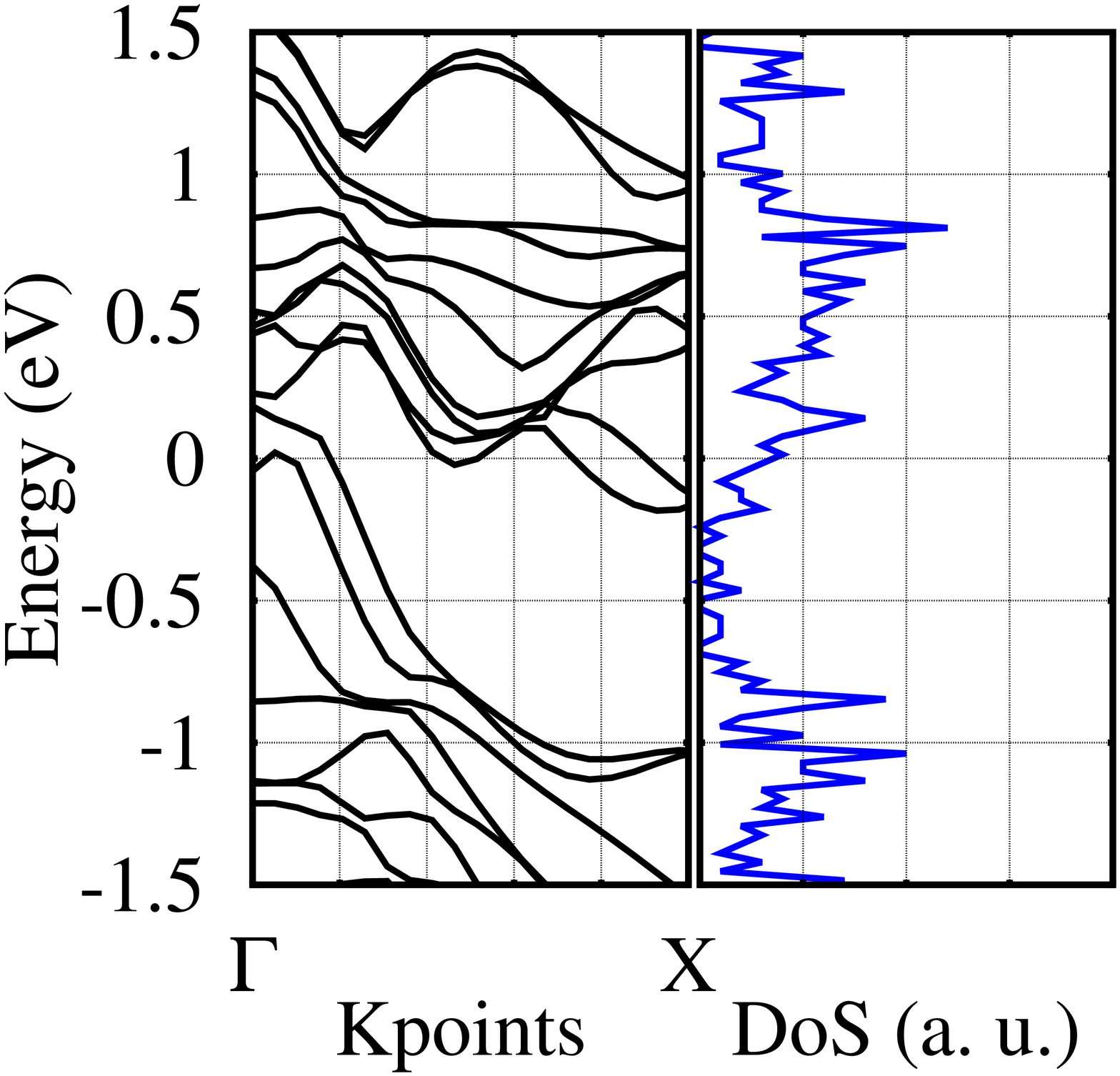}}
\subfigure[Passivated SeSeNi--NiSeSe]{\label{HSeSeNi-NiSeSe}
\includegraphics[width=0.15\textwidth]{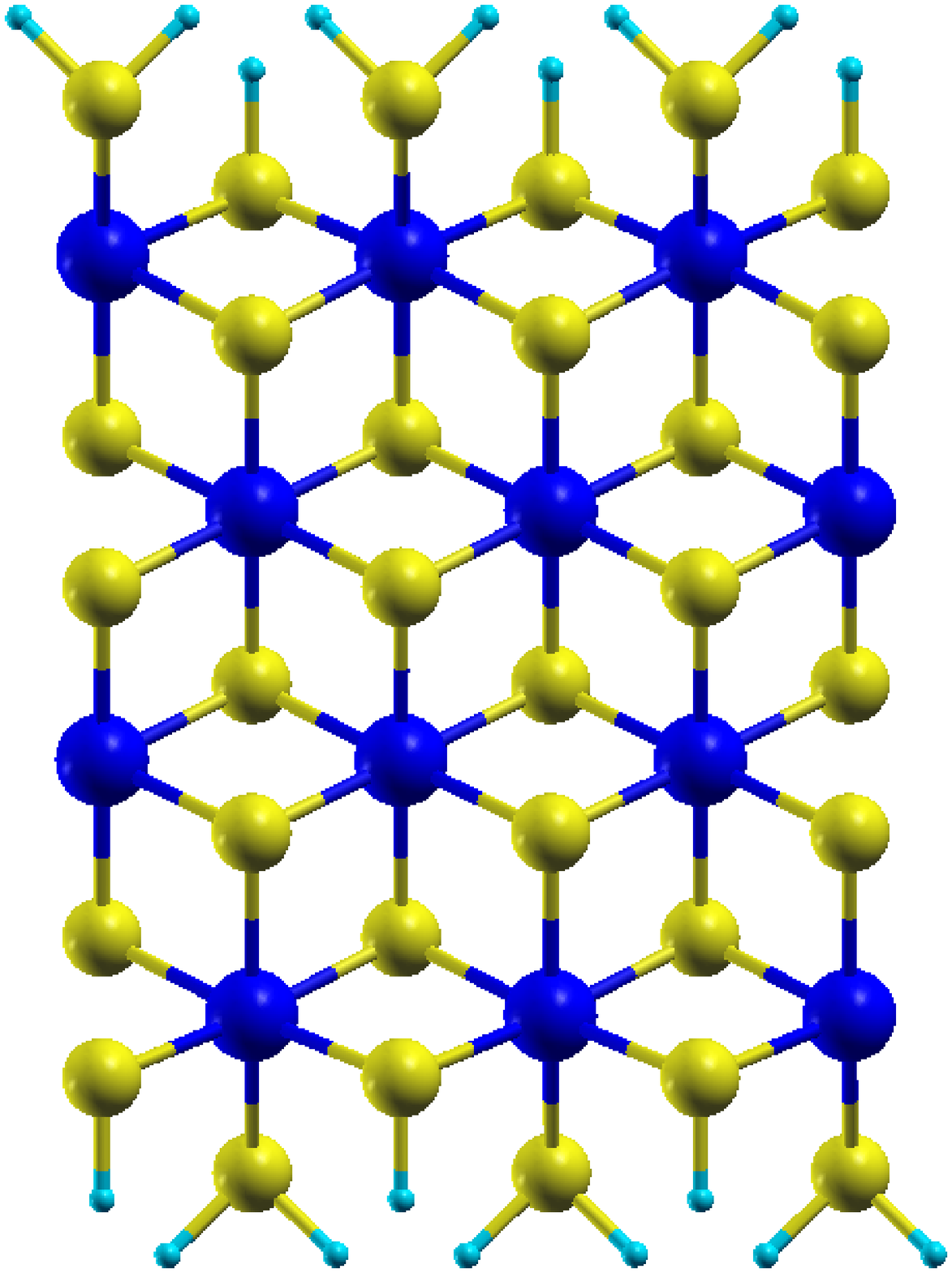}
\includegraphics[width=0.16\textwidth]{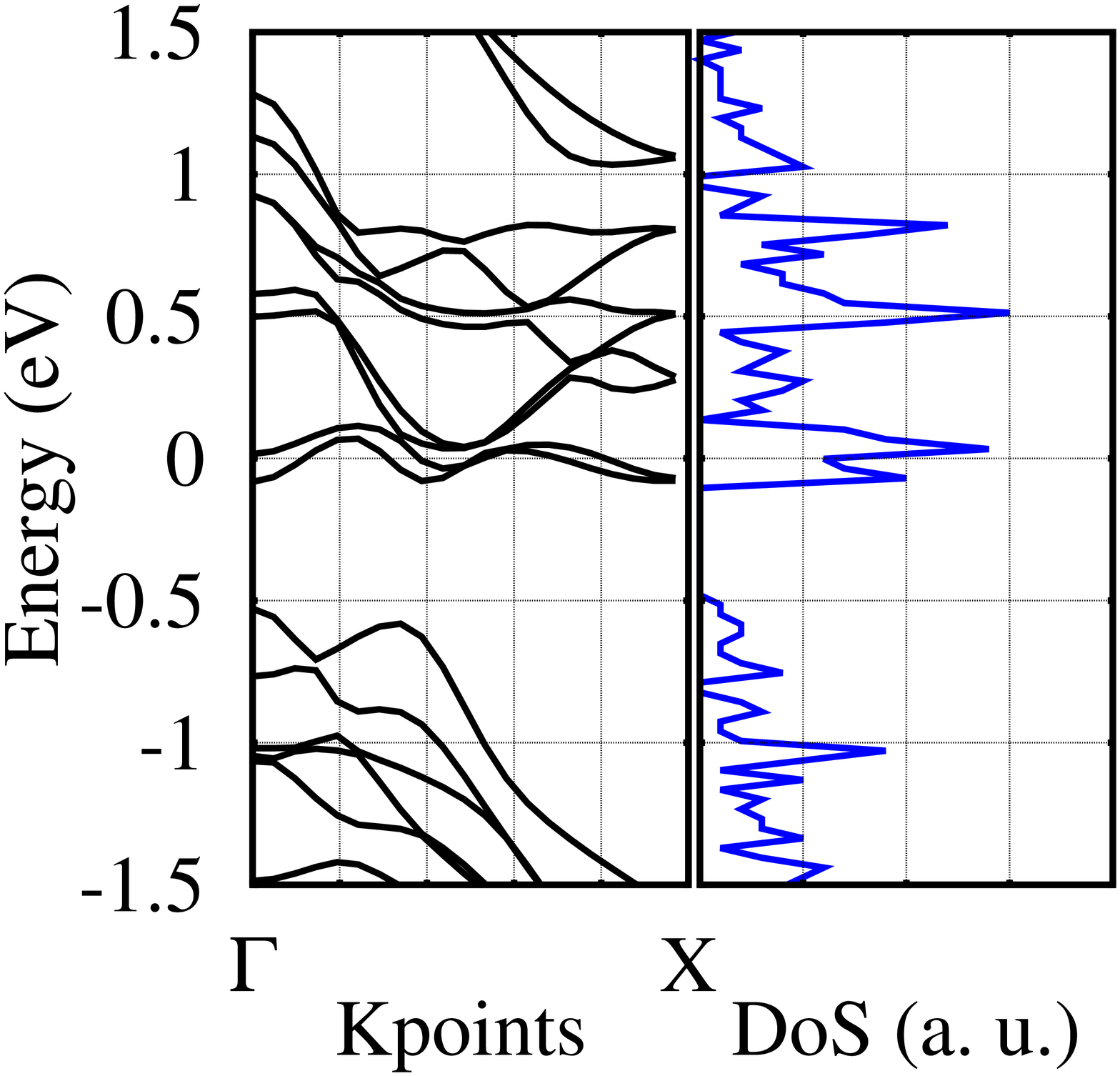}}
  
\caption{(color online) Six {\it root} ribbons of the 6 zigzag families, bare (left column) and H passivated (right column). The corresponding band structure and DOS are included in the right of each structure. \subref{HSeNiSe-SeNiSe} and \subref{HSeNiSe-NiSeSe} are  indirect semiconductors with band gaps of 0.25 and 0.30 eV respectively. Large (blue), medium (yellow) and small (aqua) circles represent Ni, Se and H atoms.}\label{Fig2}
     
\end{figure*}

 \ref{Fig1}\subref{Fig1a} shows the structure in the $xy$ and $zx$ views of the T configuration. We classified the zigzag terminations according to the three bottom atomic lines and the three top atomic lines ( \ref{Fig1b}). The ribbon that starts with three $n_1-n_2-n_3$ atoms lines ($y$-direction as reference) and ends with three $n_{z-2}-n_{z-1}-n_{z}$ atoms lines is called $n_1n_2n_3$--$n_{z-2}n_{z-1}n_{z}$ ribbon, where $n_1$,$n_2$, $n_3$, $n_{z-2}$, $n_{z-1}$, $n_{z}$ could be Se or Ni, $z$ is the total number of rows in the nanorribon. Zigzag ribbons have the restriction that $n_1n_2n_3$ and $n_{z-2}n_{z-1}n_{z}$ occurs in the cyclic order NiSeSe due to the periodicity. The six representative zigzag edge terminations are shown in (\ref{Fig2}).

In the case of armchair nanoribbons two representative families appear; we call them Ni-aligned-Ni (Fig. \ref{Fig1c}) and Ni-centered-Se (Fig. \ref{Fig1d}). We chose this nomenclature according to the atomic symmetry of the opposite edges as shown by the dotted lines in the lower panels of  \ref{Fig1}.  

Eight \textit{root} ribbons were build as case studies to characterize the edge structure and electronic properties of each family; 6 for the zigzag families and 2 for the armchair families (\ref{Fig2} and \ref{Fig3} respectively). Throughout this work we will mainly work with these 8 ribbons.

After structure relaxation of the \textit{root} ribbons, their electronic properties were investigated. All {\it root} nanoribbons were hydrogen passivated to satisfy the surface dangling bonds. The hydrogen passivated ribbons were also relaxed and their electronic properties were also investigated.  The semiconductor ribbons are expanded to $\sim$35 \AA\ in order to find the variation of the electronic band gap with the ribbon's width.

\section{Results and discussion}

\subsection{Structure and stability}

The 2D-NiSe$_2$ sheets were built and geometry optimized in the T and H configurations with a total energy for the T-structure 0.45 eV lower than for the metastable H structure.  In the T-structure the Ni-Se and Se-Se interatomic distances were 2.39 and 3.25 \AA\ respectively with a band gap of 0.11 eV. Metastable H-NiSe$_2$ presents Ni-Se and Se-Se interatomic distances of 2.41 and 2.63 \AA\ respectively, with a metallic behavior.  Phonon modes for both structures were calculated in order to guarantee the stability. Any negative frequency was found in both cases, an indication of the stability for both structures.  All these results are in good agreement with those reported by Ataca \textit{et al.}\cite{ataca2012stable}, validating our calculations and the starting system from where to build our nanoribbons.

The cohesive energy of T and H systems relative to free constituent atoms was calculated as  $E_C[NiSe_2]=E_T[Ni]+2E_T[Se]-E_T[NiSe_2]$, in terms of the total energy of NiSe$_2$, E$_T$[NiSe$_2$], and total energies of free Ni and Se atoms, E$_T$[Ni] and E$_T$[Se], respectively. The cohesive energies are 12.51 eV and 12.06 eV  for T and H structures respectively; indicating a strong cohesion relative to free atoms of the constituents. 

A high cohesion energy is important and required for stability of the compound, but more important for the synthesis is the formation energy ($E_F$) with respect to bulk systems, calculated with the expression $E_F= E_C[NiSe_2]-E_C[Ni]-2E_C[Se]$

Natural references for the formation energy of this compound are the corresponding Ni and Se.  The formation energy for T- and H-NiSe$_2$ systems are 0.81 and 0.36 eV respectively; the higher the value, the more stable the system.

Formation energies were also calculated with the experimental values of cohesive energies from Kittel {\it et al.} \cite{kittel2005introduction} as 3.20 and 2.74 eV for T and H, respectively. Both $E_F$'s are positive, implying that T-NiSe$_2$ and H-NiSe$_2$ are stable and metastable respectively, as previously reported in by Ataca {\it et al.} \cite{ataca2012stable}.

\begin{figure}
 \subfigure[Bare Ni-aligned-Ni]{\label{ArmNi-a-Ni}\includegraphics[width=0.15\textwidth]{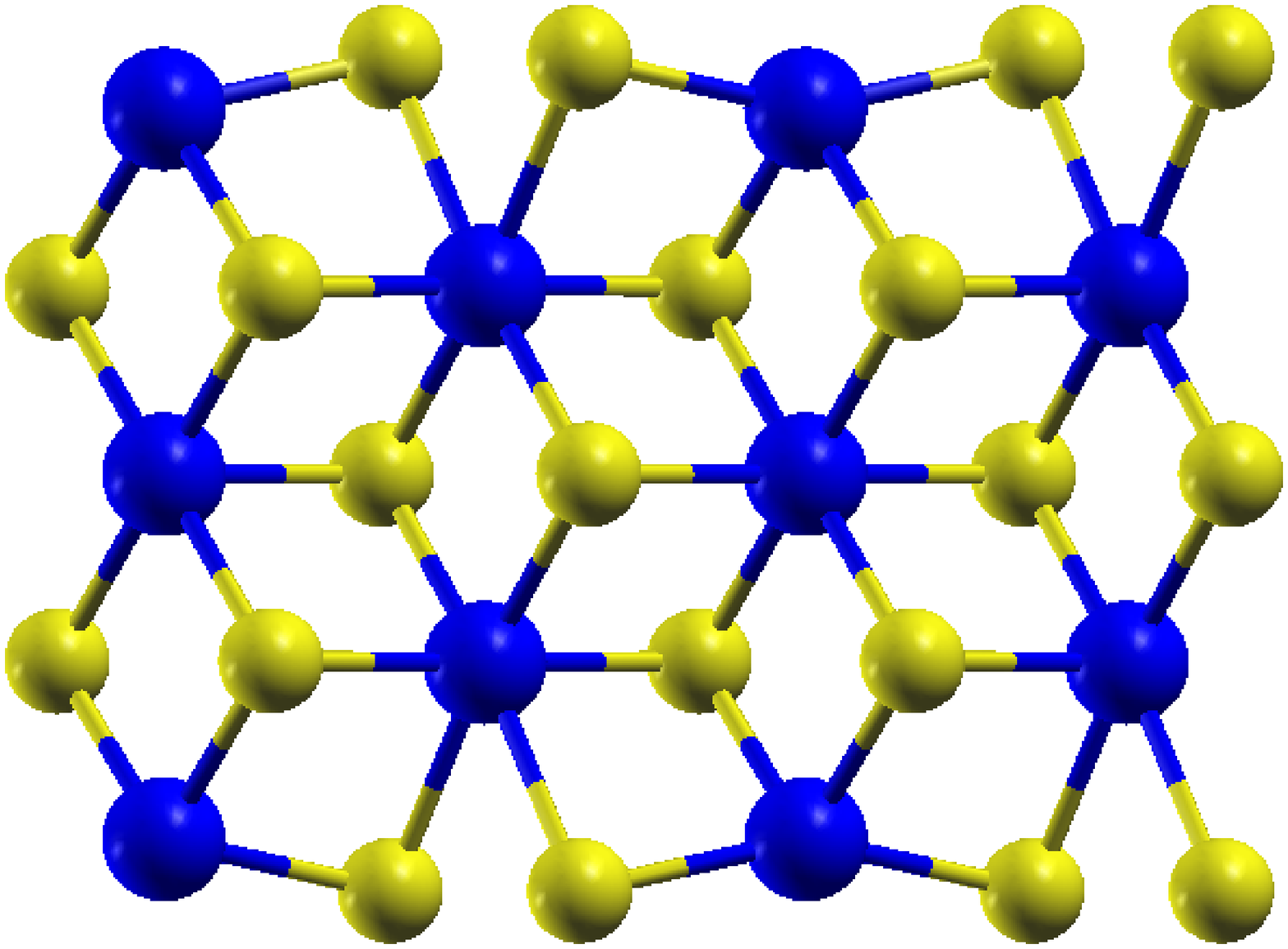}
 \includegraphics[width=0.16\textwidth]{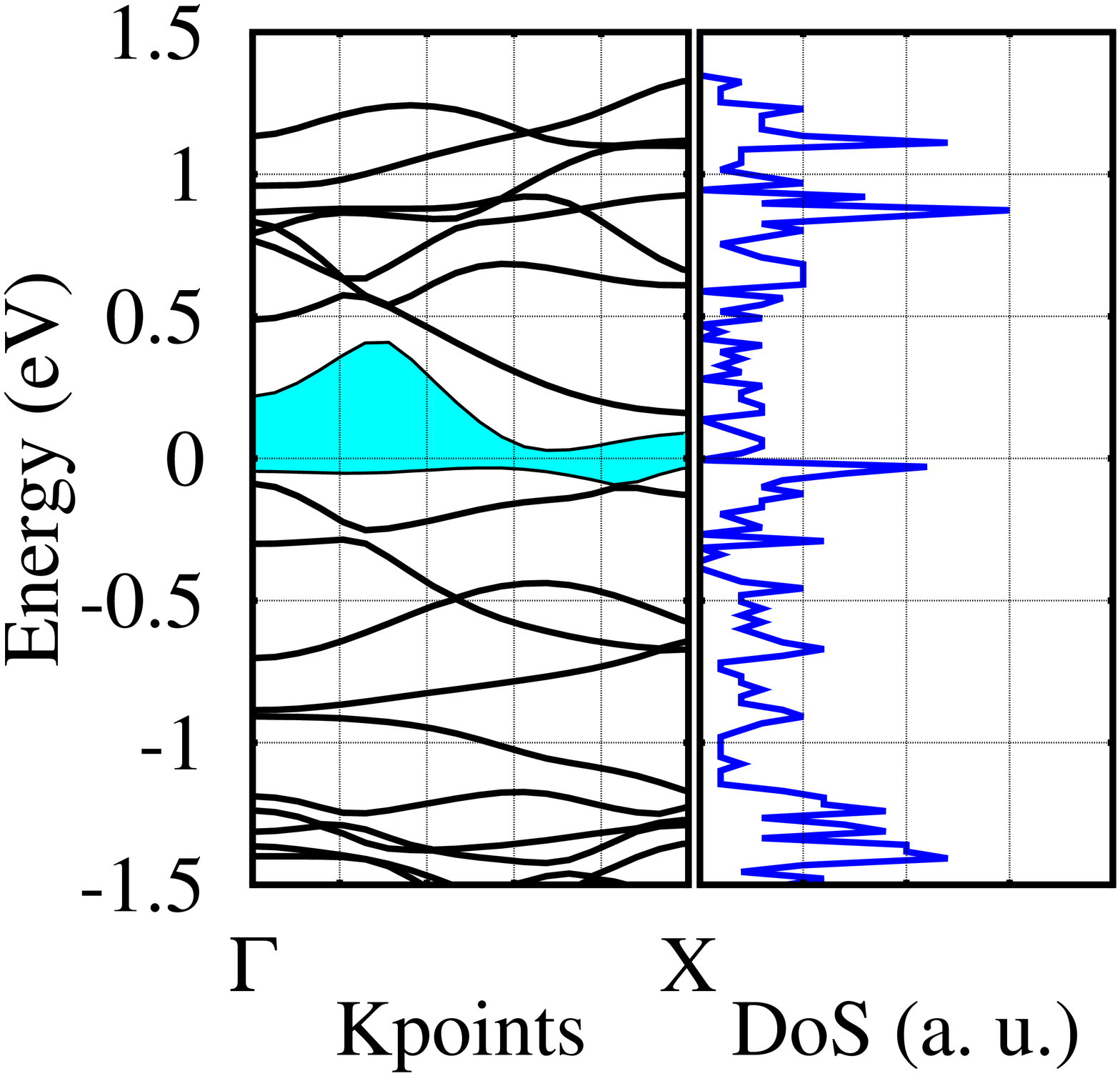}}

 \subfigure[Passivated Ni-aligned-Ni]{\includegraphics[width=0.15\textwidth]{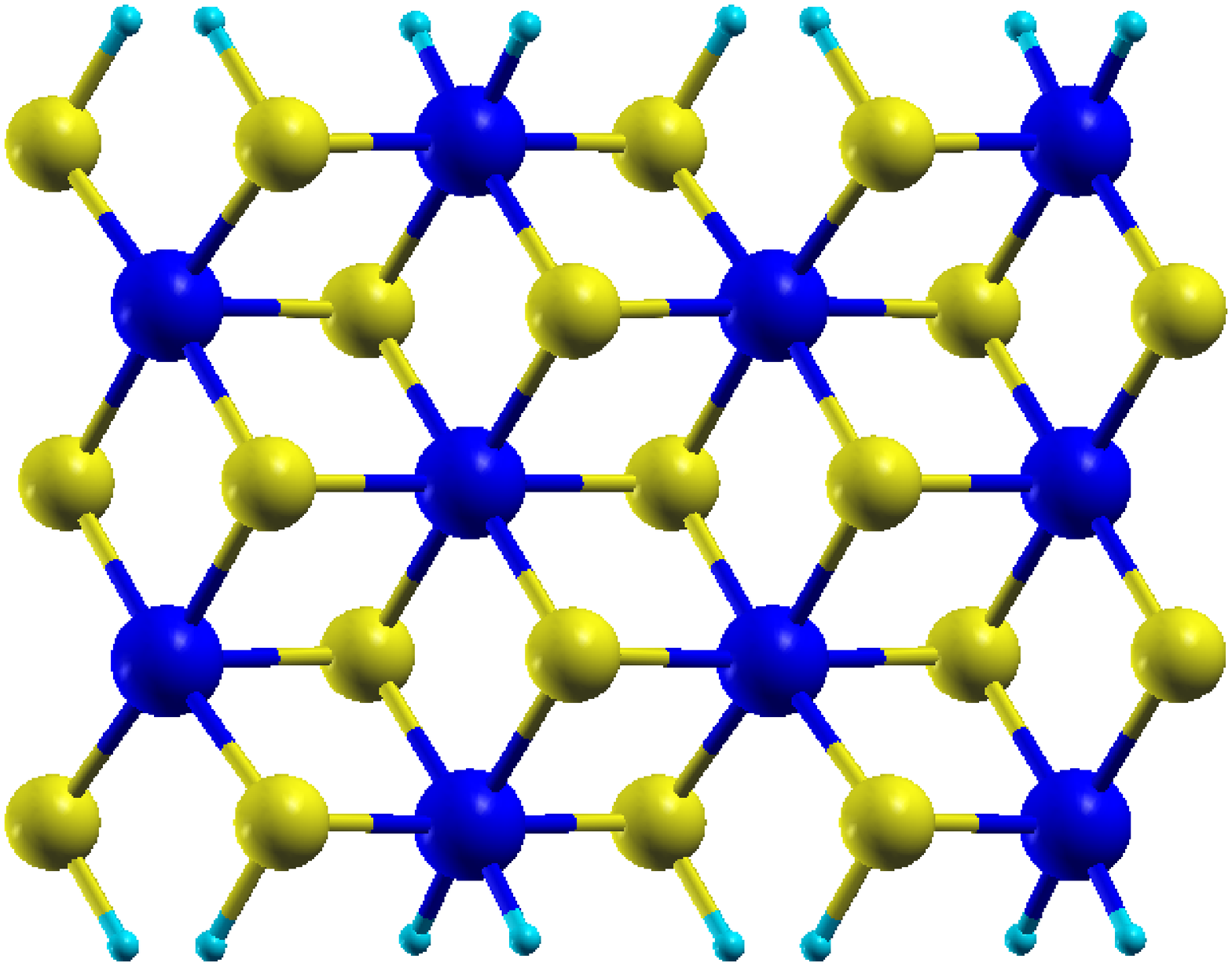}
 \includegraphics[width=0.16\textwidth]{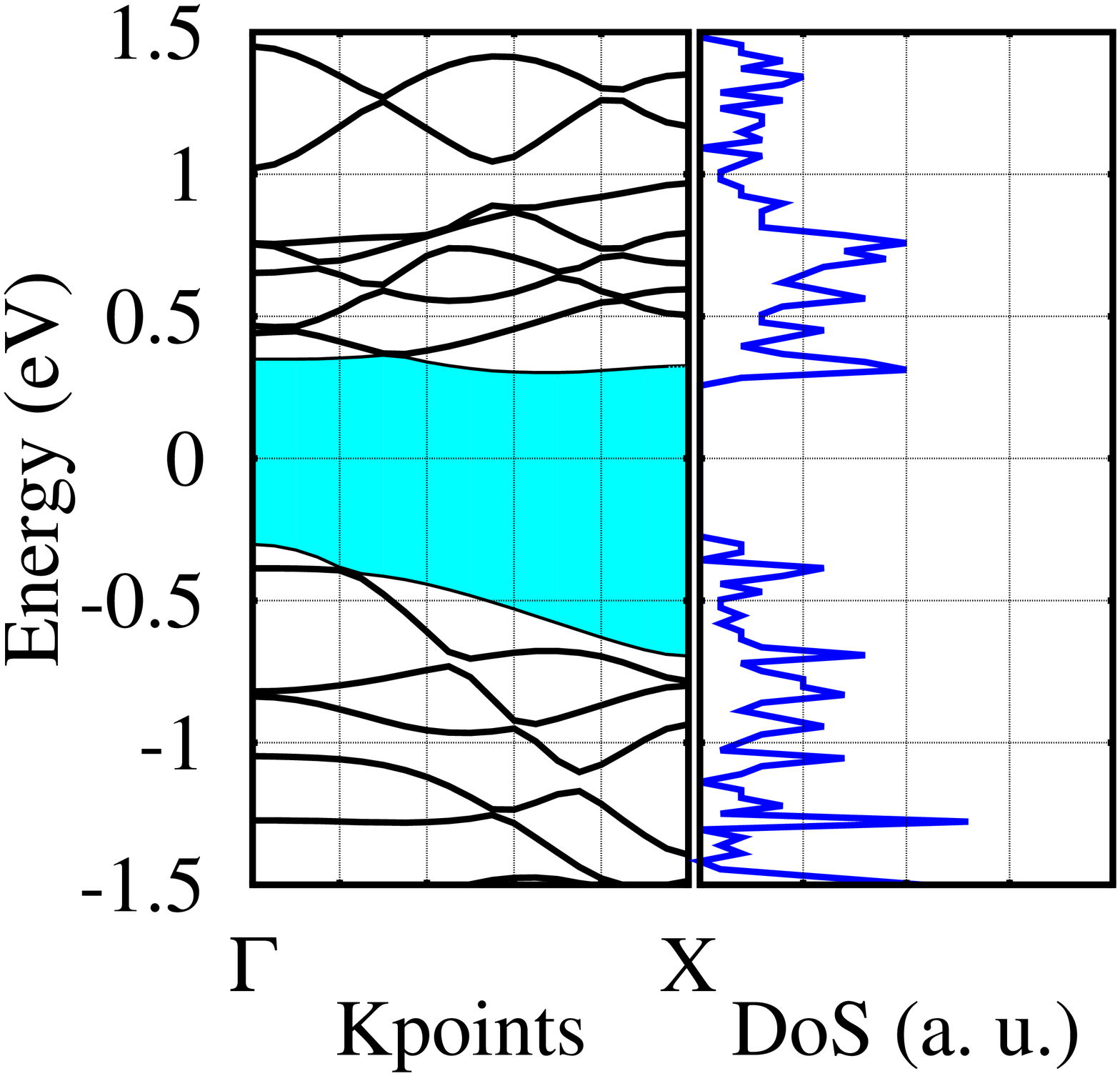}}

 \subfigure[Bare Ni-centered-Se]{\label{ArmNi-c-Se}\includegraphics[width=0.15\textwidth]{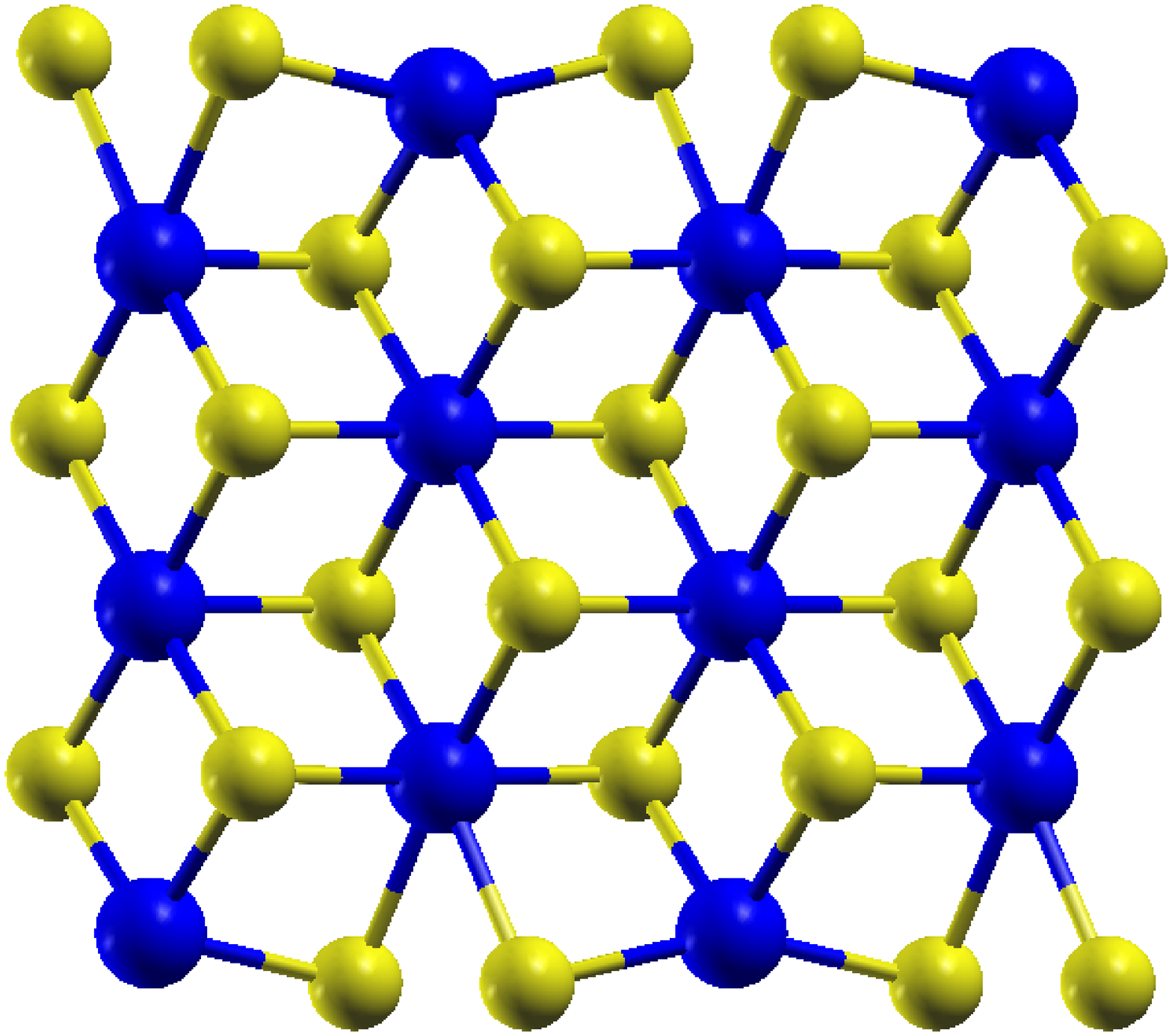}
 \includegraphics[width=0.16\textwidth]{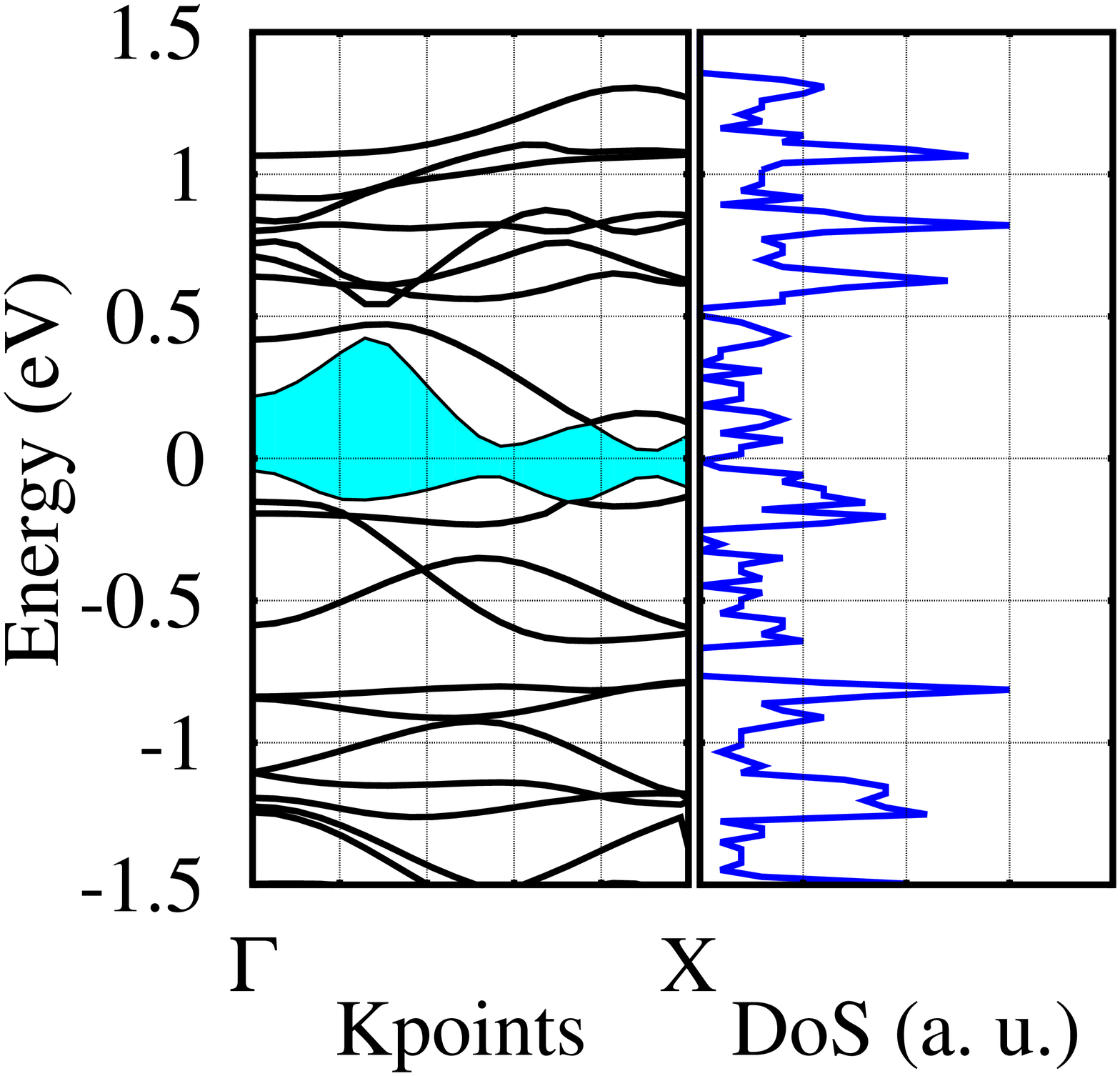}}

 \subfigure[Passivated Ni-centered-Se]{\includegraphics[width=0.15\textwidth]{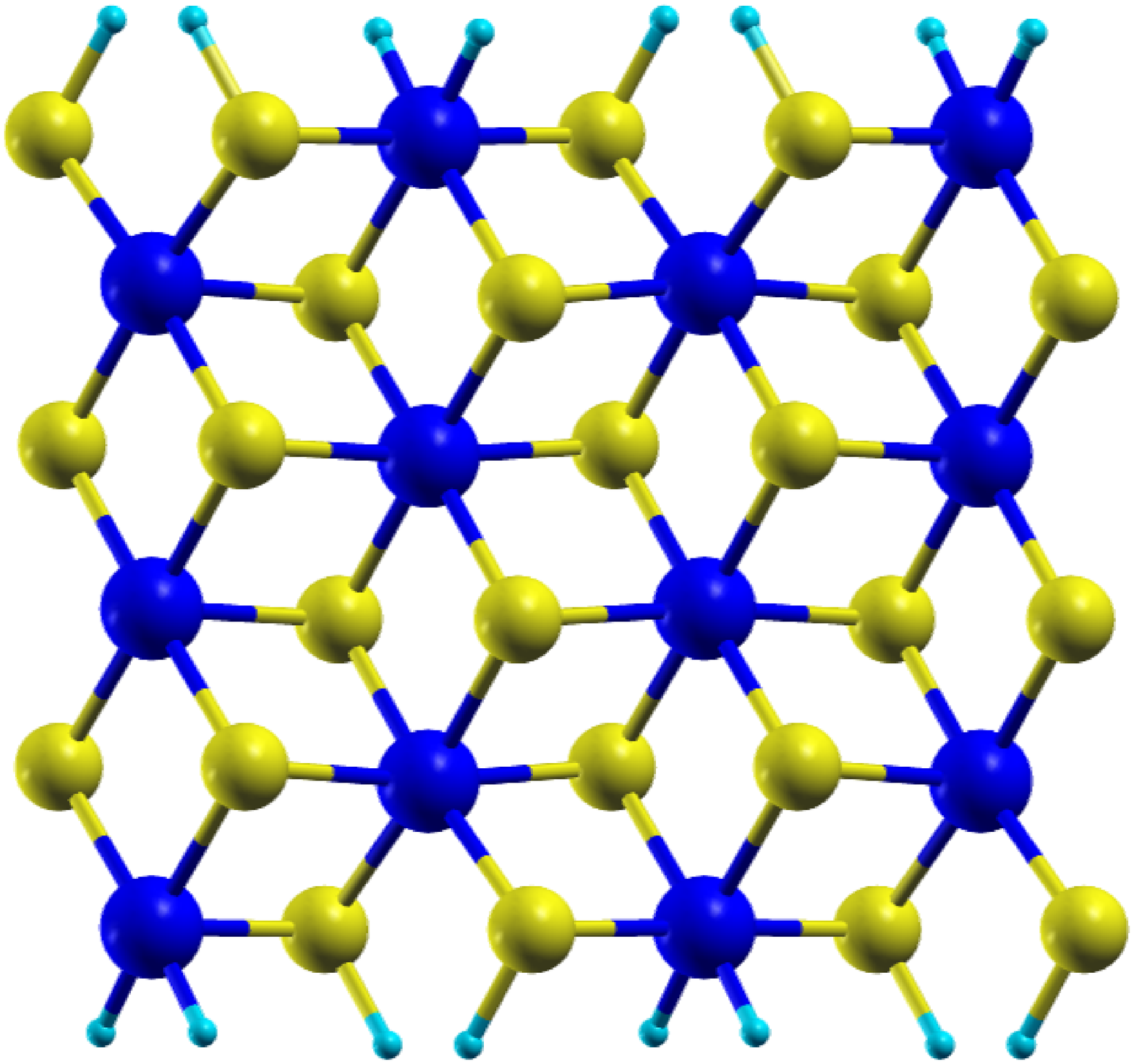}
 \includegraphics[width=0.16\textwidth]{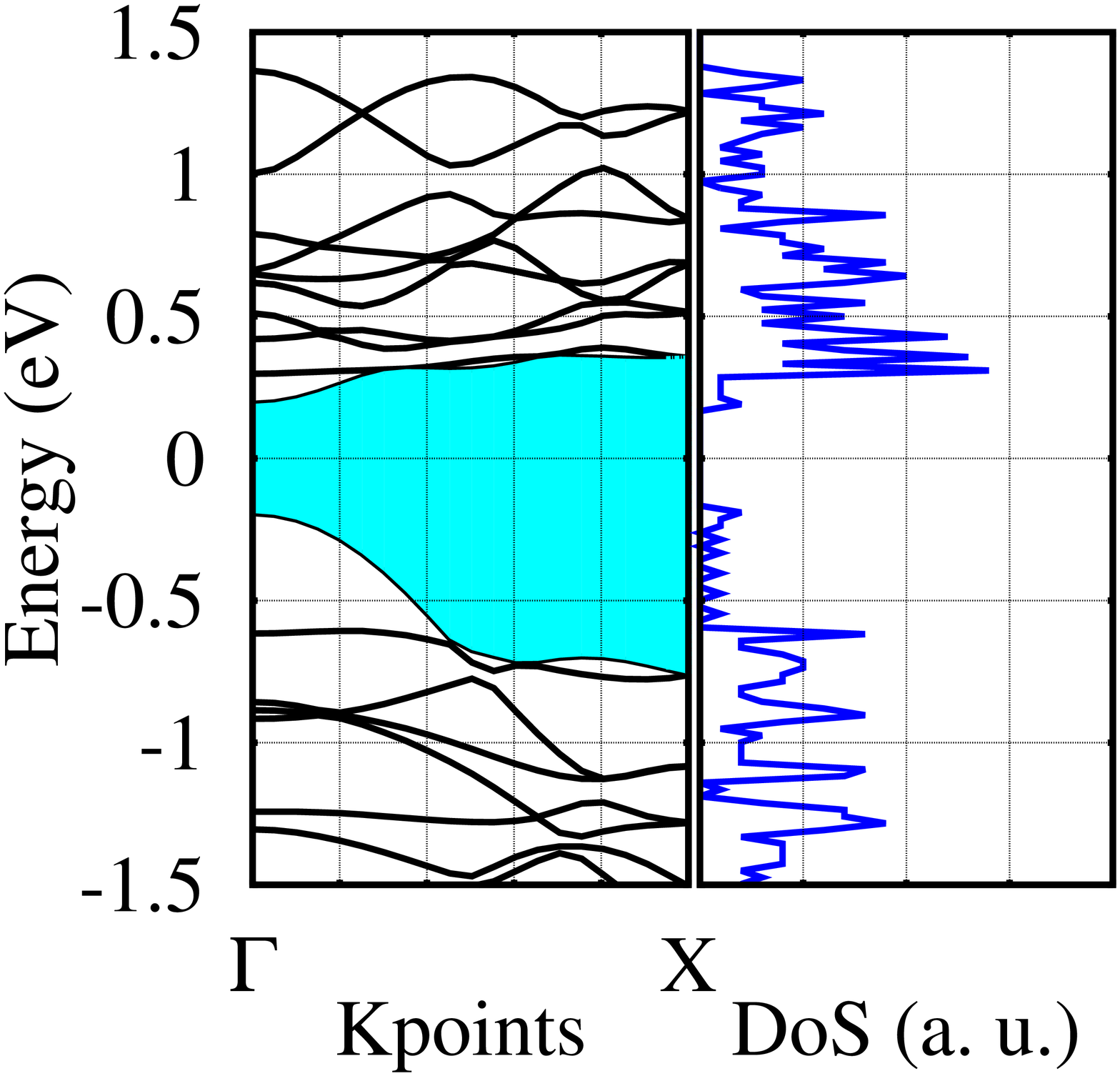}}

 \caption{(color online) The two different families for the armchair nanoribbons based on T-NiSe$_2$. Large (blue), medium (yellow) and small (aqua) circles are Ni, Se and H atoms. }\label{Fig3}
\end{figure}

We built our \textit{root} nanoribbons by cutting them from a 2D sheet of the relaxed T-2D-NiSe$_2$ 2D sheet in the zigzag and armchair directions (\ref{Fig2} and \ref{Fig3}). 

Their $E_B$ and $E_F$ energies were calculated using the expressions
$E_C=(nE_T[Ni]+mE_T[Se]+pE_T[H]-E_T[NiSe_2])/(n+m)$ and 
$E_F=(E_C[NiSe_2]-nE_C[Ni]-mE_C[Se]-pE_C[H])/(n+m)$, with $n$, $m$ and $p$ the number of Ni, Se and H atoms; $p >0$ only for H passivated ribbons. For the $E_F$ of hydrogen terminated ribbons, we use our calculated binding energy for H$_2$ of 3.25 eV/atom.

Bare ribbons present positive $E_{C}$ ranging from 3.92-4.07 eV/atom, which indicates strong cohesion relative to the constituents free atoms (\ref{Table1}). The three bare \textit{root} systems with lower cohesive energies are NiSeSe-NiSeSe, SeNiSe--NiSeSe and SeNiSe--NiSeSe, all characterized by two consecutive lines of semiconductor atoms at one or both of the edges. \ref{Table1} also includes the $E_F$ per atom for all bare ribbons. Bare SeNiSe--SeNiSe, SeNiSe--NiSeSe and Ni-centered-Se nanoribbons present an $E_F$ with values between the range of the T- and H-2D structures.

\begin{table}
\begin{center}
\begin{tabular}{cccrr}
\hline & \multicolumn{2}{c}{Cohesive energy} &\multicolumn{2}{c}{Formation energy} \\
Ribbon name& \multicolumn{2}{c}{per atom (eV)} &\multicolumn{2}{c}{per atom (meV)} \\
  &Bare & Passivated  &Bare & Passivated \\ \hline
  T-NiSe$_2$      & 4.17 &  --  & 270.0& --    \\
  H-NiSe$_2$      & 4.02 &  --  & 120.0& --    \\
   NiSeSe--SeSeNi & 4.07 & 5.90 & 24.5 & -93.7 \\
   NiSeSe--SeNiSe & 4.06 & 5.48 & 94.2 &  37.0 \\
   NiSeSe--NiSeSe & 3.97 & 5.59 & 78.1 &  67.1 \\
   SeNiSe--SeNiSe & 4.05 & 5.13 & 152.0& 147.8 \\
   SeNiSe--NiSeSe & 4.00 & 5.21 & 154.1& 121.3 \\
   SeSeNi--NiSeSe & 3.92 & 5.28 & 119.4& 101.3 \\
   Ni-aligned-Ni  & 4.01 & 5.69 & 107.2&  60.6 \\
   Ni-centered-Se & 4.04 & 5.44 & 135.4& 101.2 \\
   \hline
\end{tabular}
\end{center}\caption{Cohesive and formation energies for 2D systems, bare and H-passivated zigzag and armchair nanoribbons}\label{Table1}
\end{table}

Relaxation of the armchair ribbon edges results in an increment of the edge Ni-Se distance in 1.67 \% with respect the bulk Ni-Se distance, while the edge Se-Ni-Se angle is reduced in 24\%. 

As mentioned before, all \textit{root} ribbons were hydrogen passivated. In the T-NiSe$_2$ 2D structure, each Ni atom is bonded to 6 Se atoms and each Se atom is bonded to 3 Ni atoms. In contrast, ribbons edges present dangling bonds that were satisfied with hydrogens for passivation. We call \textit{natural} passivation the one shown in \ref{Fig2} and \ref{Fig3} for zigzag and armchair ribbons, with a hydrogen atom added to the system to complete the 6 and 3 bonds for each Se or Ni edge atoms. 

After H passivation and geometry relaxation of the ribbons, H-Se and H-Ni distances are $\sim$1.49 \AA\ and $\sim$1.47 \AA\ respectively.

\subsection{Electronic Properties}

Band structures of all {\it root} systems were obtained with the same results for spin polarized and non-spin-polarized calculations, which indicates that the systems are non-magnetic in all cases; \textit{i.e.} all the structures present perfect spin degeneration. 

We find that all \textit{root} zigzag bare-terminated nanoribbons are metallic. After \textit{natural} hydrogen passivation, only two of them have an energy band gap, the SeNiSe--SeNiSe and SeNiSe--NiSeSe. Contrary to the results in zigzag  MoS$_2$  nanoribbons\cite{botello2009metallic}, where bare and hydrogenated zigzag nanoribbons were metallic. The main difference between MoS$_2$ zigzag nanoribbons and our systems is the base 2D structure metal-chalcogenide. In zigzag nanoribbons based on H-like, there is no possibility that two consecutive chalcogenide atomic lines can exist at the same edge. While in zigzag nanoribons based on T-like structures, it is possible to have two consecutive chalcogenide (Se) atomic lines, even more, the geometry allows to have the last line of atoms in both edges composed of chalcogenides with one and two edge lines of Se atoms. 

\begin{figure}
 \includegraphics[width=0.45\textwidth]{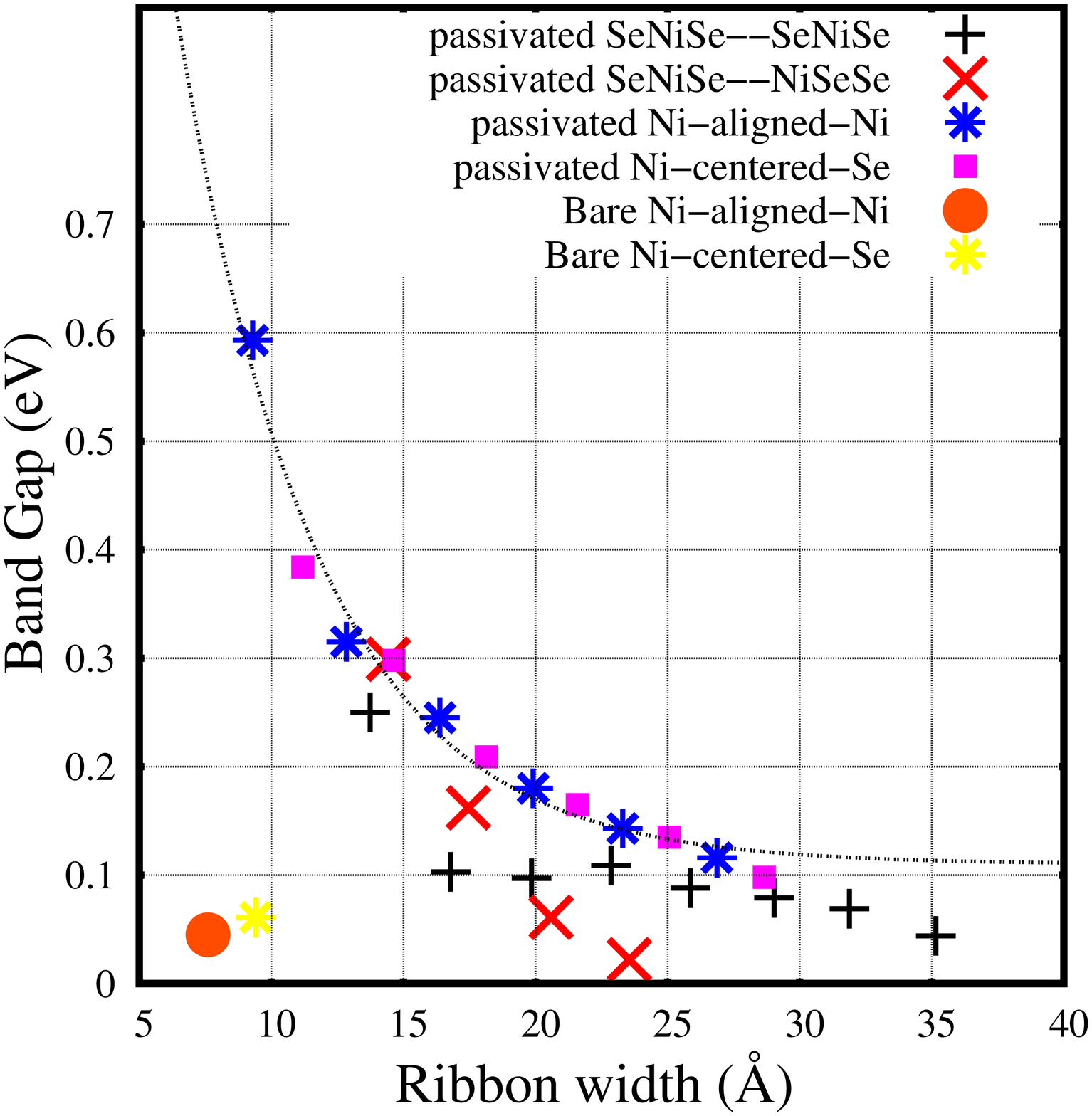}	
 \caption{(color online) The variation of band gaps of bare and hydrogen passivated ribbons as a funtion of width. The dotted line represents the asymptotic fitting of the passivated armchair.  }\label{Fig4}
\end{figure}

Zigzag nanoribbons are divided in two groups: those with an outermost Ni atomic line (NiSeSe--SeSeNi, NiSeSe--SeNiSe and NiSeSe--NiSeSe, figures \ref{Fig2}\subref{NiSeSe-SeSeNi}, \subref{NiSeSe-SeNiSe} and \subref{NiSeSe-NiSeSe} and their corresponding H passivations: \ref{Fig2}\subref{HNiSeSe-SeSeNi}, \subref{HNiSeSe-SeNiSe} and \subref{HNiSeSe-NiSeSe}), and those with all outermost Se atomic lines (SeNiSe--SeNiSe, SeNiSe--NiSeSe and SeSeNi--NiSeSe, figures \ref{Fig2}\subref{SeNiSe-SeNiSe},\subref{SeNiSe-NiSeSe} and \subref{SeSeNi-NiSeSe} and their corresponding passivations:  \ref{Fig2}\subref{HSeNiSe-SeNiSe},\subref{HSeNiSe-NiSeSe} and \subref{HSeSeNi-NiSeSe}). Surprisingly all bare and hydrogen passivated zigzag nanoribbons of the former group were metals, unlike the common finding that  semiconductor nanostructures always have a larger band gap than their 2D values \cite{deng2012effect}.  Recent studies \cite{Lukowski, Voiry}  show that centered honeycomb nanostructures enhance electrocatalytic activity due to the high concentration of metallic edges, this suggest that, NiSeSe--SeSeNi will be the best candidate as catalyst for hydrogen evolution, but also NiSeSe--SeNiSe and NiSeSe--NiSeSe will be used for the same process.
In the latter group, two of the H passivated \textit{root} ribbons are semiconductors (\textit{i.e.} the \textit{natural} H passivated SeNiSe--SeNiSe and SeNiSe--NiSeSe ribbons). The variation ribbon's width versus the band gap value is shown in  \ref{Fig4}. According to \ref{Table1}, the stablest zigag ribbon is the SeNiSe--SeNiSe, which is found to have a semiconductor behavior for all calculated widths ( \ref{Fig4}). In contrast, we found that, as the width of the SeNiSe--NiSeSe ribbon increases, the system presents a semiconductor-metal transition at $\sim$23.55 \AA. 

We also explored other possible edge hydrogen passivation densities for all \textit{root} zigzag ribbons.  The hydrogen binding energy ($E_B$) was used as an indicator of higher or lower stability, see \ref{Table2}. For example: NiSeSe--SeSeNi was hydrogen passivated with 3 (\textit{i. e.} the \textit{natural} full passivation of the dangling bonds), 2 and 1 hydrogens per edge Ni atom (H6, H3, and H2). In the case of the NiSeSe--SeNiSe ribbon, one edge was fixed with one hydrogen attached to the outermost Se atom and one hydrogen to its neighboring Ni atom, while on the other edge, the hydrogenation varied from 3, 2 and 1 H for the edge Ni atom (H5, H4, and H3). For the NiSeSe--NiSeSe ribbon, four hydrogenations were studied; one, where all dangling bonds are satisfied (the so called \textit{natural} hydrogenation with 6 H in total, H6), two, where one hydrogen is attached to the outermost Se edge atoms (H5), three, where one H atom is attached to the outermost Se edge atom and two to the outermost Ni 
edge atom (H4), and four, where the outermost Se and Ni edge atom in opposited edges are only passivated with one hydrogen each one (H3). 

The \textit{natural} hydrogenation for the SeNiSe--SeNiSe ribbon is the following: Ni atoms and outermost Se atoms were passivated with one hydrogen p/atom, satisfying the dangling bonds. Alternative passivations were: 1. Remove the H passivation of the Ni atoms, leaving it with one dangling bond (H2-Se). 2. Removing the H passivation of the Se atom,  leaving it with one dangling bond (H2-Ni). 
The latter case of the unpassivated Se atom is a clear example of a non stable system with $E_B = $  2.78 eV, which is almost 0.5 eV below the energy for H$_2$, turning the electronic properties of the fully passivated semiconductor ribbon to a metallic behavior. This means that in presence of H$_2$ any ribbon with edge Se atoms should always be H passivated.

In ribbons with two consecutive atomic rows of Se at any of the two edges, the \textit{natural} passivation includes two hydrogens attached to the outermost Se atoms and one hydrogen attached to the neighbor Se atom. The SeNiSe--NiSeSe ribbon was passivated with two (H5) and one (H4) H in the outer Se for the consecutive Se rows, while the others Se remained with one bonded hydrogen atom. Similarly, the SeSeNi--NiSeSe ribbon was passivated with two (H6) and one (H4) hydrogen atoms in the outer Se for the consecutive Se rows.

Metallic ribbons from the  SeSeNi--NiSeSe  family were build with widths from 17.08 to 33.17 \AA, with all the outer Se atoms passivated with two hydrogens, \textit{i. e.} fully satisfying the dangling bonds.  In some cases, relaxation of these ribbons resulted in the separation of SeH$_2$ molecules from both edges, and a final configuration of a SeNiSe--SeNiSe ribbon. This is a strong suggestion that if ribbons are experimentally formed by cutting 2D T-NiSe$_2$,  ribbons from the  SeNiSe--SeNiSe family are much more probable to be found than those from the SeSeNi--NiSeSe family. Also, during experiments SeH$_2$ molecules are prone to desorb from the ribbon's edges.


\begin{table}
\begin{center}
\begin{tabular}{|c|c|c|r|r|}
   \hline
   \multicolumn{5}{|c|}{Average binding energy of hydrogen (eV)}\\
   \hline
    & H6   &H4  &  H2 &\\
    NiSeSe--SeNiSe& 3.05 & 3.09 & 3.06 &\\
   \hline   
    & H5   &H4  &  H3 &\\
    NiSeSe--SeNiSe & 3.12 & 3.15 & 3.18 &\\
   \hline
    & H6   &H5  & H4 &H3 \\
   NiSeSe--NiSeSe & 3.23 & 3.34 & 3.38 &3.30 \\ 
   \hline
     & H4   &H2-Ni  & H2-Se &  \\
   SeNiSe--SeNiSe & 3.24 & 2.78 &  3.47&\\
   \hline
    & H5   &H4  & \multicolumn{2}{c|}{} \\
   SeNiSe--NiSeSe & 3.16 & 3.26 & \multicolumn{2}{c|}{}\\
   \hline
    & H6   &H4  &  \multicolumn{2}{c|}{}\\
   SeSeNi--NiSeSe & 3.21 & 3.42 & \multicolumn{2}{c|}{}\\
   \hline
\end{tabular}
\end{center} \caption{Average binding energy for H passivation of different passivation densities for the six zigzag nanoribbons.}\label{Table2}
\end{table}

Several of the $E_B$s are higher than the $E_B$ in H$_2$, and in general the energy increases as the density decreases.

It is worth mentioning that all metallic systems remained metallic and the two semiconductors became metallic when reducing the hydrogenation densities.

Our \textit{root} armchair bare-termination nanoribbons are semiconductors with a very small band gap; 0.045 eV and 0.061 eV for Ni-aligned-Ni and Ni-centered-Se, respectively. Bare terminated armchair ribbons were built with widths from 10 \AA\ to 30 \AA\ and only the \textit{root} ribbons are semiconductors. In this work, the largest electronic band gap is for hydrogen passivated armchair nanoribbons with gaps of 0.59 and 0.38 eV, for passivated \textit{root} Ni-aligned-Ni and passivated Ni-centered-Se nanoribbons respectively. In figure \ref{Fig4}, the variation of electronic band gap for hydrogen passivated is reported. The tendency of the band gap for the armchair follows the expression $E(w)=0.11+ae^{-w/\lambda}$, where a= 2.59 eV, $\lambda=$5.33 \AA\ and $w$ is the ribbon's width.

\begin{figure*}
 \subfigure[Passivated SeNiSe--SeNiSe]{\label{HSeNiSe-SeNiSeC}
 \includegraphics[width=0.20\textwidth]{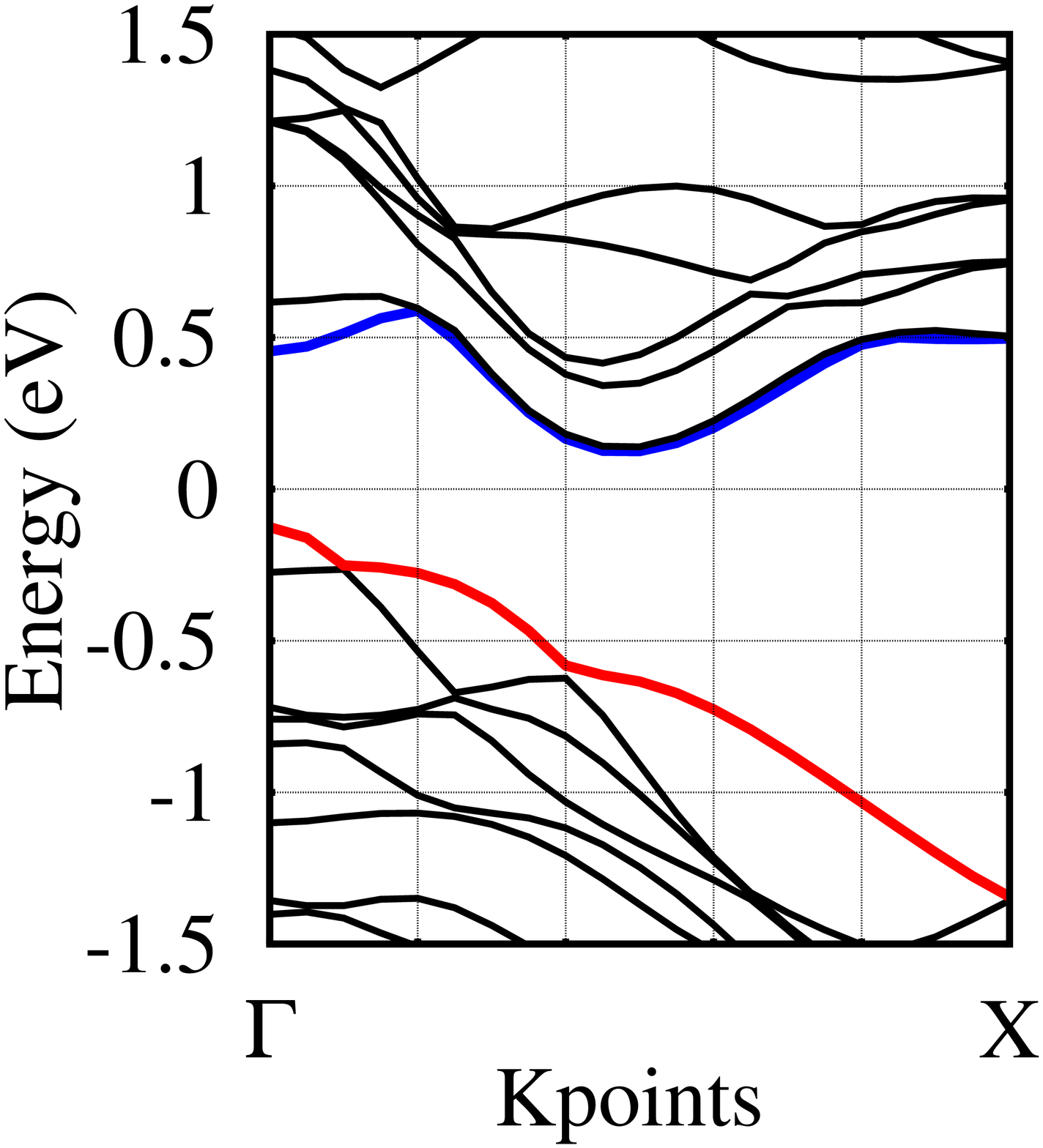}
 \includegraphics[width=0.15\textwidth]{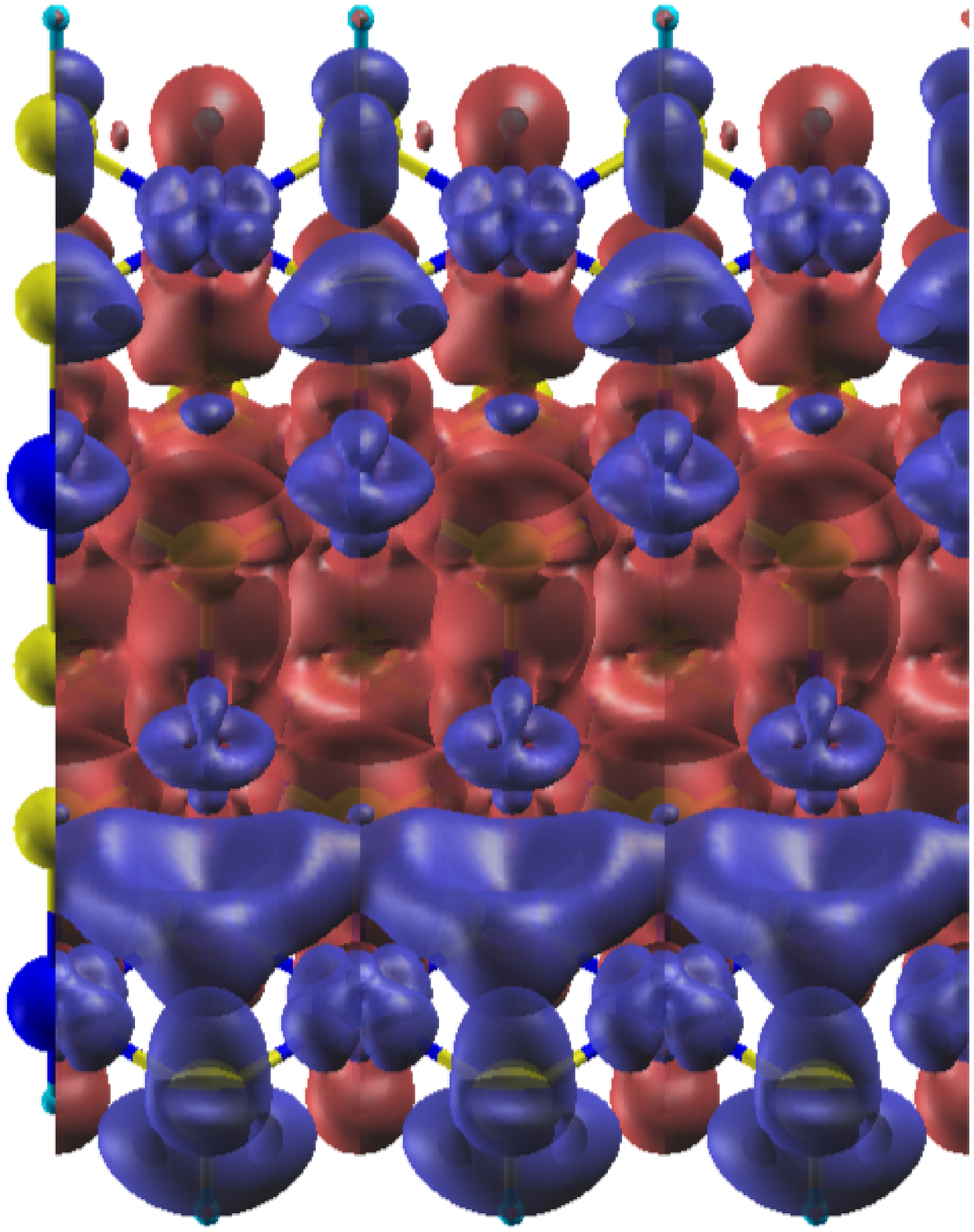}
 \includegraphics[width=0.09\textwidth,height=0.15\textheight]{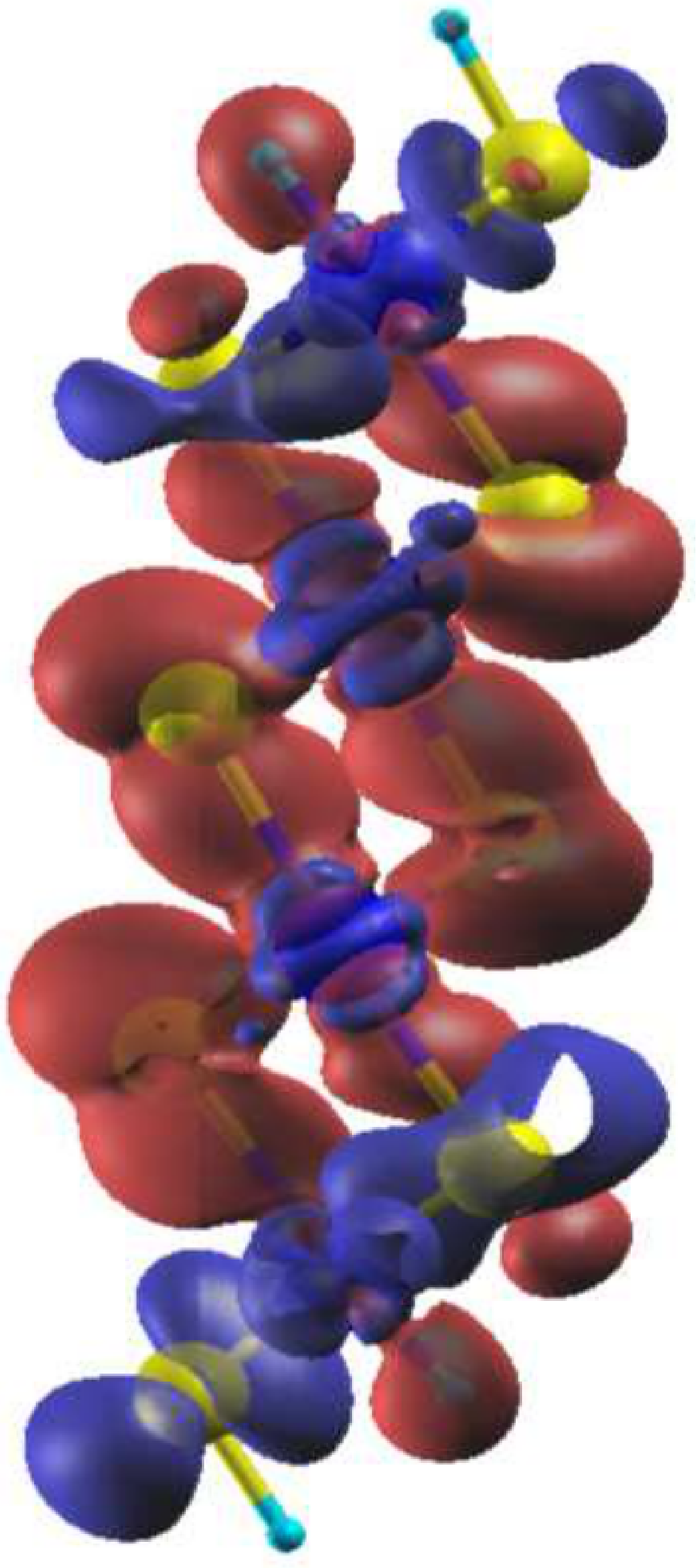}}
 \subfigure[Passivated SeNiSe--NiSeSe]{ \label{HSeNiSe-NiSeSeC}
 \includegraphics[width=0.20\textwidth]{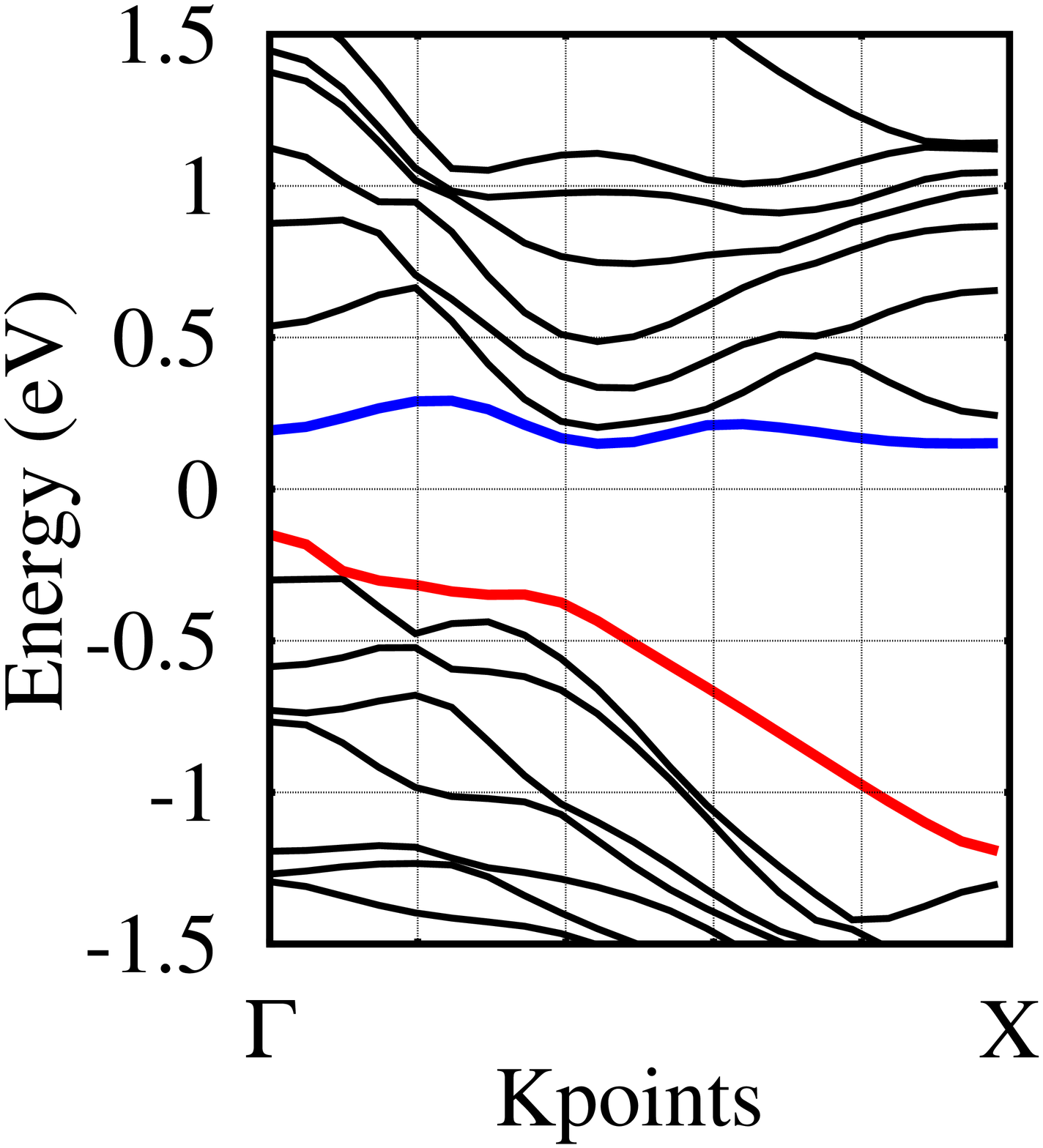}
 \includegraphics[width=0.15\textwidth]{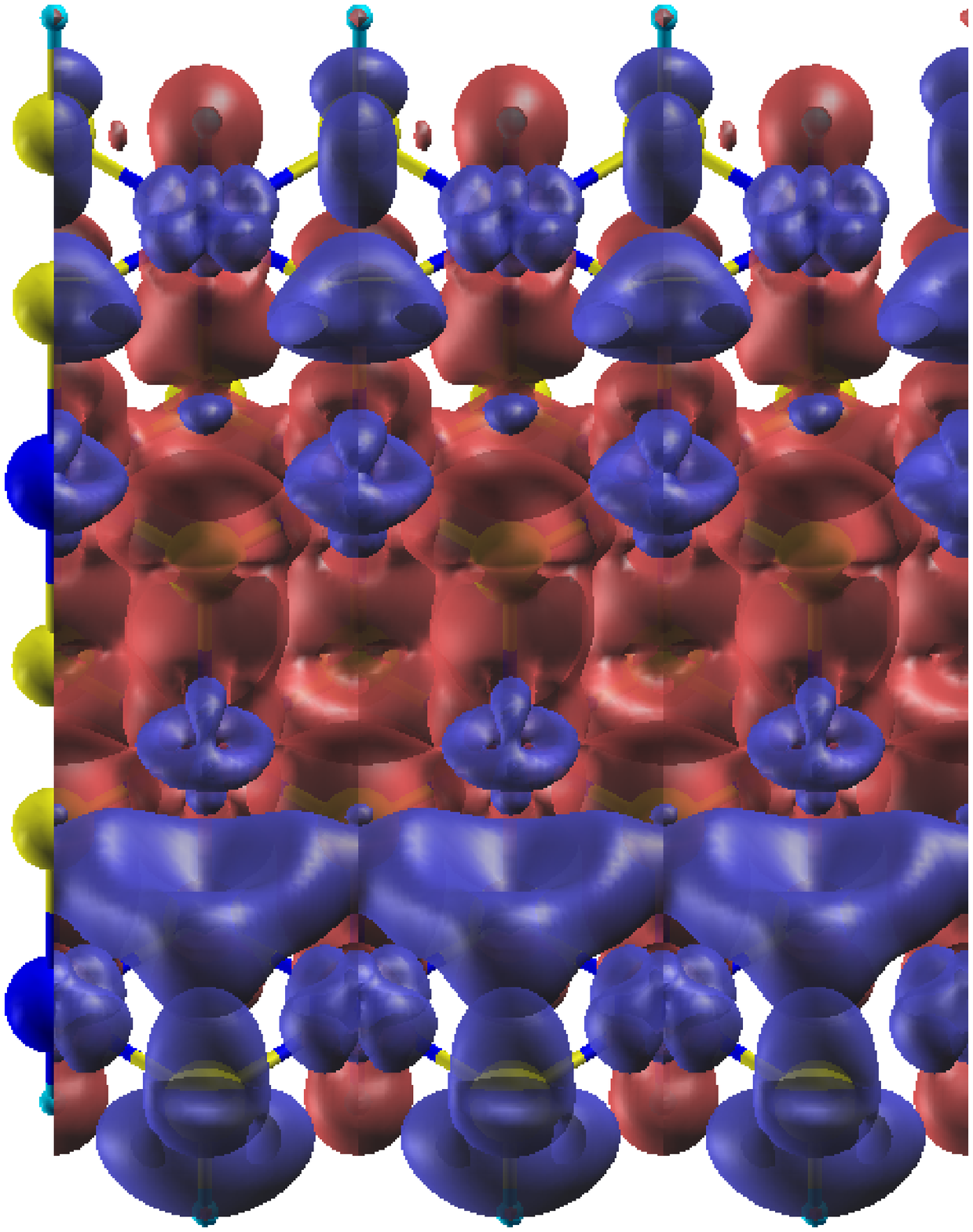}
 \includegraphics[width=0.09\textwidth,height=0.15\textheight]{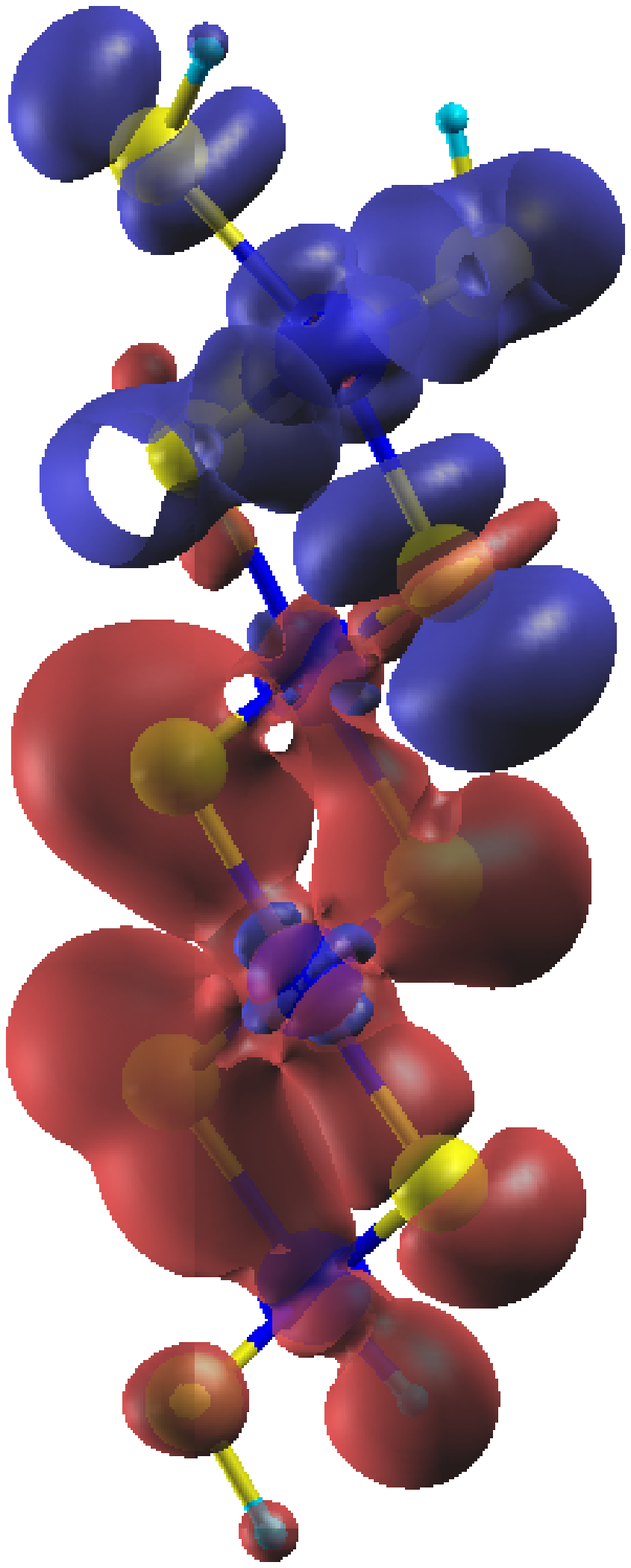}}
 \subfigure[Bare Ni-aligned-Ni]{\label{Ni-aligned-Ni}
 \includegraphics[width=0.20\textwidth]{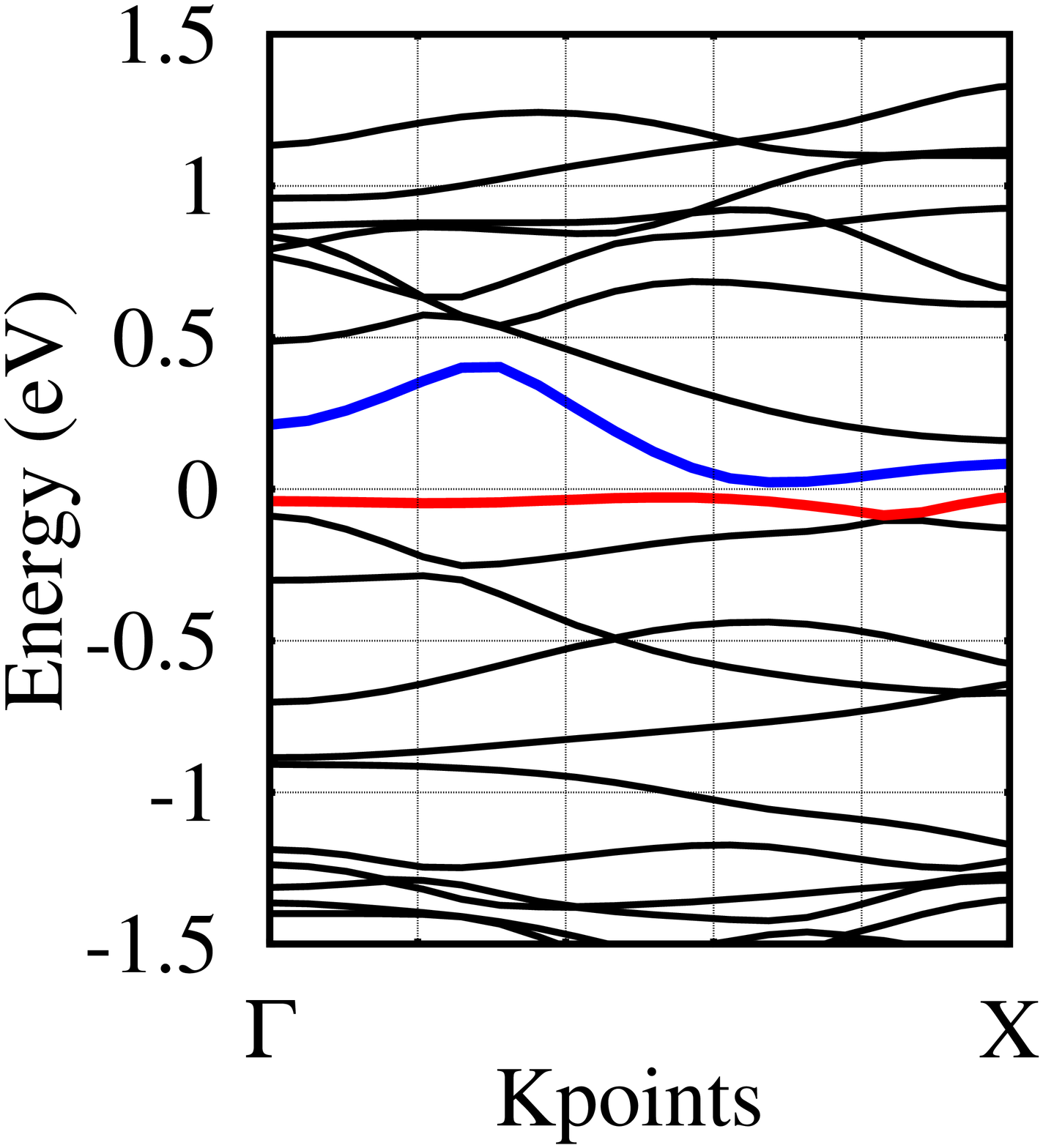}
 \includegraphics[width=0.15\textwidth]{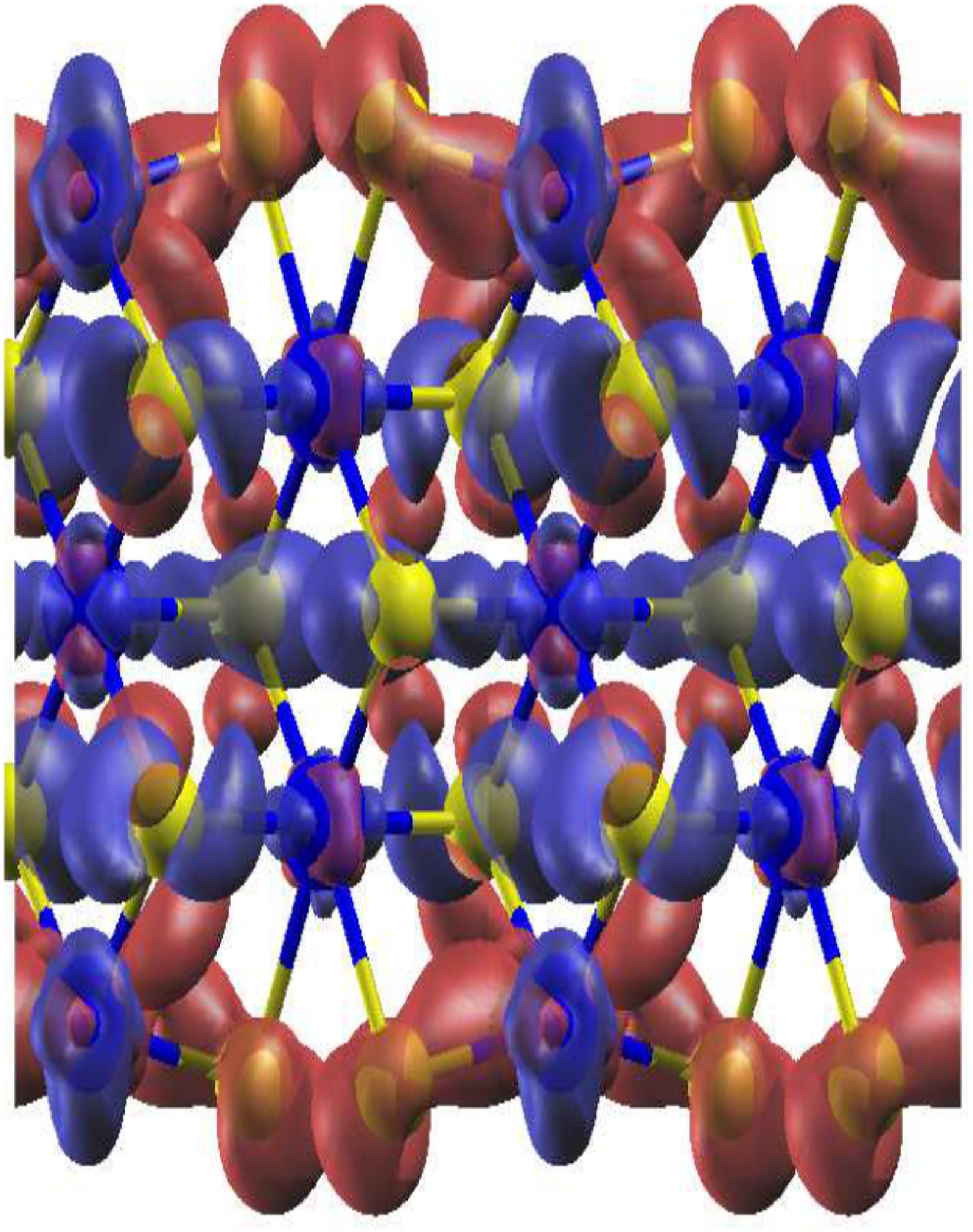}
 \includegraphics[width=0.09\textwidth,height=0.15\textheight]{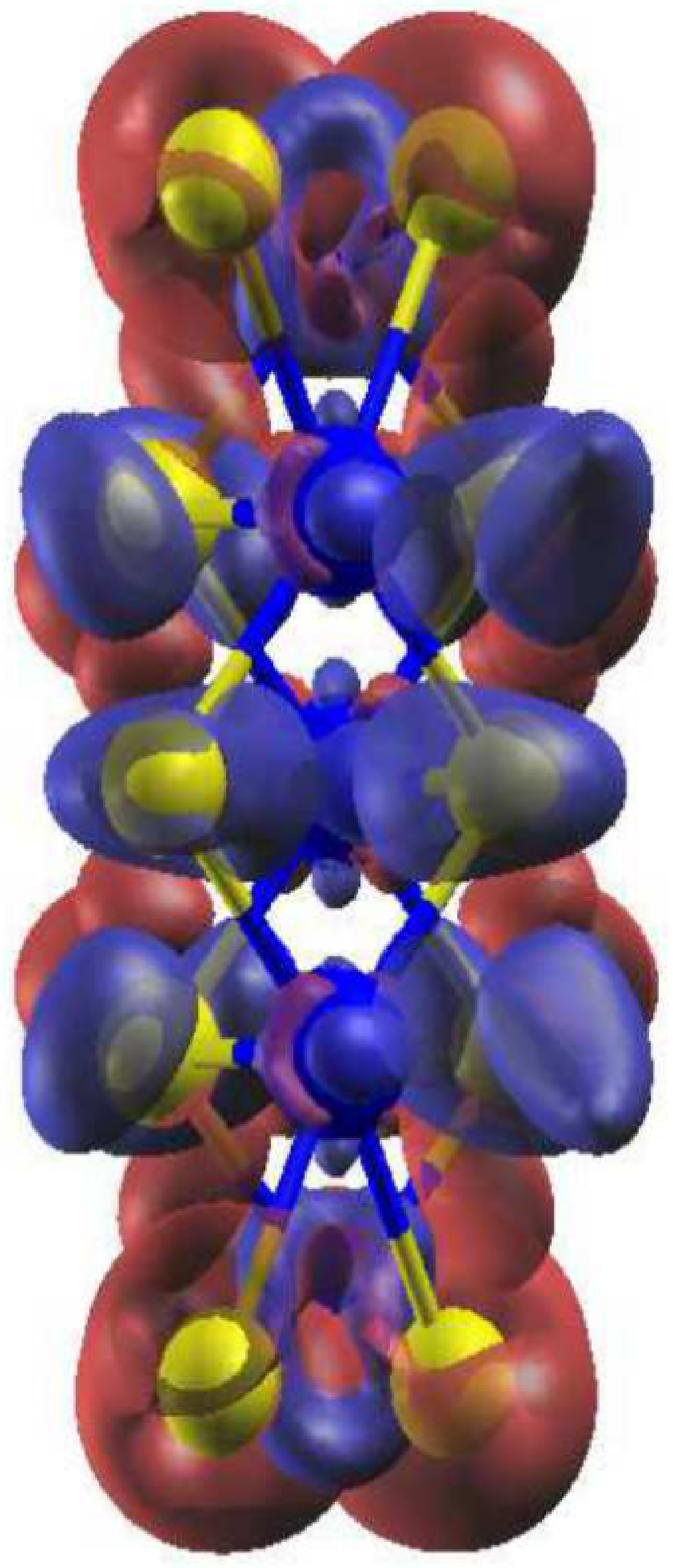}}
 \subfigure[Passivated Ni-aligned-Ni]{ \label{HNi-aligned-Ni}
 \includegraphics[width=0.20\textwidth]{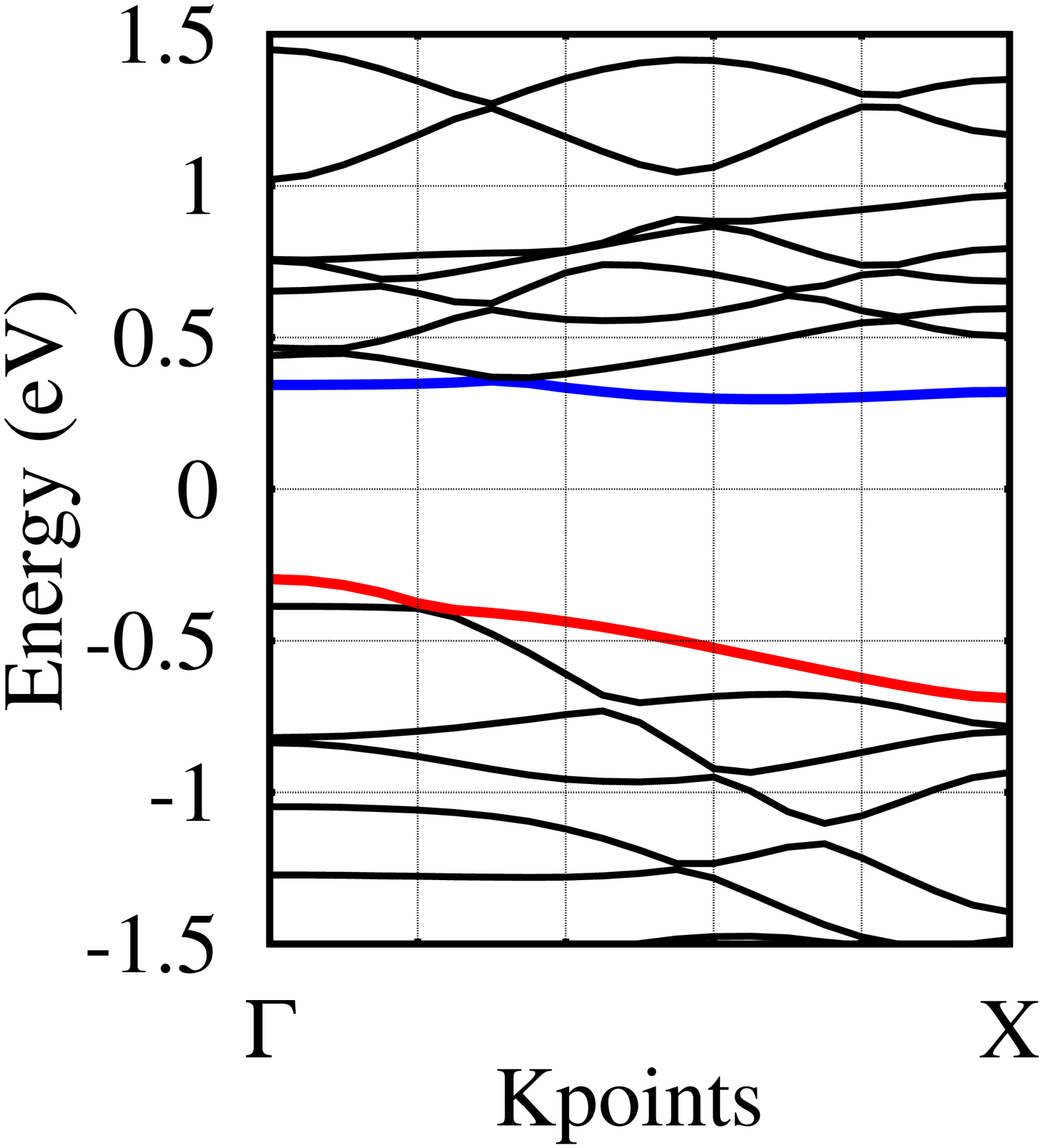}
 \includegraphics[width=0.15\textwidth]{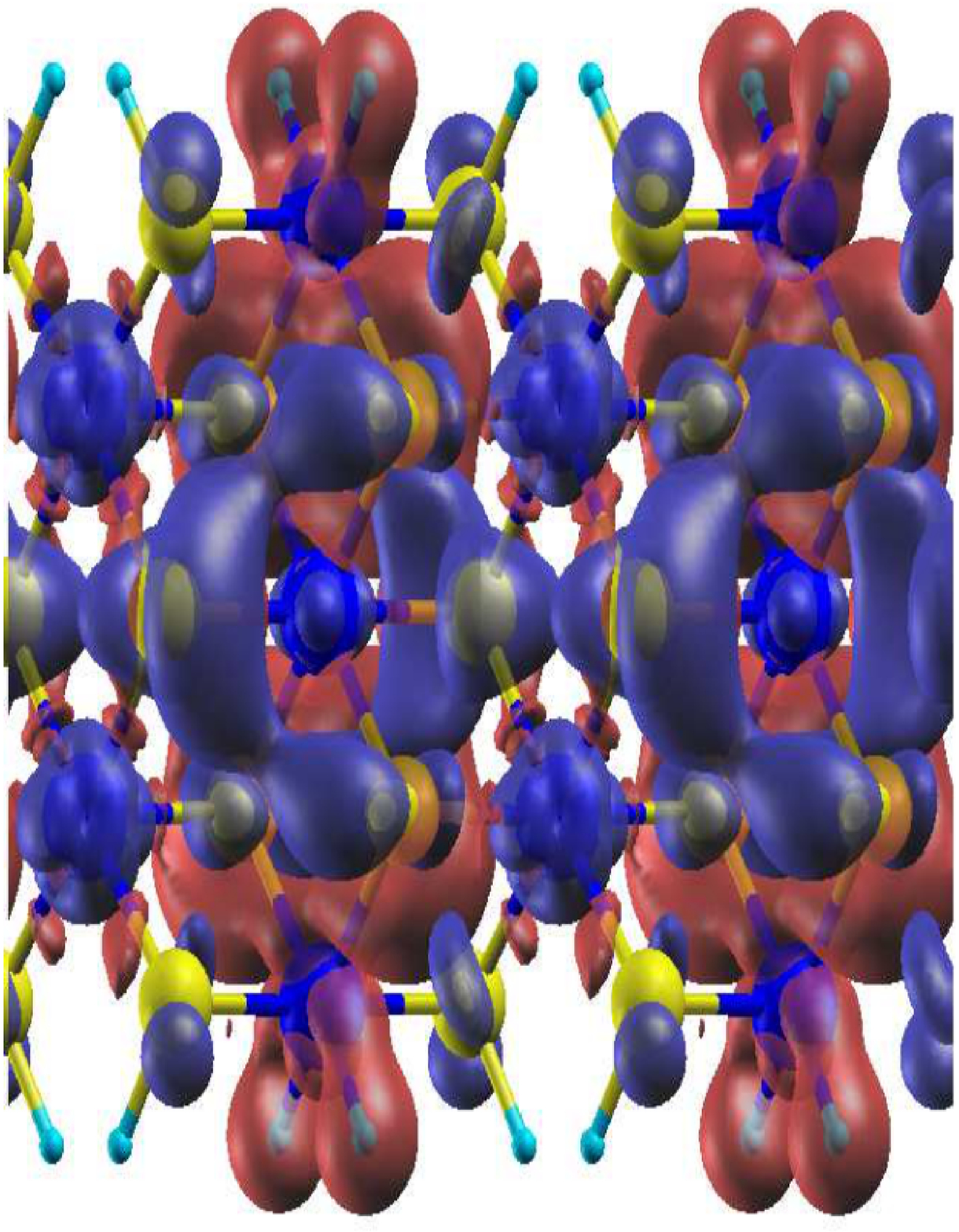}
 \includegraphics[width=0.09\textwidth,height=0.15\textheight]{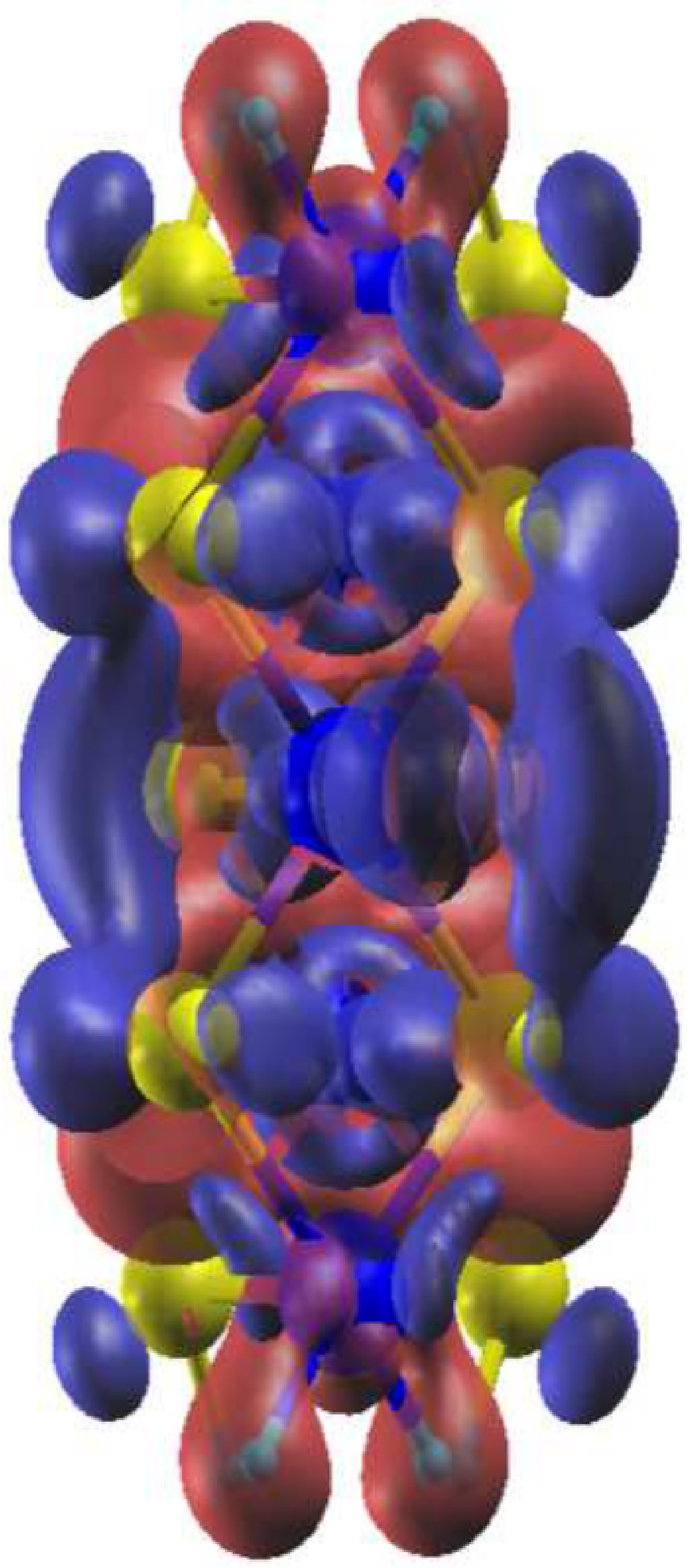}}
 \subfigure[Bare Ni-centered-Se]{ \label{Ni-centered-Se}
 \includegraphics[width=0.20\textwidth]{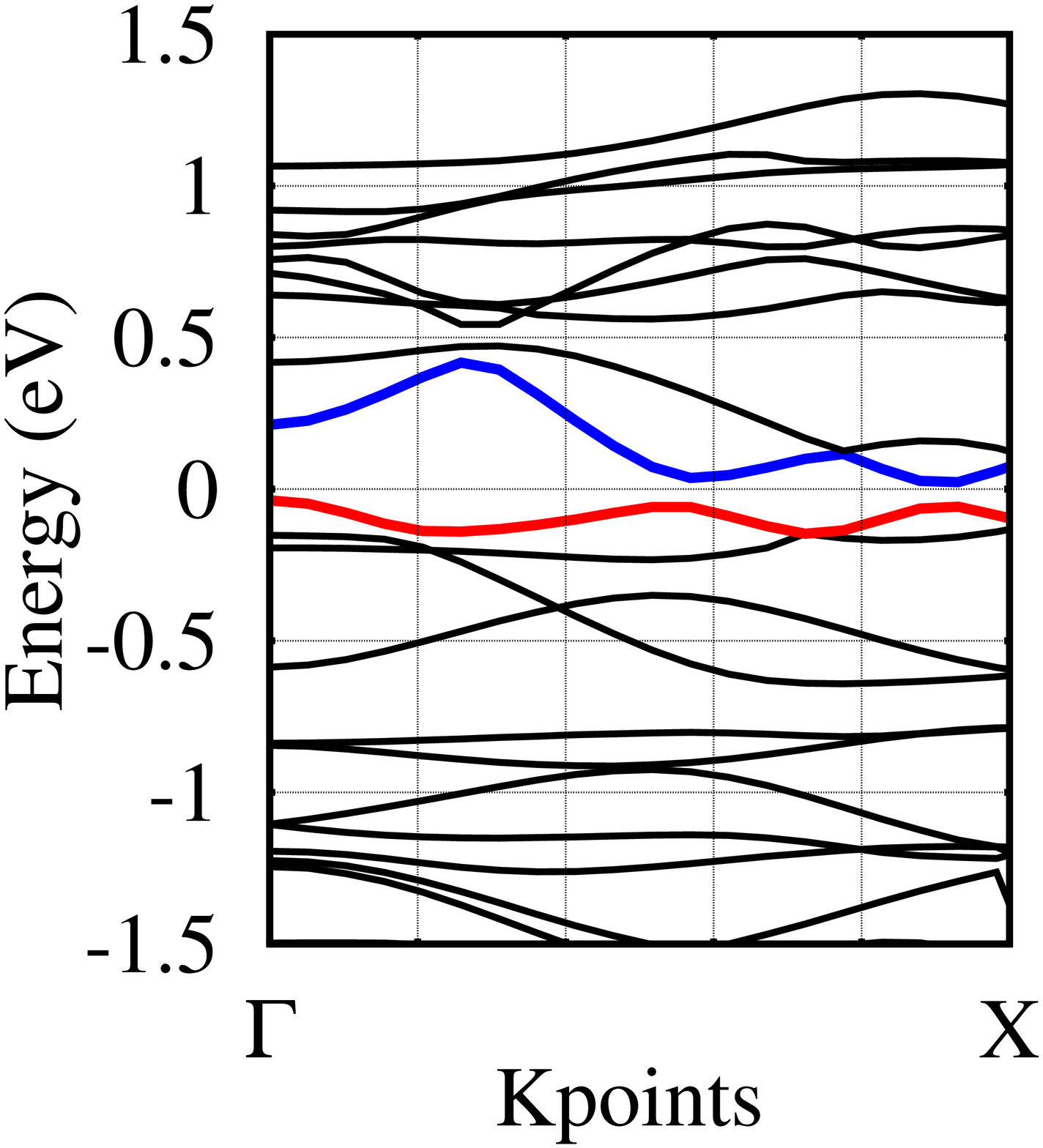}
 \includegraphics[width=0.15\textwidth]{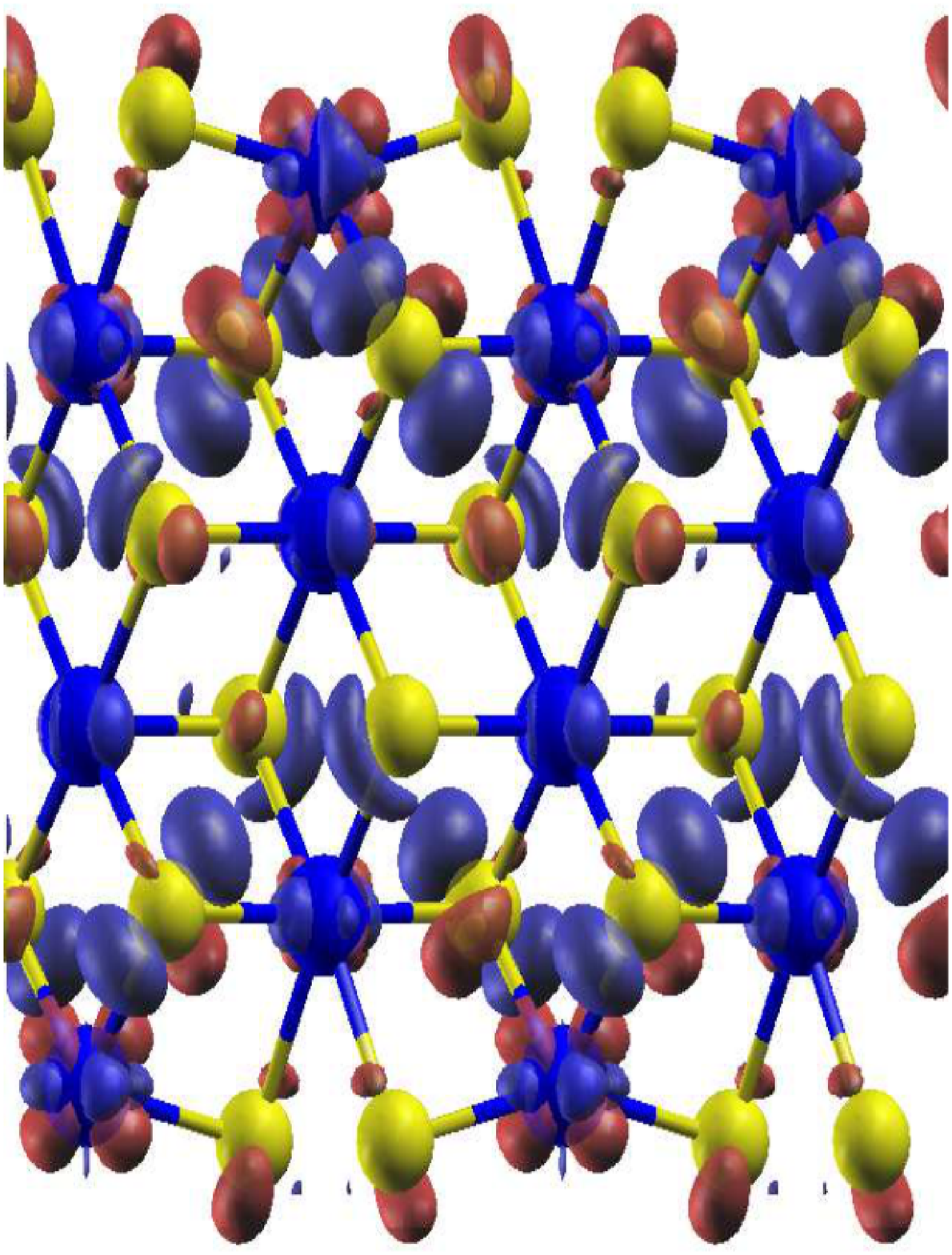}
 \includegraphics[width=0.09\textwidth,height=0.15\textheight]{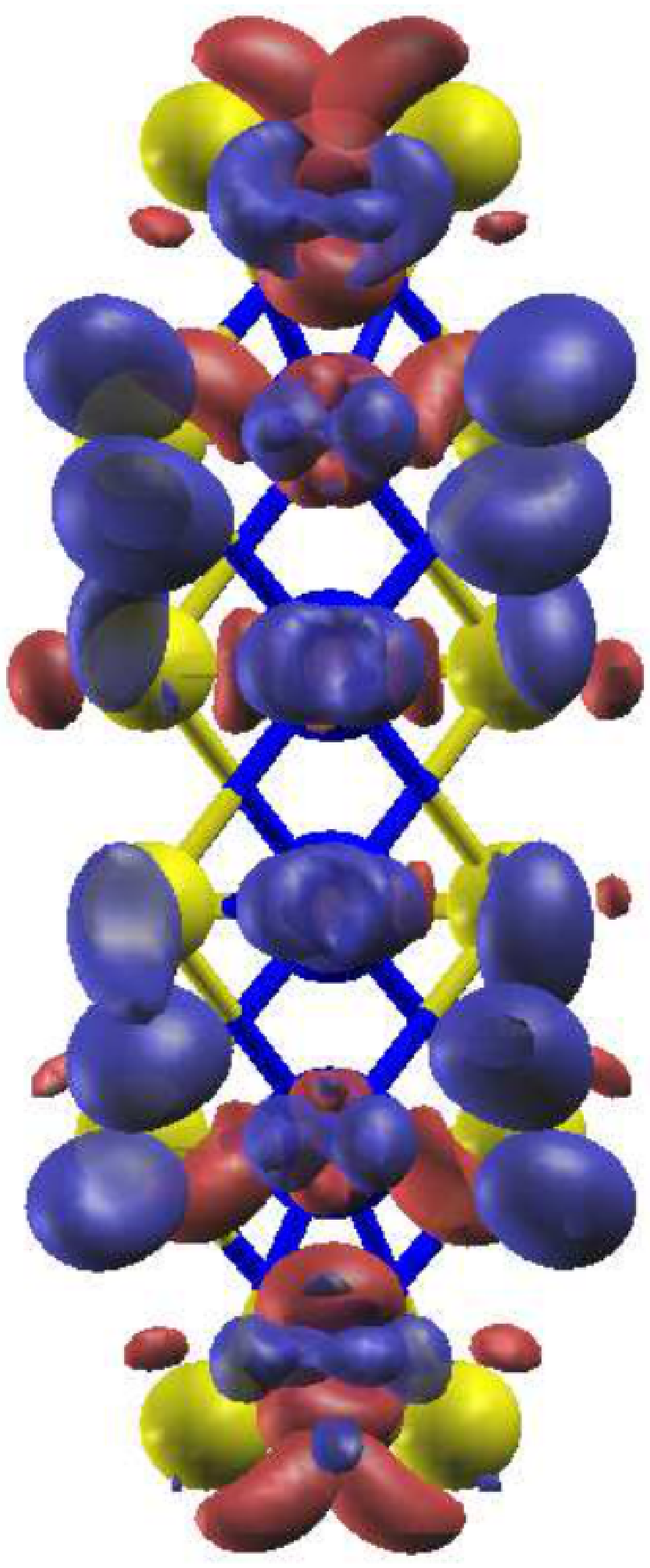}}
 \subfigure[Passivated Ni-centered-Se]{  \label{HNi-centered-Se}
 \includegraphics[width=0.20\textwidth]{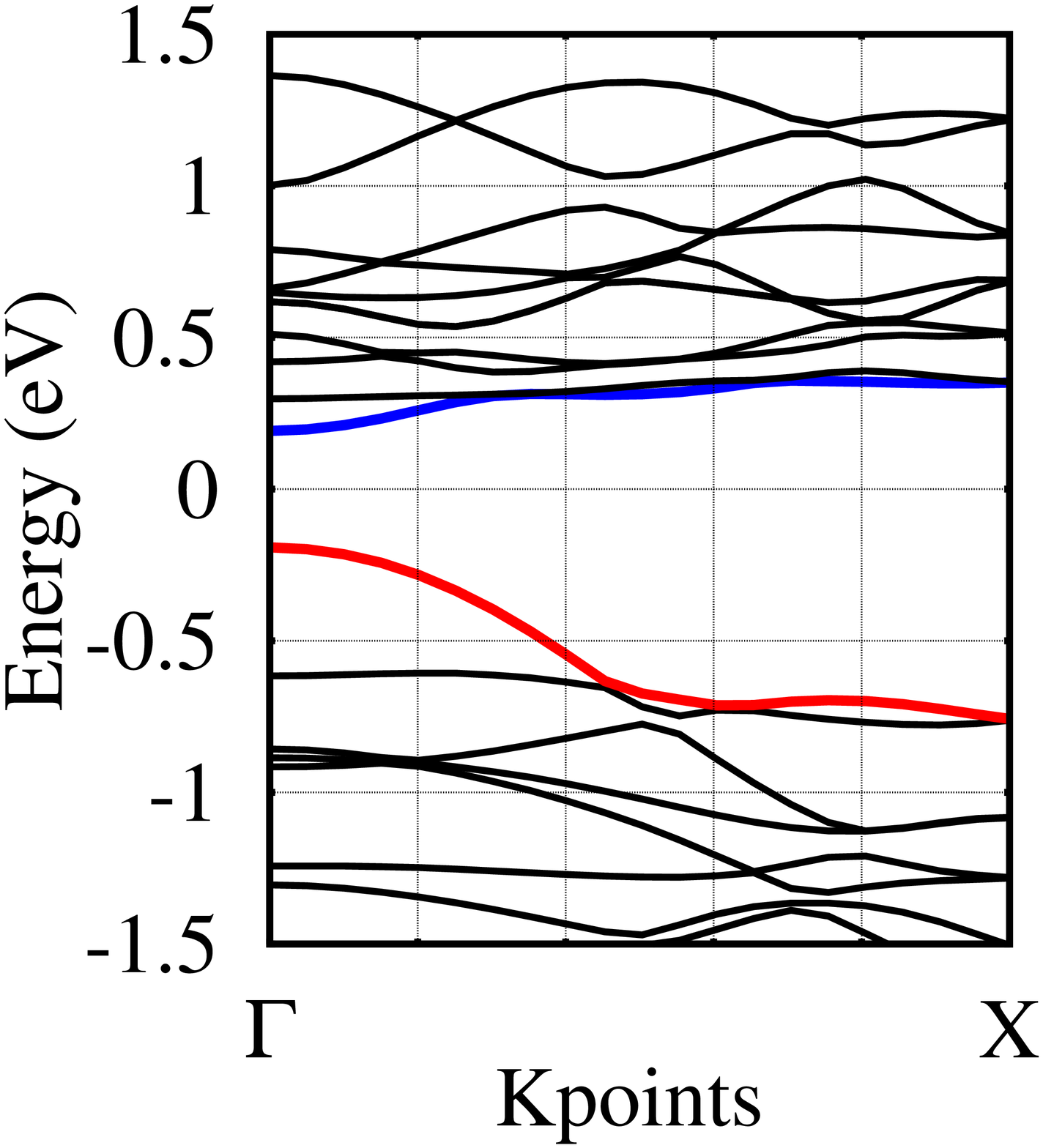}
 \includegraphics[width=0.15\textwidth]{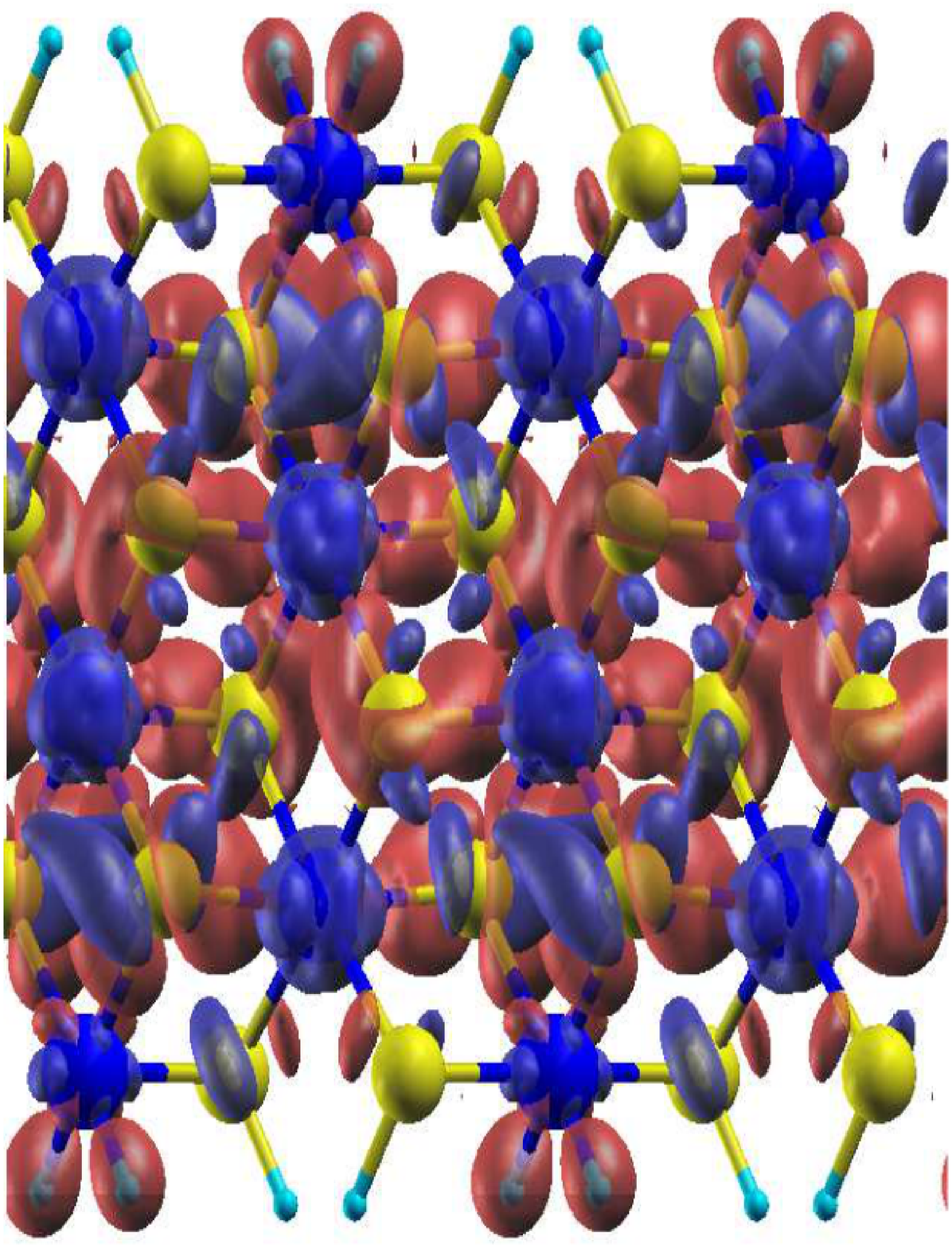}
 \includegraphics[width=0.09\textwidth,height=0.15\textheight]{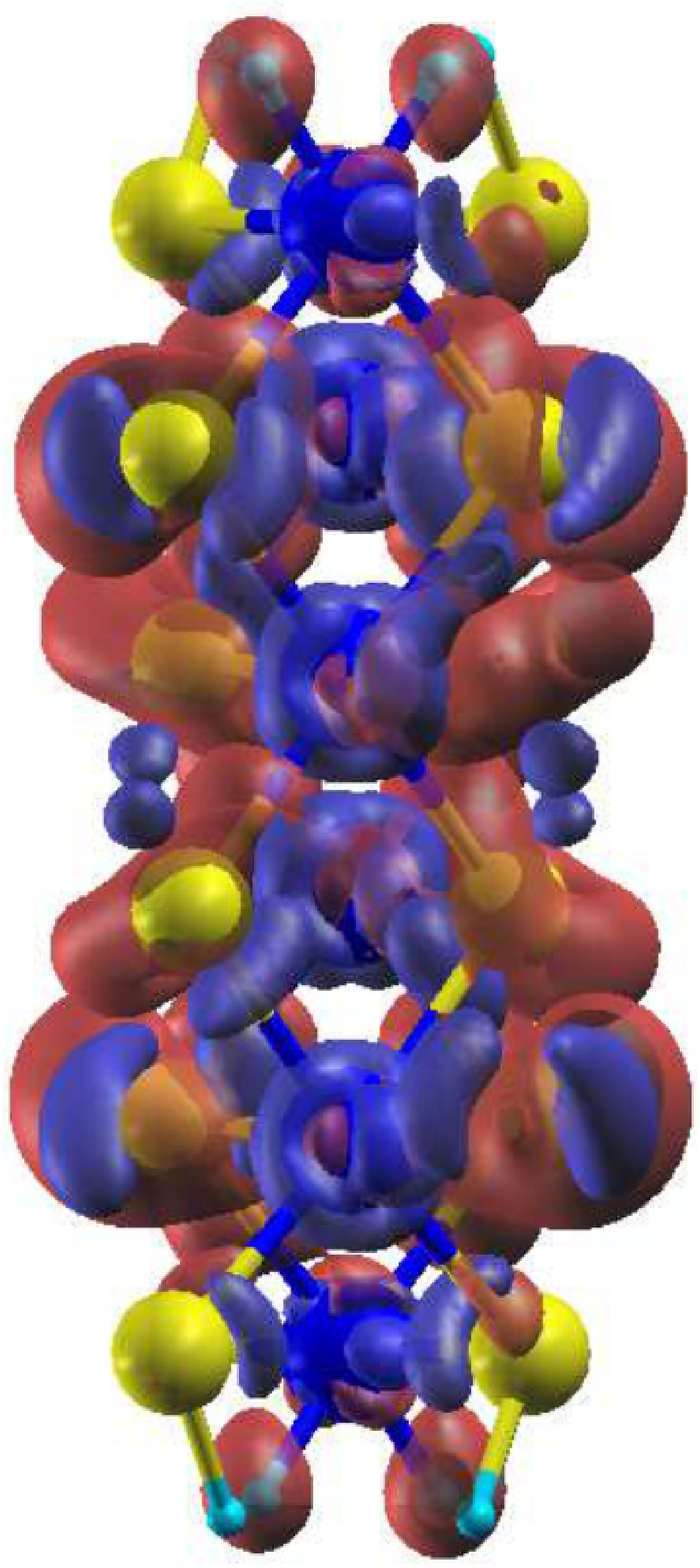}}
\caption{(color online) In each panel we present: the band structure with the maximum valance band (MVB) (red solid) and minimum conduction band (MCB) (blue solid), MVB (red) and MCB (blue) charge density in plane {\it xy} and plane {\it yz}. The isovalue for the surface is taken as 0.01 electron/\AA$^3$.  Large (blue), medium (yellow) and small (aqua) circles are Ni, Se and H atoms.}\label{Fig5}
\end{figure*}

 \ref{Fig5}  shows the band structure of the semiconductor \textit{root} systems with the highest occupied and the lowest unoccupied bands highlighted in red and blue respectively. The  highest occupied and lowest unoccupied orbitals are also illustrated in the figure with the same color code. The SeNiSe--SeNiSe \textit{root} ribbon presents a symmetrical distribution of both orbitals with respect of the $x$ axis. In contrast the SeNiSe--NiSeSe system present an accumulation of the top valence orbital on the ribbon's bottom SeNiSe edge while the bottom conduction orbital localizes on the ribbon's top NiSeSe edge. Both systems are indirect semiconductors.

Regarding the armchair ribbons, the band structure and highest occupied and lowest unoccupied orbitals are also shown. In this case all the ribbons present symmetry of the orbitals with respect of the $x$ axis and the band gaps are also indirects.

\section{Summary}

This work pioneers the \textit{ab-initio} studies of centered honeycomb nanoribbons, defining the representative ribbon's families and opening the filed to study other systems in the T structure. 
In this work, we use first principles calculations to explore the different atomic arrangements for NiSe$_2$ nanoribbons in the stable T crystalline structure. Zigzag bare-nanoribbons geometrical reconstruction leads to metallic systems, while just two hydrogen passivated nanoribbons are semiconductors. NiSeSe--SeSeNi has the largest metallic edge, suggesting that this ribbon will be the perfect candidate as catalyst for hydrogen evolution. SeNiSe--SeNiSe is the stablest zigzag nanoribon. Studies of edge  H passivation densities reveals that SeSeNi- edges are prone to convert to SeNiSe by a SeH$_2$ desorption mechanism.
Only thin bare-armchair nanoribbon are semiconductors with a very small band gap (0.045 and 0.061 eV), thicker ones are metals. When armchair ribbons are hydrogen passivated, the band gaps increase considerably to values up to $\sim$0.6 eV. The study of the variation in electronic band gap with the width was done and we report the asymptotic tendency to T-NiSe$_2$ band gap (0.11 eV) for armchair passivated nanoribbons. 

\acknowledgement
The authors thank DGTIC-UNAM. JARR acknowledges support from DGAPA-UNAM. GGN acknowledges support from DGAPA-UNAM project number 102513. This work was done under the support of FICSAC with the program: \textit{Atraction of visiting academics 2012} and by the Physics and Mathematics Department of UIA.

\bibliography{jpcc}

\end{document}